\documentclass[final]{article}

\usepackage{amsmath}    % need for subequations
\usepackage[english]{babel}
\usepackage{latexsym}
\usepackage{amssymb}
\usepackage{multirow}
\usepackage{graphicx}   % need for figures
\usepackage{float}
\usepackage{verbatim}   % useful for program listings
\usepackage{color}      % use if color is used in text
\usepackage{subfigure}  % use for side-by-side figures
\title{ Stochastic Modeling and Simulation of Ion Transport through Channels
\thanks{This work was partially
supported by  MIUR, the Italian Ministery of Education, University
and Research, within the project PRIN 2009RNH97Z-002- 2009: From
the microscale to macroscale in stochastic systems of interacting
individuals in population dynamics.}}

\author{D. Morale\footnotemark[2]\ \footnotemark[3]
,  M. Zanella  \footnotemark[4]  ,  V. Capasso
\footnotemark[3], 
 W. J\"ager \footnotemark[5]}

\begin{document}
\maketitle
%\slugger{mms}{xxxx}{xx}{x}{x--x}%slugger should be set to mms, siap, sicomp, sicon, sidma, sima, simax, sinum, siopt, sisc, or sirev

\renewcommand{\thefootnote}{\fnsymbol{footnote}}

\footnotetext[2]{  Department of Mathematics, University of Milan,
Via C. Saldini 50, 20133 Milan, Italy (daniela.morale@unimi.it) }

\footnotetext[4]{Department of Mathematics and Computer Science,
University of Ferrara, Via N. Machiavelli 35, 44121 Ferrara,
Italy}

\footnotetext[3]{ADAMSS, University of Milano,  Via C. Saldini 50,
20133 Milan, Italy  }

\footnotetext[5]{ Interdisciplinary Center for Scientific
Computing, Im Neuenheimer Feld 368, 69120 Heidelberg, Germany}

\begin{abstract}

Ion channels are of major interest and form an area of intensive
research in the fields of biophysics and medicine since they
control many vital physiological functions. The aim of this work
is     on one  hand  to propose a fully stochastic and discrete
model describing the main characteristics of a multiple channel
system.  The  movement of the ions is coupled, as usual,  with a
Poisson equation for the electrical field; we have considered, in addition, the influence of exclusion forces. On the other hand, we
have   discussed about the nondimensionalization of the stochastic
system by using real physical parameters, all supported by
numerical simulations. The specific features of both cases of
micro- and nanochannels  have been taken in  due
consideration  with particular attention to the latter case in
order to show that it is necessary to consider
a discrete and stochastic model for ions movement inside the channels.
\end{abstract}
%\begin{keywords}Stochastic differential equations, ion channels,
%stochastic processes, nano-micro channels, nondimensionalization
%\end{keywords}

%\begin{AMS} 60J99 , 60K35,     82C31, 82C22,     60H10
%\end{AMS}

\pagestyle{myheadings} \thispagestyle{plain}

\markboth{ D. Morale, M. Zanella, V. Capasso, W. J\"ager
}{Stochastic Modelling and Simulation in Ion Channels}

\section{Introduction}

  Ion channels are of major interest and form an area of intensive
research in the fields of biophysics and medicine, since they
control many vital physiological functions. Since certain aspects
of ion channel structure and function are hard or impossible to
address in real  experiments, mathematical models form a useful
addition, support and possible guidelines for future
investigations.

 \medskip

Ion channels are membrane proteins that catalyze the transfer of
ions down their electrochemical gradients across the plasma
membrane \cite{YJ}. They cannot perform thermodynamic work; that
is, they are not able to move an ion against its electrochemical
gradient. As a consequence, the direction of ions traveling
through an open channel is solely dependent on the electrochemical
gradient. Another characteristic of ion channels is that   the
rate at which ions move through these proteins is very high;   the
throughput of an ion channel can be fast up to 100 million ions
per second \cite{YJ}. On a fundamental level, the activity of ion
channels can change the membrane potential of the cell or alter
the concentrations of ions inside the cell. These processes are
basic to the cell physiology; therefore, it is easily imagined
that ion channels may be involved, directly or indirectly, in
virtually all cellular activities. While the role of some ion
channels is known, as for example for  calcium or  sodium
channels,  in other cases the physiological role of an ion channel
is unknown, and sometimes it is discovered when a disease arises.

\medskip

 { We  are aware  of the need of  analyzing many more features  of  a  system of ion channels, so that  the  model presented here  is not meant to be an
exhaustive description of all relevant phenomena which enter the
dynamical behavior of ion channels. Though, it can be seen as a
first step anticipating more complex, hence more realistic,
models.}

\medskip

 {A summary of the main features of ion channel
mechanism
 is the following \cite{YJ,ZT}: }

\medskip

a.  {\emph{The electrochemical gradient}}: This determines the
direction along which ions will flow through an open ion channel
and is a combination of two types of gradients: a concentration
gra\-dient and an electrical field gradient.  The electrical field
gradient takes into account the charge on the ion. The
relationship between the electrical potential and the magnitude of
the concentration gradient that is created is intuitive: the
stronger the electrical potential, the greater the concentration
gradient.

b.  {\emph{The  current-voltage relationship}}: The current is given by the movement of charges, and in cells the flow of ions through ion channels can be measured.

c.  {\emph{The anatomy of a typical ion channel}}: The ion
channels are proteins with a hole in their middle
  {Their walls contain ionizable
side chains  producing high charge densities \cite{DAD,EIS1,EIS}.}

d. {\emph{The ion selectivity}}: The capacity of the membrane to be
permeable to a class of ions.   The mechanism by which ion
channels pick and choose certain ions is still not fully
understood.  {Selectivity is produced by physical
forces that depend on the ion charge and size and chemical
interactions with parts of the channel protein that form what is
called 'the selectivity filter' \cite{EIS1,EIS}.}

e.  {\emph{The ion gating}}: The opening and closing
 mechanism of ion channels.

 {f.  {\emph{The solvent}}: Ions move in  water;
hence  a  realistic model should eventually  include  the fluid dynamics of the system. }

\medskip

 {In the present work we shall essentially consider
only the movement of  ions subject  to ion-ion interactions,
interaction with the geometry of the channels, and the coupling
with the electrical field which fluctuates, due to the strong
interaction with ions themselves. By taking all this into account, we have  already achieved a  high  level  of complexity of the system. It is clear that further steps will
be in the direction of including the selectivity and gating
processes, which are essentials in the channel dynamics.
Selectivity and gating are also influenced by the enormous
densities of fixed charges since channels are   small and walls
  contain ionizable side chains  and   carbonyl oxygens which produce an
effective charge density \cite{EIS1,EIS}. As mentioned later on,
we do not enter into such a level of description, that may be included in the
 modeling later on;  here we consider a mathematical schematization
 of the walls of the channels with which ions interact via specific potentials.
 Furthermore,  here
solvent effects enter through the dielectric and  the diffusion
coefficients only, but a more complete model should also include a
mathematical description of the solvent water. A proposed model
may be found in \cite{EIS2010}, where the authors couple a
continuum model for the ion density with a Navier--Stokes system
for the water treated as an incompressible fluid. It is also clear
that   salt water contains many families of particles, as
sodium, potassium, calcium, and chloride ions with different
physical characteristics, as, for example, different diameters.
For the sake of simplicity,  here we study the dynamics of   a single family of
ions, i.e. the diameter of ions is the same for all.    }

\medskip

There is widespread literature on molecular
dynamics simulation methodology  for studying ionic permeation in
ion channels, where authors use fully atomistic simulations of the
entire systems. The reader may refer, for example, to
\cite{AKC,CKC1,CKC2,DAD,Erban,ISR1,ISR2,MCKC,NSSE1,NFvSMH,smithsansom}.
However, Brownian dynamics may represent an attractive
computational approach for simulating the permeation process
through ion channels over long time-scales without having to treat
a system in all atomic details explicitly. The approach consists
of the generation of random trajectories of   ions as a function of time, by
numerically integrating stochastic equations for the motion using
some effective potential function to calculate the microscopic
forces operating among them \cite{Erban,ISR2}. This reduces the
dimensionality of the problem, although already very high, making
BD less computationally intensive than the corresponding molecular
dynamics  simulations. While various approaches for microscopic
modeling, either based on equations for the motion with forces
accounting for finite sizes \cite{CAKC,DKB,MCKC,MCKC2} or on
exclusion processes, have been investigated in detail, there are
still open problems in the transition to macroscopic models based
on partial differential equations. Some recent literature involves
some heuristic derivations of macroscopic dynamics
\cite{BSW,NSSE2,Schuss_nadler_eisenberg_2001}. Often away from
equilibrium standard equations with linear terms are used, whose
appropriateness remains unclear. In order to model some exclusion
principle, in \cite{BSW} the following modified
Poisson--Nernst--Planck equations have been obtained via a
heuristic derivation from an exclusion discrete process;
\begin{eqnarray*}
-\epsilon \Delta V &=&e\left ( \sum_j z_j c_j +f\right)\\
\frac{\partial}{\partial t} c_i&=& \nabla \cdot \left(D_i
(1-\rho)\nabla c_i + c_i \nabla\rho+ e z_i \mu_i c_i (1-\rho)
\nabla V \right))
\end{eqnarray*}
where $c_i$ is the concentration , $z_i$ the charge, $D_i$ the diffusion coefficient, and
$\mu_i$ the mobility of the ions of type $i$; $\rho=\sum_i c_i$ is
the total concentration, $V$ the electrical potential, $\epsilon$
the permittivity, and $e$ the elementary charge. Hence, the
authors introduce a size exclusion effect. The derivation of the
continuum model is not completely rigorous, so there still is a
lot of work that needs to be done in such a direction.
 A different approach used to deal with the finite
size of particle is via a Fermi like distribution, as in
\cite{EIS2014}.

\medskip

 As mentioned in \cite{NSSE1,NSSE2,NSS} while macroscopic
conservation laws clearly govern the macroscopic behavior of
ensembles of channels, their application to a single protein
channel able to contain at most one or two ions is questionable.
 The same authors in
\cite{Schuss_nadler_eisenberg_2001} consider a stochastic
treatment of the ions. Their aim is a reduction of the complexity
of solving the multidimensional  Fokker--Planck system associated
with the stochastic  Langevin system of all ions, coupled with the
Poisson equation for the electrical potential associated with all
ions. They perform their analysis upon the assumption of
stationary joint probability density of all ions,  and under the
additional simplification that  positive ions of the same species
are indistinguishable and interchangeable. Finally, they decoupled
position and velocity by considering the Smoluchowski limit of
large friction.   Among the problems that are left open,
they mention the size exclusion effect of ions and the
multiple channel case.  Here we  propose a first attempt in this
direction.
\medskip

The aim of the present work is, first of all, to consider a fully
stochastic model able to describe the main characteristics of a
multiple channel system, in which ion movement in the bath and
throughout   channels is described via a system of stochastic
differential equations,   coupled with a Poisson equation, which
becomes itself stochastic due to the dependence upon the ion
random  positions.  The treatment of multichannel systems becomes
important in the description when the number of channels
increases. In such living devices, ions are not  points and cannot overlap; as a consequence,
 atomic scale distances must be included explicitly in this multiscale approach \cite{EIS1,EIS2013}.
 Hence, exclusion forces are considered and modeled via Pauling and Lennard--Jones potentials. Pauling forces take into account  the interaction among  ions, while Lennard--Jones potential takes into account the interaction of moving ions with
the  boundary of  the channels, in particular their reflection at
the channels' boundary.   One of the main problems  so widespread
 in literature \cite{BSW,CKC1,EIS0,MRS,NSSE1,NSSE2,NSS} has been to link such a
completely stochastic model with an averaged continuum  model
described by partial differential equations.  Actually it is well
known that it becomes  reasonable to consider a continuous model
when laws of large numbers may be applied \cite{DM1,DM2,DM3}, i.e.
when the number of ions increases to infinity or the population is
large enough that an approximation may be applied. It is clear
that, while  this is always the case in the bath, this  is  not
always true in the channels; indeed, as already mentioned, in the
specific  case of nanochannels the dimension of the  ions is of
the same order  as  the one of a  channel, so that in a channel
the description of the system has to remain discrete and
stochastic. As a consequence, in the limit of infinite
(sufficiently large) number of ions, a problem of coupling their
dynamics outside (averaged continuum) and inside (discrete
stochastic) channels arises. We may refer to such a model as a
hybrid model, as in \cite{Moro2004}. The present work means to be
a first step in this direction; via a simulation analysis we
build up a  discrete stochastic model and understand the
role of each parameter.  As also mentioned in \cite{EIS2013}, some systems have macroscopic effects
that depend on atomic details. Indeed, at a later  stage our plan is to analyze
the asymptotic behavior of the system as the number  of  ions
increases to infinity or typical parameters characterizing the
system, as e.g. the size pore, vanish to zero. In literature we
may see a widespread interest on the latter topic, but  most of
the results refer to averaging problems of continuum deterministic
Poisson--Nernst--Planck equations \cite{Schmuck2013,Schmuck2014}.
The treatment of a fully stochastic case is still open.

\medskip

 Here  the issue of nondimensionalization of the
stochastic system and the choice of the right space-time rescaling
to catch the main features of the dynamics  has been addressed.
Since we have many parameters, we discuss also how we may
standardize some of them, and the role played by the others. Even
though the issue of nondimensionalization is very well known in
literature \cite{BSW}, the best of our knowledge there is no such explicit calculation, analysis, or similar discussion.
 The proposed space rescaling is strictly related
to the size of the channels and of the ions, while the time
rescaling depends on the possibility to treat the stochastic part
of the Langevin equation, described by the Wiener process.
  We discuss the role of different rescalings by
comparing the outcome of the dynamics obtained by numerical
simulations. In parti\-cu\-lar, we compare two different
rescalings by referring to specific length measures of the
channels. Both cases of micro- and nanochannels are considered.

\section{The Model}

We consider   a domain $\Omega=  \Omega_+ \cup \Omega_-\cup
\Omega_M\in \mathbb R^d,$ which consists of two regions of intra-
and extra-cellular medium,   $\Omega_+$ and $\Omega_-$, divided by
the membrane domain $\Omega_M$,   composed of many similar
channels; let us denote by   $\Gamma_{D_+}$ and    $\Gamma_{D^-}$
the superior and   inferior boundary of the bath, respectively.
  Typically, for a proper match with the biological
  case, we
  consider $d=3$, while  in order to
  reduce the complexity of the study, the simulation domain is such that
  $d=2$.
%the interfaces of the channels with the intra and extra cellular
%domains are $S ^+$ and $S ^-$, respectively, and the walls of the
%channels and the interface between the membrane and
%$\Omega_{\pm}$, denoted by   $C^M$ and by $S^M$.

\begin{figure}[h!]
\begin{center}
\includegraphics[scale=0.3]{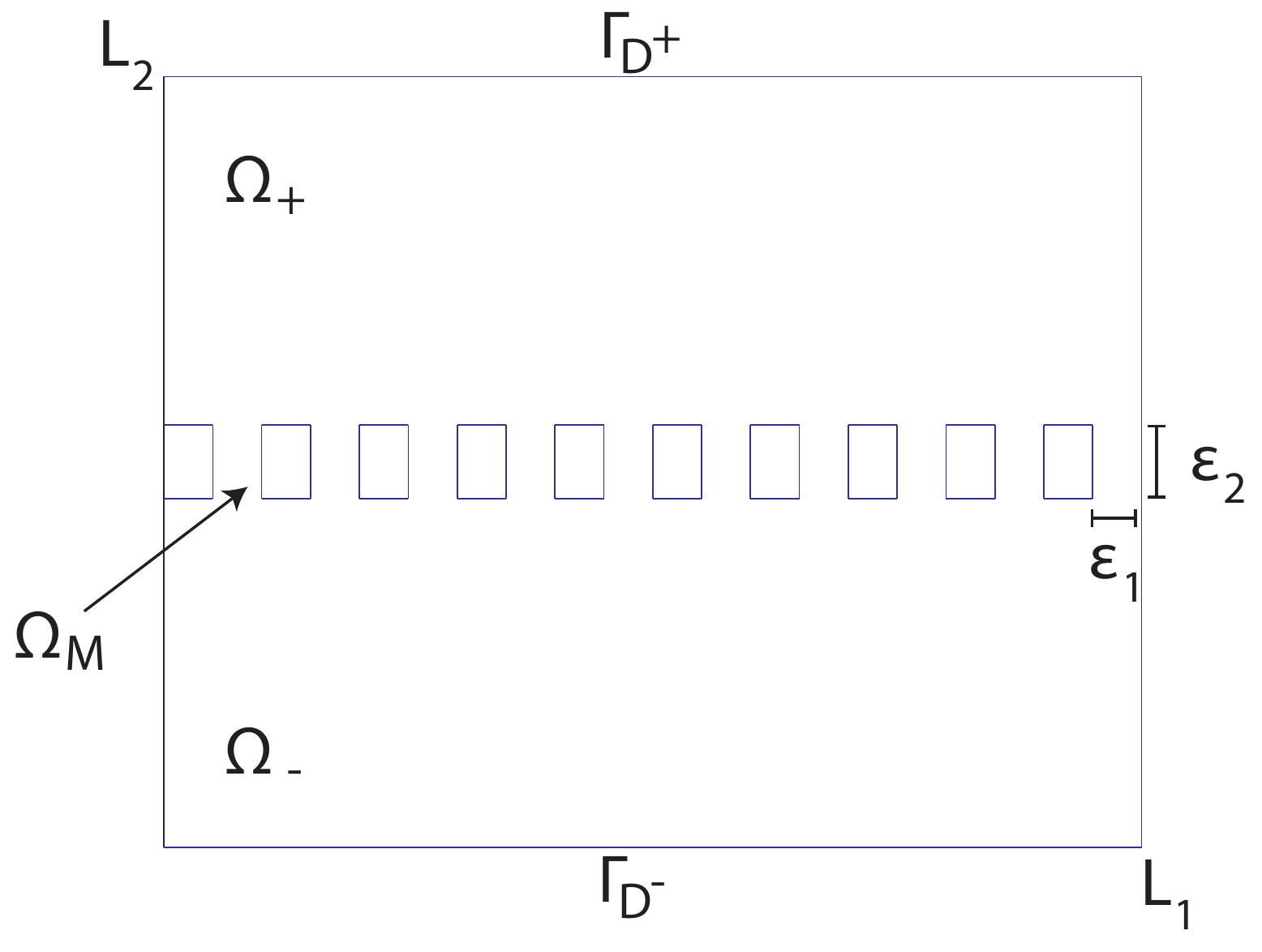}

\caption{Computational domain in $\mathbb R^2$, di dimension
$L_1\times L_2$ with dimension of the membrane $\epsilon_2\times
L_1$.}\label{domain_epsilon} \end{center}

\end{figure}

We suppose that the membrane   has dimension $ L_1 \times L_1
\times \epsilon_2\in \mathbb R^3$; the dimension of each pore is $
\epsilon_1\times \epsilon_1 \times \epsilon_2\in \mathbb R^3$;
hence, in the membrane we have  $\lfloor L_1/2\epsilon_1\rfloor
\in \mathbb{N}$ pores. So we have sketched each channel as a three-dimensional parallelepiped with typical dimensions $\epsilon_1$
and $ \epsilon_2$ ( $\epsilon_1 < \epsilon_2$). See Figure
\ref{domain_epsilon} for the two-dimensional case.

Ions move freely in the water, which is treated as a continuum, and their
motion through channels is driven by  an electrical field,
described as a gradient of an electrical potential.

\medskip

 We need to
describe the ions, their dynamics, the electrical field, and the
reciprocal interactions.

\subsection{The Variables}
\medskip

\emph{The Ions.} In general, we might suppose that  $K\in
\mathbb{N}$ is the total number of types of ions; out of them, the
first $K-1$ are free species and the $K$th is a species confined
in the membrane region, which creates the permanent charge of the
channels. As a consequence, in the domain we have a number   $N+J
=\sum_{k=1}^{K-1} N^{k} +J$   of ions,  where $N^k$ is the
number of ions of the $k$th species and $J$ the number of fixed
confined ions. The first $N$ ions are moving in $\Omega$ and each
of them is characterized by its Langevin coordinates, i.e.
$\left(X^{k,j}(t), V^{k,j}(t)\right)\in \Omega\times \mathbb R^3$,
being, respectively,  {\it the location and the velocity} of the
$j$th ion, of the  $k$th species, with $j=1,\ldots, N^{k}(t)$.
The latter $J$   fixed charges are characterized by their position
$Y^j \in \Omega_M, j=1,\ldots, J.$

\medskip

Here we consider a simplified model with $K=2$, that is the case
where only one population of ions is free of moving in $\Omega$
and the other one is composed by fixed charges, few in number,
located in the upper region of the channels \cite{H}.

\medskip

Hence, let  $N$ be the total number of moving ions in the  domain
$\Omega\subset\mathbb R^3$. Each ion is characterized by its
Langevin coordinates, i.e.
$$Z^j=\left(X^j, V^j\right)\in
\Omega\times \mathbb R^3, \quad  j=1,\ldots, N,$$ being,
respectively   the location and the velocity  of the $j$th ion
out of $N$. Furthermore, let $Y^j, j=1,\ldots, J$  be the position
of the $j$th fixed charge out of $J$.

\medskip

Each ion is also characterized by its occupied volume of order
${\tilde\epsilon}\,^3$, being $\tilde\epsilon$ its diameter. If
$\tilde\epsilon \sim \epsilon_1$, we are in the case of nanochannels, i.e.  a finite small number of ions may enter each channel, while if $ \tilde\epsilon \ll \epsilon_1 $, we are in the
case of microchannels.

\medskip

We consider the following counting measure defined on  the
location-velocity space $\Omega \times \mathbb R^3$
\begin{eqnarray}
\mu_Z(t)&=&\sum_{j=1}^{N}\epsilon_{ Z^j(t)}
=\sum_{j=1}^N\epsilon_{(X^j(t),V^j(t))}\in\mathcal{M}{(\Omega\times \mathbb R^3)},\label{def:mu_Z}\\
\nu_Y&=&\sum_{h=1}^{J}\epsilon_{ Y^h} \in\mathcal{M}{(\Omega)}.
 \label{def:mu_Y}\end{eqnarray}
 As a consequence, from \eqref{def:mu_Z}, the moving ion spatial counting process is given by
 \begin{eqnarray}
\nu_X(t)&=&\mu_Z(t)(\cdot \times \mathbb R^3) = \sum_{j=1}^{N}
\epsilon_{ X^j(t) }\in\mathcal{M}{(\Omega)}.\label{def:nu_Z}
\end{eqnarray}

Finally, we may also consider the following empirical measures
\begin{eqnarray}
X(t)&=&\frac{1}{N}\nu_X(t)\in\mathcal{M}_P{(\Omega)};\label{def:X_Z}\\
Y&=&\frac{1}{J}\nu_Y\in\mathcal{M}_P{(\Omega)}.\label{def:Y_Z}
\end{eqnarray}
In \eqref{def:nu_Z}-\eqref{def:Y_Z}
$\mathcal{M}{(\Omega)}$ and $\mathcal{M}_P{(\Omega)}$ denote the
spaces of (discrete) measures  and probability measures on
$\Omega$, respectively.

\medskip

 \emph{The
Electrical Field.} The electrical field $E(t,x)$ (in $V/m$) is
described by means of an electrical potential $\Phi (t,x)$ (in $V=
kg\,m^2/s^2\, C$), which is a continuous variable on
$\Omega_T=[0,T]\times\Omega$.

 \subsection{The Dynamics}
A  strong coupling between the variation of the distribution
charges and the variation of the electrical potential is shown.

\medskip

\emph{The Electrical Field.} The electrical potential $\Phi$ is
the solution of the Poisson equation
\begin{equation}\begin{split}
\label{eq:poisson} \textbf{}-\nabla \cdot \left(\alpha_w \nabla
\Phi (t,x)  \right) &= z~q~(K_{\kappa_1} * X(t))(x) +
z_F~q~(K_{\kappa_2} * Y)(x),
\end{split}\end{equation}
where  $\alpha_w $ is the  dielectric constant for the water (in
$F/m = C/Vm$), $z$ is the valency of the free ions, $z_F$ is the
valency of the fixed ions, and $q$ is the charge of ions (in $C$).
The smoothing functions $K_{\kappa_i}\in C_b^2(\mathbb R^3),
i=1,2$,   are such that
\begin{eqnarray}\label{eq:V_and_W}
K_{\kappa_i}(x)=\frac{1}{ \kappa_i} U_{i}(x), \quad x\in \Omega,
\end{eqnarray}
where, for any $i=1,2$, the function $U_i\in C_b^2(\mathbb R^3)$
has compact support equal to $\kappa_i$, i.e.
$|\textrm{supp}(U_i)|=\kappa_i$. In such a way, the quantities at the right-hand side of \eqref{eq:poisson} has the right dimension of a
concentration, i.e. quantity of charge per unit volume.
Furthermore, the convolution terms in \eqref{eq:poisson} may be
seen as an empirical concentration, justifying the dependence on
$N$,   via the empirical measure $X(t)$, given by \eqref{def:X_Z}.

\medskip

For any $t\in[0,T]$, the boundary conditions are given by
\begin{equation}\begin{split}
\label{boundary_condition}
\Phi(t,x)&=\Phi_1, \quad x\in \Gamma_{D^+};\\
\Phi(t,x)&=\Phi_2, \quad x\in \Gamma_{D^-};\\
\frac{\partial \Phi(t,x)}{\partial \nu}&=0, \qquad x\in
\partial\Omega\setminus(\Gamma_{D^+}\cup\Gamma_{D^-}).
\end{split}\end{equation}

\medskip

\emph{The  Moving Ions.}   The $N\in \mathbb N$ free ions move in
the surrounding water along a direction determined by the
electrochemical gradient; they interact with each other and with
the walls of the membrane. An external source of randomness is
introduced in the acceleration field, so that the evolution of the
ions state $Z^j(t)=(X^j(t),V^j(t))$, for
 $j=1,\ldots, N$
 is described by the following  Langevin model:
\begin{eqnarray} \label{Langevin_equation_nonscaled_1}
dX^j(t)&=& V^j(t)   dt,   \\
m dV^j(t)&=& -\left[m ~\gamma V^j(t) + z~q~\nabla \Phi (t,X^j(t)) \right.
\label{Langevin_equation_nonscaled_2}\\
&&\hspace{0.4cm} + F_{I}  \nabla \left( H_P^*[\nu_X(t)+\nu_Y]
\right)(X^j(t)) \nonumber \\
&&\hspace{0.4cm} \left. + F_{I}  \nabla H_{LJ}  (d_{\partial
\Omega_M}(X^j(t)) \right]\,dt +\sigma dW_t^{j}.\nonumber
 \end{eqnarray}

In System \eqref{Langevin_equation_nonscaled_1}-\eqref{Langevin_equation_nonscaled_2},
$m$ is the effective mass of the ions (in $kg$), $\gamma$ is the
friction coefficient per unit of mass (in $s^{-1}$), describing
the effect of the surrounding water molecules; they are related to
the diffusion coefficient $D$ (in $m^2/s$) deriving from random
collisions with water as shown by the Stokes--Einstein relation
\cite{RAAACA}
\begin{equation}
m\gamma = \frac{K_B T}{D},
\end{equation}

where $K_B$ is the Boltzmann constant ($1.30 \cdot 10^{-23} J/K
=1.30 \cdot 10^{-23} kg\, m^2/ s^2 \, K$) and $T$ is the absolute
temperature (in $K$).

\medskip

The stochastic effect is described by a multivariate  Wiener
process $({W}^1_t,\ldots,W_t^N)_{t\ge0}$ with independent
components. The parameter
 $\sigma $ (in $kg\, m /s\sqrt{s}$) is the  diffusion coefficient, acting upon
 $m V^j(t)$, for any $j=1, \ldots, N;$ it must satisfy the
 following relation with $\gamma$:
 \begin{equation}\label{eq:mgamma_sigma}
 m \gamma =
 \frac{1}{2K_B T} \sigma^2  = C_{KT} \sigma^2 .
 \end{equation}
Finally, the constant  $F_{I}$   (in $N$) represents the magnitude
of short range forces at contact. One estimate is given in
\cite{MCKC} and is of the order $10^{-10} N$.
\medskip

In   System
\eqref{Langevin_equation_nonscaled_1}-\eqref{Langevin_equation_nonscaled_2}
we also introduced an ion-ion interaction through the Pauling
potential $H_P$. Following \cite{MCKC}, we might consider
\begin{equation}\label{eq:H_paulii}
H_P(r)=\frac{(r_1+r_2)^{10}}{9r^9}.
\end{equation}

It represents a repulsive potential which arises from the overlap
of the electron clouds of the ions, $r$ is the ion-ion distance,
and $r_i, i=1,2,$ are the van der Waals radii of ions. Here, since
we consider only one single family of ions, $r_1=r_2$, we define
the operator $H_P^*[\nu_Z(t)+\nu_Y] (x) $ as the following:
\begin{equation}
H_P^*[\nu_X(t)+\nu_Y](x):=\sum_{k=1}^N H_P(|X^k(t)-x|)
+\sum_{j=1}^J H_P(|Y^j-x|),
\end{equation}
where $|x|$ is the norm of $x$ and $|x-y|$ denotes the distance
between $x,y \in \mathbb R^3$, that is
$$
|x-y|=\sqrt{\sum_{i=1}^3 (x_i-y_i)^2}.
$$

The walls of the membrane are made  of fixed
particles of a repulsive nature with respect to  the moving ions.
A mathematical schematization  to include the boundary conditions
has been made via a hard-wall potential, like   a truncated
shifted Lennard--Jones potential activated when the ions are at
fixed distance from the boundary
\cite{NFvSMH,EIS2014_comm,EIS2014,peletier}.   In \eqref{Langevin_equation_nonscaled_2} the quantity $d_{\partial
\Omega_M}(x)$ denotes the distance of a point $x\in \Omega$ from
the boundary of the membrane $\partial \Omega_M$, that is
\begin{equation}
d_{\partial\Omega_M}(x)=\min_{y\in\partial\Omega_M}\left|x-y\right|.
\label{def:rho}
\end{equation}
The truncated shifted Lennard--Jones potentials, as suggested in
\cite{NFvSMH}, has the following form:
\begin{equation}
H_{LJ}(r)=\left\{
\begin{array}{ll}
    \widetilde H_{LJ}(r)- \widetilde H_{LJ}(r_c), & \hbox{ }  r\le r_c;  \\
    0, & \hbox{ }   r> r_c,\\
\end{array}
\right.     \label{eq:tJL_potential_troncato}
\end{equation}
where $r$ is the distance between two particles and $\widetilde H$
is the Lennard--Jones potential which includes attraction-repulsion
effects
\begin{equation}
\widetilde H_{LJ}(r)= \varepsilon_{LJ} \left[ \left(\frac{ \tilde
\epsilon}{r}\right)^{12} - 2 \left(\frac{ \tilde
\epsilon}{r}\right)^{6}\right].
       \label{eq:tsJL_potential}
\end{equation}
In \eqref{eq:tsJL_potential}, $\varepsilon_{LJ}$ is  the well depth (in $m$) and a measure of how strongly the two particles attract each other; $ \tilde
\epsilon$ is the distance at which the intermolecular potential
between the two particles is zero; it gives a measurement of how
close two nonbonding particles can get and is thus referred to as
the van der Waals radius. Here we treat $\tilde \epsilon$ as the
diameter of the ions, i.e. it is of same order of the van der
Waals radius. Table \ref{table:vandewalls} shows the size of some
families of ions.

\medskip

%%%%%%%%%% TABELLA IONI  %%%%%%%%%%%%%%

\begin{table}[h!]\footnotesize
\begin{center}
\begin{tabular}{|c|c|}
\hline Ions & vdW radius\tabularnewline \hline \hline $Ca^{2+}$ &
$114\,\,pm$\tabularnewline \hline $Na^{+}$ &
$116\,\,pm$\tabularnewline \hline $K^{+}$ &
$152\,\,pm$\tabularnewline \hline $Cl^{-}$ &
$167\,\,pm$\tabularnewline \hline
\end{tabular}
\\[3mm]
 \caption{Van der Waals radii $\widetilde \epsilon$
 for some families of ions in picometers ($10^{-12}m$).}
  \end{center}
 \label{table:vandewalls}
\end{table}

%%%%%%%%%%%%%%%%%%%%%%%%%%%%%%%%%%%%%%%%%%%%%%%%%%%%%%%%%%%%%%%%%%%%%

%

\medskip

If we consider the truncated Lennard--Jones potential
\eqref{eq:tJL_potential_troncato} with a cutoff $x_c=\tilde
\epsilon$, only the repelling part of the potential $\widetilde
H_{IJ}$ is taken into account; indeed,
\[
\widetilde H_{LJ}\left(\tilde\epsilon\right)=-\varepsilon_{LJ},
\]
and (\ref{eq:tJL_potential_troncato}) becomes
\begin{eqnarray}\label{eq:troncata2}
{H}_{LJ}(r)&=& \left( \widetilde H_{LJ}(r)+\varepsilon_{LJ}\right)
1_{(0,  \tilde\epsilon]} (r) \\
&=& - \varepsilon_{LJ} \left\{1- \left[ \left(\frac{\tilde
\epsilon}{r}\right)^{12} -  2\left(\frac{\tilde
\epsilon}{r}\right)^{6}\right]\right\}1_{(0,\tilde\epsilon]} (r).
\nonumber\end{eqnarray}

The choice of $\varepsilon_{LJ}$ influences how much the potential
$H_{LJ}$ tends to hard-sphere potential $H_{hs}$. For the
truncated Lennard--Jones potential the choice $\varepsilon_{LJ}=1$
describes    relatively ``soft'' molecules; that is, molecules can
partially overlap during collisions, whereas if they are closer
than $\tilde\epsilon$ they start to repel each other. By increasing
$\varepsilon_{LJ}$, a hard-sphere behavior is mimicked.

\medskip

\medskip

%%%%%%%%%%%%%%%%%%%%%%%%%%%%%%%%%%%%%%%%%%%%%%%%%%%%%%%%%%%%%%%%%%%%%%%%%%%%%%%
\begin{table}[h!]\footnotesize
\begin{center}
\begin{tabular}{|c|c|c|}
\hline Parameter & Value & Measure units \\ \hline \hline
$q$ & $1.6\times10^{-19}$ & $C$\\
\hline
$m$ & $6.5 \times  10^{-26}$ & $Kg$\\
\hline
$D$ & $1.33\times10^{-9}$ & $m^{2}\; s^{-1}$\\
\hline
$\epsilon_1, \epsilon_2$ & $10^{-9}-10^{-7}$ & $m$\\
\hline
$F_I$ & $2\times10^{-10}$ & $N$\\
 \hline
$K_B$ & $1.38\times10^{-23}$ & $J\; K^{-1}$\\
\hline $T$ & $300$ & $K$\\\hline
$\alpha_{w}$ & $7.08\times10^{-10}$ & $C\; V^{-1}\; m^{-1}$\\
 \hline
 $\alpha_{0}$ & $8.85\times10^{-12}$ & $C\; V^{-1}\;
m^{-1}$\\
\hline
\end{tabular}
\\[3mm]
 \caption{List of the parameters used in the model. For the one
related to ions, refer to the case of $K^+$.}
 \end{center}\label{table:parameters}
\end{table}

%%%%%%%%%%%%%%%%%%%%%%%%%%%%%%%%%%%%%%%%%%%%%%%%%%%%%%%%%%%%%%%%%%%%%
\medskip

In Table \ref{table:parameters} the value of the model parameters
 is listed, in the case of $K^+$. We observe  that their order of magnitude
  may be very different; in particular,   mass $m \sim O(10^{-26}),$ while the friction term
 $m \gamma\sim O(10^{-12}),$ the charge $q\sim O(10^{-19})$, and the interaction
  $\sim O(10^{-10})$.

\medskip

It is clear that in order to better manage  System
\eqref{eq:poisson} and
\eqref{Langevin_equation_nonscaled_1}-\eqref{Langevin_equation_nonscaled_2},
a nondimensionalization procedure is needed.

\medskip

\section{Nondimensionalization of the system}
\label{sec:nondim}

We consider the nondimensionalization starting with a time-space
rescaling, via some  typical   scale parameters $(t_0, \epsilon)$.

\subsection{Time, Space, positions, and Velocities}

Let  $(x_s,t_s)$ be the scaled space-temporal coordinates, such
that
\begin{equation}\label{eq:time_space_rescaling}
t = t_0 ~ t_s \qquad  x = \epsilon ~ x_s,
\end{equation}
where $\epsilon\in \mathbb R $ and $t_0\in \mathbb R_+$ are
scaling parameters that will be chosen later on. As a consequence
of \eqref{eq:time_space_rescaling} the time dependent position and
velocity are scaled as follows:
\begin{equation}\begin{split} \label{scaling_space_time}
x_s(t_s) &= \epsilon^{-1} x(t_0^{-1} t)\\
v_s(t_s) &= \frac{t_0}{\epsilon} v(t_0^{-1} t).
\end{split}\end{equation}

We denote by $\Omega^s=\Omega^s_+\cup \Omega^s_-\cup \Omega^s_M$
the rescaled domain with the three different regions, and by
$\Gamma^s_{D^+}, \Gamma^s_{D^-}$ the superior and inferior
boundary of the rescaled bath, respectively.

\medskip

Given the Langevin coordinate of the moving ions at time $t_s$,
for $j=1,\ldots, N,$ $$Z_s^j(t_s)=\left(X_s^j(t_s),
V_s^j(t_s)\right)\in \Omega^s\times \mathbb R^3, $$ and the
position of the fixed ones $Y_s^i$, $i=1,\dots,J$, let us denote by
$\nu_{X}^s(t_{s}), \nu_{Y}^s,$ the corresponding counting
processes at time $t_s$, that is
$$
\nu_X^{s}(t_{s})={\displaystyle
\sum_{j=1}^N\varepsilon_{X_{s}^{j}(t_{s})}} ,  \qquad
\nu_{Y}^s={\displaystyle \sum_{i=1}^J\varepsilon_{Y_{s}^{i}}},
$$
and  by $X^s(t_{s}), Y^s$ the corresponding empirical measures
$$X^s(t_{s})=\frac{1}{N}\nu_X^{s}(t_{s}), \qquad  Y^s=\frac{1}{J} \nu_{Y}^s.$$

\subsection{The Electrical Field}
 One might obtain a dimensionless equation for the potential $\Phi$
in several ways. Scaling the electrical potential means finding a
scaling parameter $\widetilde \Phi$ such that the potential field $\Phi_{s}(t_s,x_s)$ in the new coordinates is
\begin{equation}
\Phi_{s}(t_s,x_s)=   \frac{1}{\widetilde \Phi} \Phi (t_0 t_s,
\epsilon x_s). \label{eq:scaling_potential}
\end{equation}
Proper techniques of semiconductor
literature can be used for ion channels \cite{EIS2012,MRS}.  Indeed, being ion channels provided of few fixed charges inside the channels, the applied difference of potential
is approximately close to the thermal voltage. This fact
represents a bridge between ion channels literature and the studies
of semiconductors \cite{SH}.  A way to rescale $\Phi$ is  the following:
$$\widetilde \Phi=\frac{K_BT}{q}\approx 0.39  \;J/C.
$$ If the
variation of $\Phi$ is  smaller than $ {K_BT}/{q}$, the diffusion
prevails, otherwise the advection
dominates. The chosen scaling corresponds to a limit in which both
advection and diffusion are in balance.

\subsection{Rescaling the Equations}

We will consider a nondimensionalization of \eqref{eq:poisson} and
\eqref{Langevin_equation_nonscaled_1}-\eqref{Langevin_equation_nonscaled_2},
based on the transformation proposed in
\eqref{scaling_space_time}-\eqref{eq:scaling_potential}.
 Details of
the nondimensionalization may be found in the appendix.

\medskip

\paragraph{The Poisson Equation}

The scaled   Poisson  equation  is
\begin{equation}\label{eq:rescaled_poisson}\begin{split}
-\textrm{div}_{x_{s}}{\left(\alpha_{s}\nabla_{x_{s}}\Phi_{s}\left(t_{s},x_{s}\right)\right)}
&=  \lambda^\epsilon z
\left(K_{\kappa_1^s}^{s}*X^{s}(t_s)\right)(x_{s})    +
\lambda^\epsilon z_{F}
 \left(K_{\kappa_2^s}^{s}*Y^{s}\right)(x_{s}).
\end{split}\end{equation}
In \eqref{eq:rescaled_poisson}, the kernels
 $K_{\kappa_i^s}, i=1,2$ have been scaled as follows:
\begin{equation}\label{eq:rescaled_potential_K_W}
K_{\kappa_i^s}(x_{s}) =\frac{1}{\kappa_i^s} U_{i}^{s}(x_{s})=
\frac{1}{\kappa_i^s} U_{i}\left(\epsilon x_{s}\right),
\end{equation}
and  the relative electrical permittivity $\alpha_s$ is such that
\begin{equation}\label{eq:rescaled_alpha}
\alpha_{w}=\alpha_{s} \alpha_{0},
\end{equation}
where $\alpha_{w}$ is the typical electrical permittivity of the
water,   and $\alpha_{0}$ is the electrical permittivity of the
free space. For any $i=1,2$, the parameter $\kappa_i^s$ is obtained
by the rescaling of the support $\kappa_i$ of $U_i$, that is $
\kappa_i=\epsilon^3 \kappa_i^s. $ See, again, Table
\ref{table:parameters} for their physical dimensions.

\medskip

Finally, the parameter $\lambda^\epsilon$ is given by
\begin{equation}\label{eq:lambda}
\lambda^\epsilon=\dfrac{q}{\alpha_{0}\tilde{\Phi}\epsilon}=\dfrac{q^{2}}{\alpha_{0}K_BT}\frac{1}{\epsilon}.
\end{equation}
Hence, it depends on the scaling length $\epsilon$. Note that,
coherently with our purpose, the coefficient $\lambda^\epsilon$ is
dimensionless, in fact we have
\begin{equation}
\dfrac{C^{2}}{\dfrac{C^{2}}{m\cdot J}\; JK^{-1}\; K\; m}=1.
\end{equation}

Consistently with \eqref{eq:scaling_potential}, in the new
reference scale the boundary conditions \eqref{boundary_condition}
become
\begin{equation}\begin{split} \label{boundary_cond}
\Phi_s(t_s,x_s)&=\dfrac{\Phi_1}{\tilde{\Phi}}=\dfrac{q}{K_B
T}\Phi_1=\Phi_{1,s},
\qquad x_s\in \Gamma^{s}_{D^+}; \\
\Phi_s(t_s,x_s)&=\dfrac{\Phi_2}{\tilde{\Phi}}=\dfrac{q}{K_B
T}\Phi_2=\Phi_{2,s},
\qquad x_s\in \Gamma_{D^-}^{s}; \\
\dfrac{\partial \Phi_s}{\partial \nu}(t_s,x_s)&=0, \qquad x_s \in
\partial\Omega_{M,s}\setminus(\Gamma_{D^{+}}^{s}\cup\Gamma_{D^-}^{s}).
\end{split}\end{equation}

\paragraph{The Langevin system}
The nondimensionalized Langevin system is, for any $j=1,\ldots,N$,
\begin{eqnarray}
dX_{s}^j\left(t_{s}\right)&=&V_{s}^j\left(t_{s}\right)dt_{s}, \label{eq:labgevin_rescale_1_first}\\
m_{s}dV_{s}^j\left(t_{t_{s}}\right)
&=&-\left[\lambda_{1}m_{s}\gamma_{s}V_{s}^j\left(t_{s}\right)\right.
\label{eq:labgevin_rescale_2_first}
 +
\lambda_{2}\nabla_{x_s}\Phi_{s}\left(t_{s},X_{s}^j(t_{s})\right)    \\
&& \hspace{0.3cm}\left.+\lambda_{3}\nabla_{x_s}{H_{P}^{s,*}}
\left[\nu_X^{s}(t_s)
+\nu_Y^{s}\right](X_s^j(t_{s}))\right.\nonumber\\
&&\hspace{0.3cm}\left.+\lambda_{3}\nabla_{x_s}H_{LJ}^{s}\left(d_{\partial{\Omega_M^{s}}}(X_s^j(t_{s}))\right)
\right]dt_{s}\nonumber\\
&&+\lambda_{4}\sigma_{s}dW_{t_s}.\nonumber
\end{eqnarray}
In the previous system $m_s, \sigma_s $, and $ \gamma_s$ are such
that $ \gamma=\bar{\gamma}\gamma_{s},
\sigma=\sigma_{s}\bar{\sigma},   m= M m_s,
$
and
\begin{equation}\label{eq:sigma_rescaled}
\sigma_s=\sqrt{2m_s \gamma_s}, \quad
\bar{\sigma}=\sqrt{K_BTM\bar{\gamma}}.
\end{equation}
The functions $H_P^s,H_{LJ}^s $ are the scaled  Pauling and
Lennard--Jones potential
\begin{eqnarray}\label{eq:rescale_LJ}
\widetilde{H}_{LJ}^{s}\left(x_{s}\right)&=&\varepsilon_{LJ}^{s}
\left[\left(\dfrac{2R^{s}_{vdW}}{x_{s}}\right)^{12}-2\left(\dfrac{2R^{s}_{vdW}}{x_{s}}\right)^{6}\right],
\\\label{eq:rescaled_pauling}
H^{s}_{P}\left(r_s\right)&=&\dfrac{\left(r_{1,s}+r_{2,s}\right)^{10}}{r_s^{9}}.
\end{eqnarray}

 Finally,  the dimensionless coefficients
$\lambda_{i},i=1,\dots,4$ in the Langevin system are
\begin{equation}%\begin{split}
\label{eq:rescaled_parameter} \lambda_{1} = t_{0}\bar{\gamma},
\qquad  \lambda_{2}= \dfrac{t_{0}^{2}K_BT}{M\epsilon^{2}}, \qquad
\lambda_{3}=\dfrac{F_{I} }{M}\dfrac{ t_{0}^{2}}{\epsilon},\qquad
\lambda_{4}=\dfrac{t_{0}\sqrt{t_{0}}}{M\epsilon}\sqrt{K_BTM\bar{\gamma}}.
 \end{equation}

So the nondimensionalized system has five  parameters
\eqref{eq:lambda} and \eqref{eq:rescaled_parameter}, depending on
the time and space characteristic lengths $t_0$ and $\epsilon$.

\subsection{Reduction of the Parameters}

In order to reduce the parameters, first of all we impose
$\bar{\gamma}=t_0^{-1}$, so that also $\lambda_{1}=1$.
Furthermore, we choose the scaling parameter $t_0$ such that the
random term is a Wiener process with diffusion coefficient
$\sigma_s$; this is equivalent imposing that is $\lambda_{4}=1$.
It follows that
\begin{equation}
t_{0}=\epsilon\sqrt{\dfrac{M}{K_{B}T}};
\label{eq:scaling_parametr_t_0}
\end{equation}
as a consequence, $ \lambda_{2} =1.$  Hence, by
\eqref{eq:scaling_parametr_t_0}, we have standardized three of the
five parameters.

\medskip

In conclusion, the nondimenzionalized system is given by the
Poisson equation \eqref{eq:rescaled_poisson}-\eqref{eq:lambda} and
the following Langevin system:
\begin{eqnarray}
dX_{s}^j\left(t_{s}\right)&=&V_{s}^j\left(t_{s}\right)dt_{s},\label{eq:scaled_langevin_1} \\
m_{s}d_{t_{s}}V_{s}^j\left(t_{t_{s}}\right) &=&-\left[
m_{s}\gamma_{s}V_{s}^j\left(t_{s}\right)\right.
\label{eq:scaled_langevin_2}  +
 \nabla_{x_s}\Phi_{s}\left(t_{s},X_{s}^j(t)\right)   \\
&&\hspace{0.3cm}\left.+\lambda_{3}^\epsilon\,\nabla_{x_s}{H_{P}^{s,*}}
\left[\nu_X^{s}(t_s)
+\nu_Y^{s}\right](X_s^j(t))\right.\nonumber\\
&&\hspace{0.3cm}\left.+\lambda_{3}^\epsilon\,\nabla_{x_s}H_{LJ}^{s}\left(d_{\partial{\Omega_M^{s}}}(X_s^j(t))\right)
\right]dt_{s}\nonumber\\
&&+\sigma_{s}dW_{t_s},\nonumber
\end{eqnarray}
with
\begin{equation}\label{eq:parameters_lambda_3}
\lambda_{3}^\epsilon=F_{I}\dfrac{M\epsilon^{2}}{K_BTM\epsilon}=\dfrac{F_{I}}{K_BT}\epsilon.
\end{equation}

\subsection{The Choice of the Spatial Scale}

Experimental methods for determining the physical dimensions of
ion channels have been widely investigated during the last few decades
\cite{H}. One of the most studied classes of channels is the one
relative to potassium ions $K^+$, which exhibit common
permeability characteristics. A precise description of their
structure is presented in \cite{DAD}. They are mostly composed by
a highly selective porous protein located into a lipid bilayer,
the total length of the pore is $\epsilon_2 \sim 45 \AA$ and its
diameter $ \epsilon_1$ varies along the channel, assuming its
maximum size into a cavity of $\sim 10 \AA$ placed in the middle
of the membrane. Then the ions $K^+$ move through the pore and
remain hydrated. Besides the class of biological channels there
exist artificial nanochannels; see \cite{CJSMNG} and \cite{VSS},
radii and lengths of which range from $100\cdot 10^{-9}$m to
$10^{-9}$m. As a consequence we may consider  two different
scaling regimes taking into account the possible real dimension of
the pore, i.e. either  $ \epsilon_1\sim 10^{-9}$ or $
\epsilon_1\sim 10^{-7}$.
\\ \\
Let us consider now the magnitude of  $\lambda^\epsilon$ and
$\lambda_3^\epsilon$ in the stochastic system, using the data for
a typical ion channel in Table \ref{table:parameters}; from
\eqref{eq:lambda} and \eqref{eq:parameters_lambda_3}, we obtain
$$
\lambda^\epsilon \simeq \frac{7\cdot 10^{-7}}{\epsilon},\qquad
\qquad \lambda_{3}^\epsilon \simeq 4.8\cdot10^{10}\epsilon.
$$

  Thus, if we nondimensionalize with the
typical size of the neck of the channel $\epsilon_1 = O(10^{-9} m)
$, i.e. $\epsilon= \epsilon_1$, we obtain
\begin{equation}
\lambda^{\epsilon}= \lambda^{ \epsilon_1}=7 \cdot10^2, \qquad
\lambda_3^{\epsilon}=\lambda_3^{\epsilon_1}=4.8\cdot 10,
\label{eq:parameters_L_vanderwalls-9}
\end{equation}
 whereas if we nondimensionalize with   $ \epsilon_1 =
O(10^{-7} m)$,  the involved parameters become
\begin{equation}
\lambda^{\epsilon}=\lambda^{ \epsilon_1}= 7, \qquad
\lambda_3^{\epsilon}=\lambda_3^{\epsilon_1}=4.8\cdot 10^3.
\label{eq:parameters_L_vanderwalls_-7}
\end{equation}

Next step is to compare the two nondimensionalized systems in
order to catch the different   features of the dynamics.

%%%%%%%%%%%%%%%%%%%%%%%%%%%%%%%%%%%%%%%%%%%%%%%%%%%%%%%%%%%%

\section{Numerical Experiments}

 Simulations of systems
\eqref{eq:rescaled_poisson} and
\eqref{eq:scaled_langevin_1}-\eqref{eq:scaled_langevin_2} are
performed in  the MATLAB environment, in which we have simulated
explicitly the Langevin system
\eqref{eq:scaled_langevin_1}-\eqref{eq:scaled_langevin_2}, while
we used the package PDETool for the discretization of the Poisson
equation \eqref{eq:rescaled_poisson}.

\medskip

\paragraph*{The Computational Domain}
As already mentioned, we consider numerical results for $d=2$.
Denoted by $\epsilon_1$ and $\epsilon_2= k \, \epsilon_1$, $k\in
\mathbb N$ the neck and depth of the channel, respectively, the
not rescaled domain,  shown in Figure \ref{domain_epsilon}, has
dimensions given by $L_1=    2 n_c \epsilon_1, $ and $L_2= (2 m +
1) \epsilon_2, m\in \mathbb N$.  $n_c\in \mathbb N$ is the number
of channels in the membrane.

\begin{figure}[h!]
\begin{center}
\includegraphics[scale=0.25]{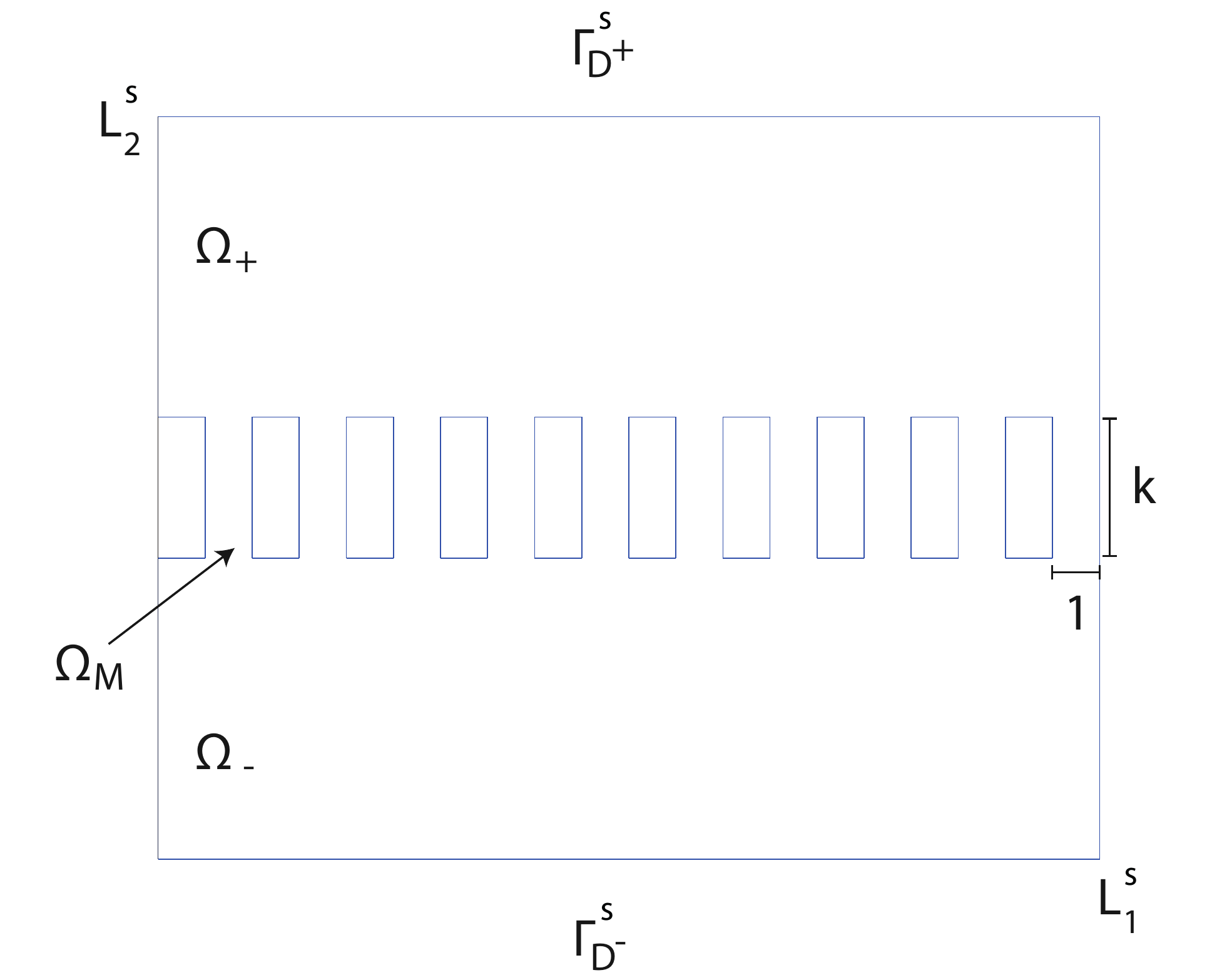}
\includegraphics[scale=0.25]{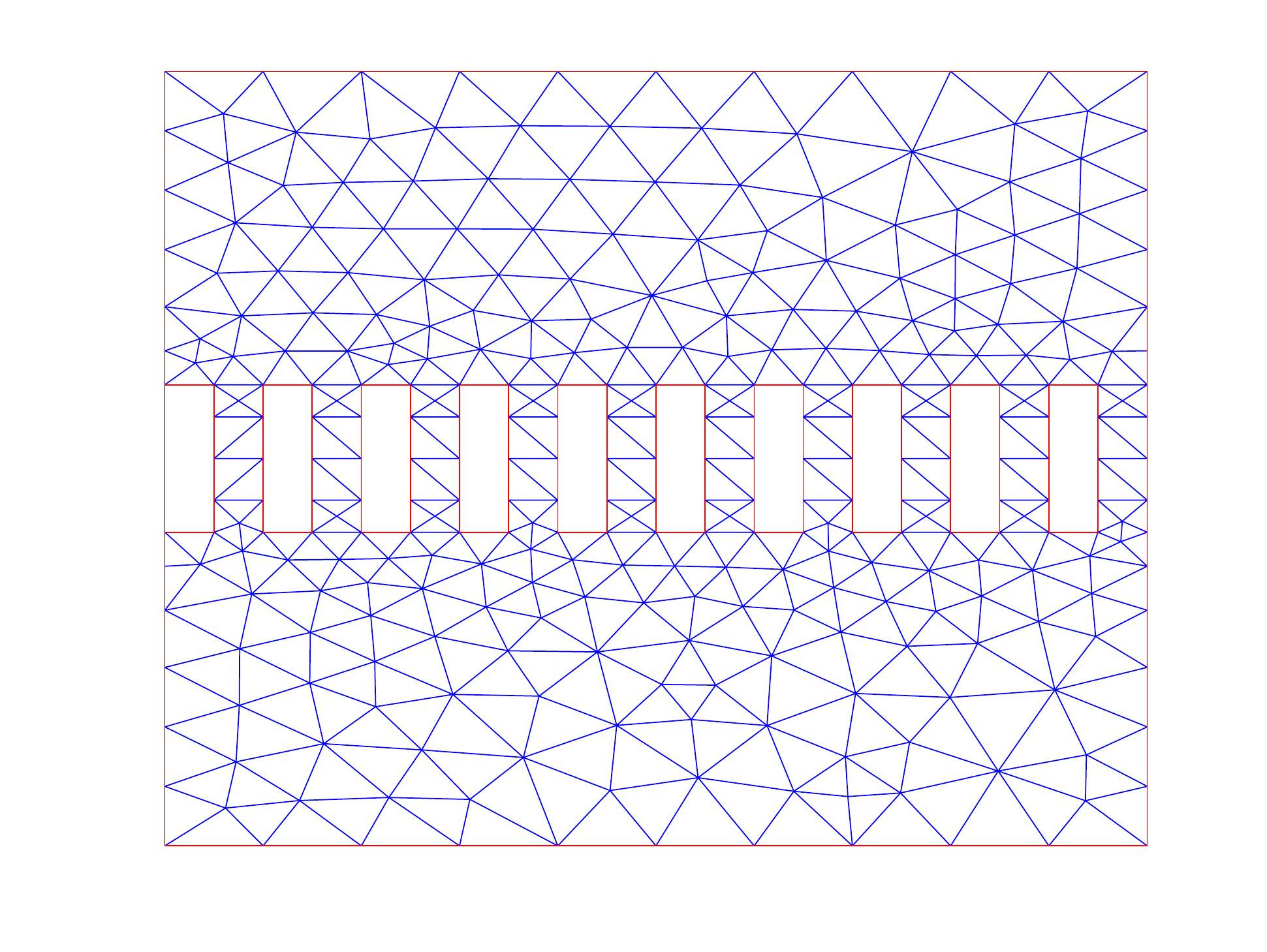}
  \\(a)\hspace{3.5cm}  (b)\end{center}
  \caption{(a) The rescaled
computational domain. (b) Adaptive mesh for the computational domain.} \label{fig:rescaled_domain_mesh}
\end{figure}

The rescaled domain after nondimensionalization is shown in Figure
\ref{fig:rescaled_domain_mesh}-(b). The domain has dimension
$L^s_1\times L^s_2 = 2n_c \times (2m+1) k$, while the channel has
unit base dimension. In the simulation we consider $k=4$.
 Furthermore, we consider
an adaptive triangular mesh as in Figure
\ref{fig:rescaled_domain_mesh}-(b).

\paragraph*{The Boundary Conditions}
\medskip
 For the electrical potential, we consider the
boundary conditions \eqref{boundary_condition}, with
$\Phi_{1,s}=10^h$, $h\in \mathbb N$, and $\Phi_{2,s}=0$, that is, for
  \begin{equation}\begin{split} \label{eq:boundary_cond_effettive_epsilon}
\Phi_s(t_s,x_s)&= 10^h, \qquad x_s\in   \Gamma_{D+}^{s};\\
\Phi_s(t_s,x_s)&= 0,  \quad\qquad x_s\in  \Gamma_{D-}^{s};\\
\dfrac{\partial \Phi_s}{\partial \nu}(t_s,x_s)&=0,\qquad\quad
x_s\in\partial\Omega^s\setminus\Gamma_{D-}^{s}\cup\Gamma_{D-}^{s}.
\end{split}\end{equation}

\begin{figure}[h]
\begin{center}
\includegraphics[scale=0.25]{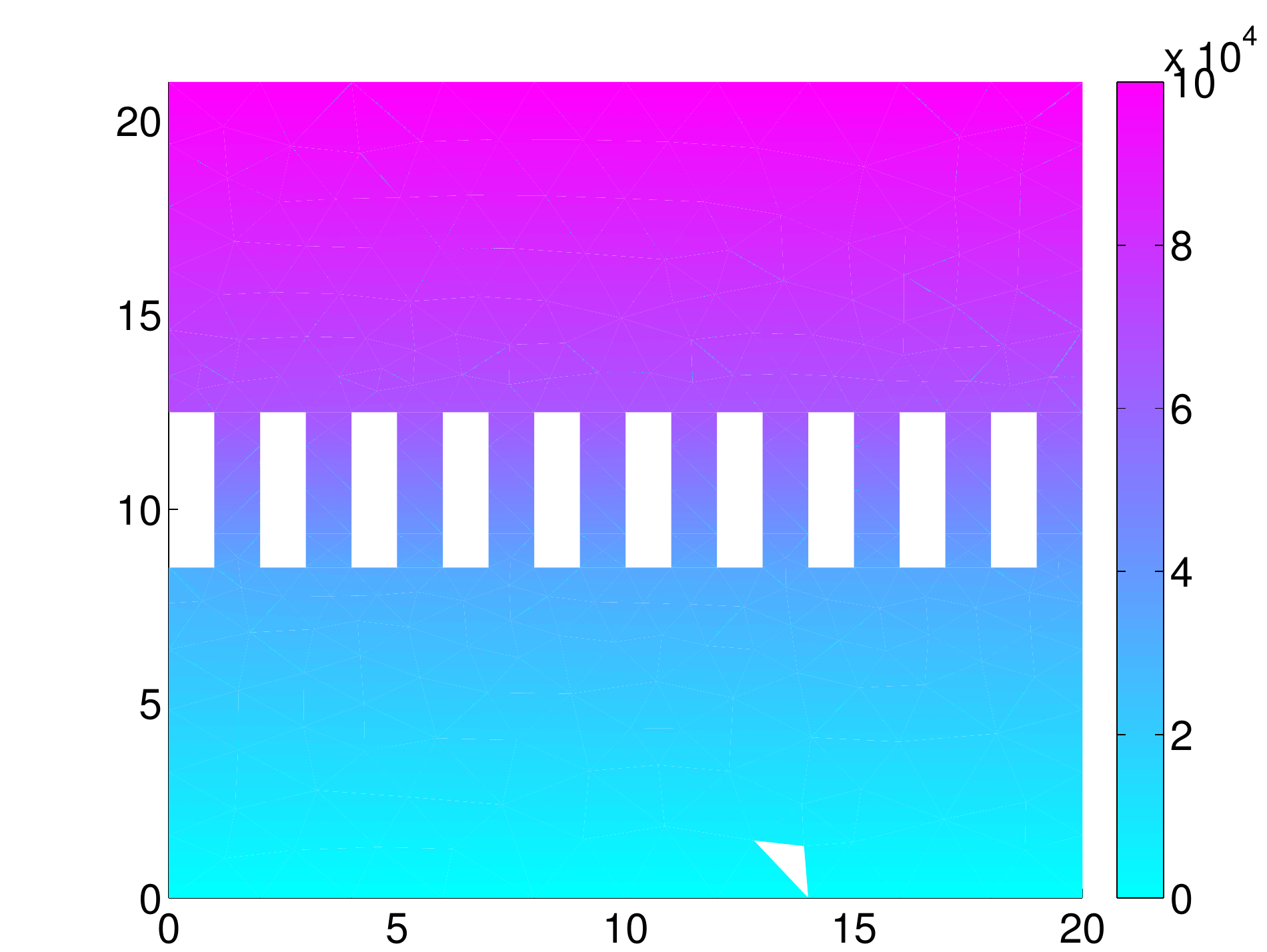}
\end{center}
\caption{Electrical field with  boundary conditions
\eqref{eq:boundary_cond_effettive_epsilon}.}
\label{fig:electric_field}
\end{figure}

The consequent initial electrical field is shown in Figure
\ref{fig:electric_field}, in the case $h=4$.

\medskip

For the moving ions, in order to simulate ion flow through the
system, periodic boundary conditions upon $\partial \Omega$  are
applied. In such a way, we may mimic an infinite reservoir and
ions are allowed  to enter and exit the simulation box.

\medskip

\paragraph*{The Fixed Ions}
Two fixed charges have been located in the upper layer of each
channel,  in order to simulate the behavior of the receptors
present in protein channels. As a consequence, the total number of
fixed charges is $J= 2 n_c $.
\medskip

\paragraph*{The  Ions Dimensions}
We consider two different possible dimensions. \\The first one
refers to a radius of the same scale of the pore, for example
$\tilde\epsilon\sim 1.52 \cdot 10^{-10}$, as in the case of $K+$.
In general we consider $\tilde\epsilon\sim  \epsilon_1/4$.
\\
In order to simulate   possible cases in which the ion dimension
is much smaller than the channel one, we consider also the case
$\tilde\epsilon\sim \epsilon_1/100\sim 10^{-8}$, too.

\medskip

\paragraph*{The Initial Conditions}
At the initial time $t=0$ a number of $N$ ions are uniformly
distributed in the upper bath $\Omega_+$ with normally distributed
velocities (see \cite{MSA}).
\medskip

\paragraph*{Spatial Distribution of Moving Ions }
From the definition of  the spatial distribution of the moving
ions \eqref{def:nu_Z}, we consider a regularized version via a
convolution of the empirical measure \eqref{def:X_Z}  by a
bivariate normal density $f$  of zero mean and diagonal
covariance matrix, and variances $ \sigma_1^2,  \sigma_2^2, $ that
is
\begin{equation}\label{smooth_spatial_density}
f * X(x)=\dfrac{1}{N}\displaystyle\sum_{i=1}^{N}f(x_i- x), \quad
x\in \Omega.
\end{equation}
In the following, we consider $\sigma_1^2=\sigma_2^2=10^{-1}$.

\subsection{Case 1: dimension of the channel  $\mathbf{\epsilon_1 \sim
O(10^{-9} m)}$ }

We simulate the coupled system \eqref{eq:rescaled_poisson},
\eqref{eq:scaled_langevin_1}, and \eqref{eq:scaled_langevin_2} by
taking into account the first scaling factor introduced in Section
\ref{sec:nondim}, that is $\epsilon=\epsilon_1=10^{-9}$. We
consider $N=10^3$ free moving ions and boundary conditions
\eqref{eq:boundary_cond_effettive_epsilon} with $h=4$.
Furthermore, in the membrane we take $n_c=10$ channels. The time
increment is $dt=10^{-3}$.

\medskip

\medskip

\begin{figure}[h!]
\begin{center}
\subfigure[
t=0]{\includegraphics[scale=0.25]{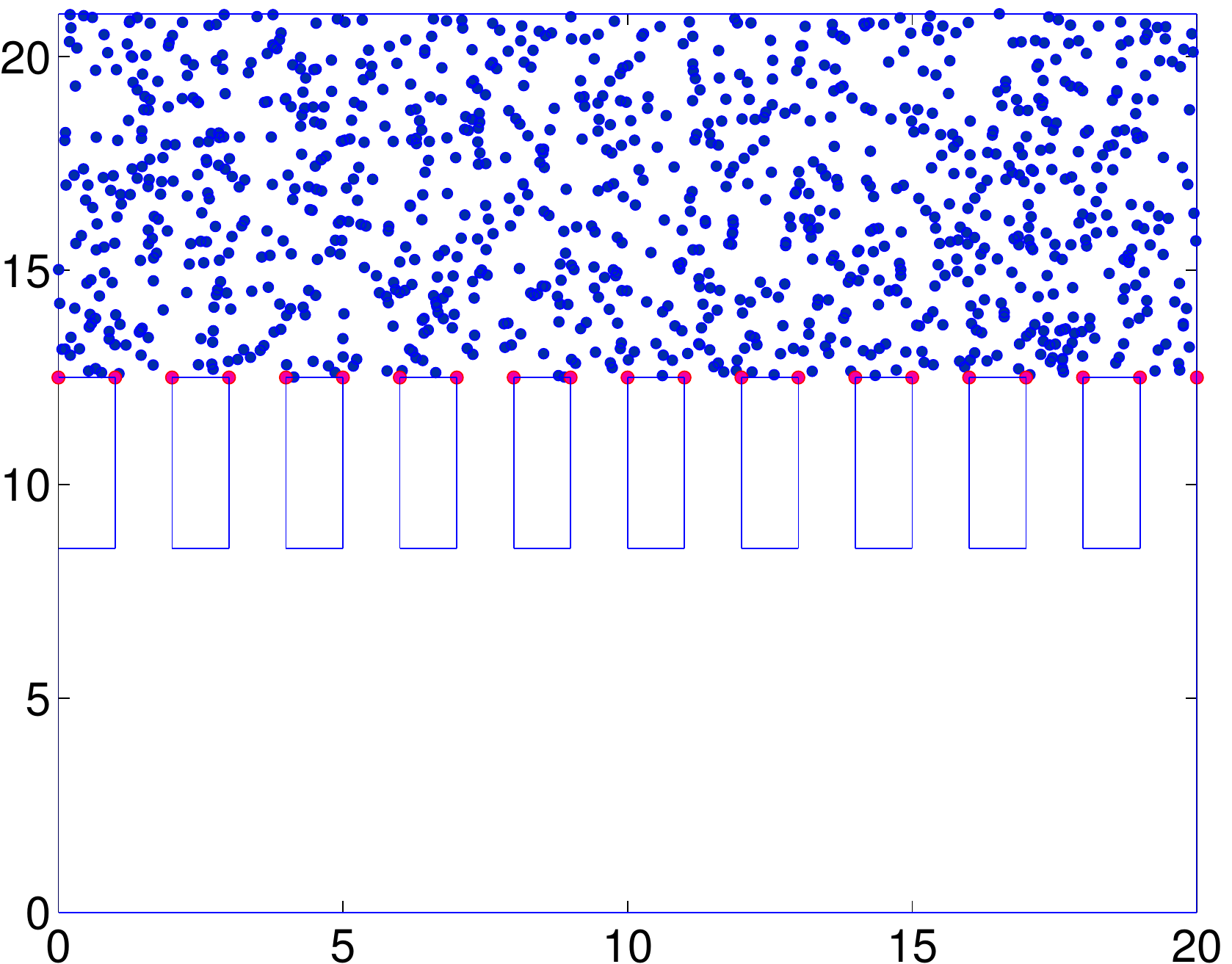}}
\subfigure[
t=3]{\includegraphics[scale=0.25]{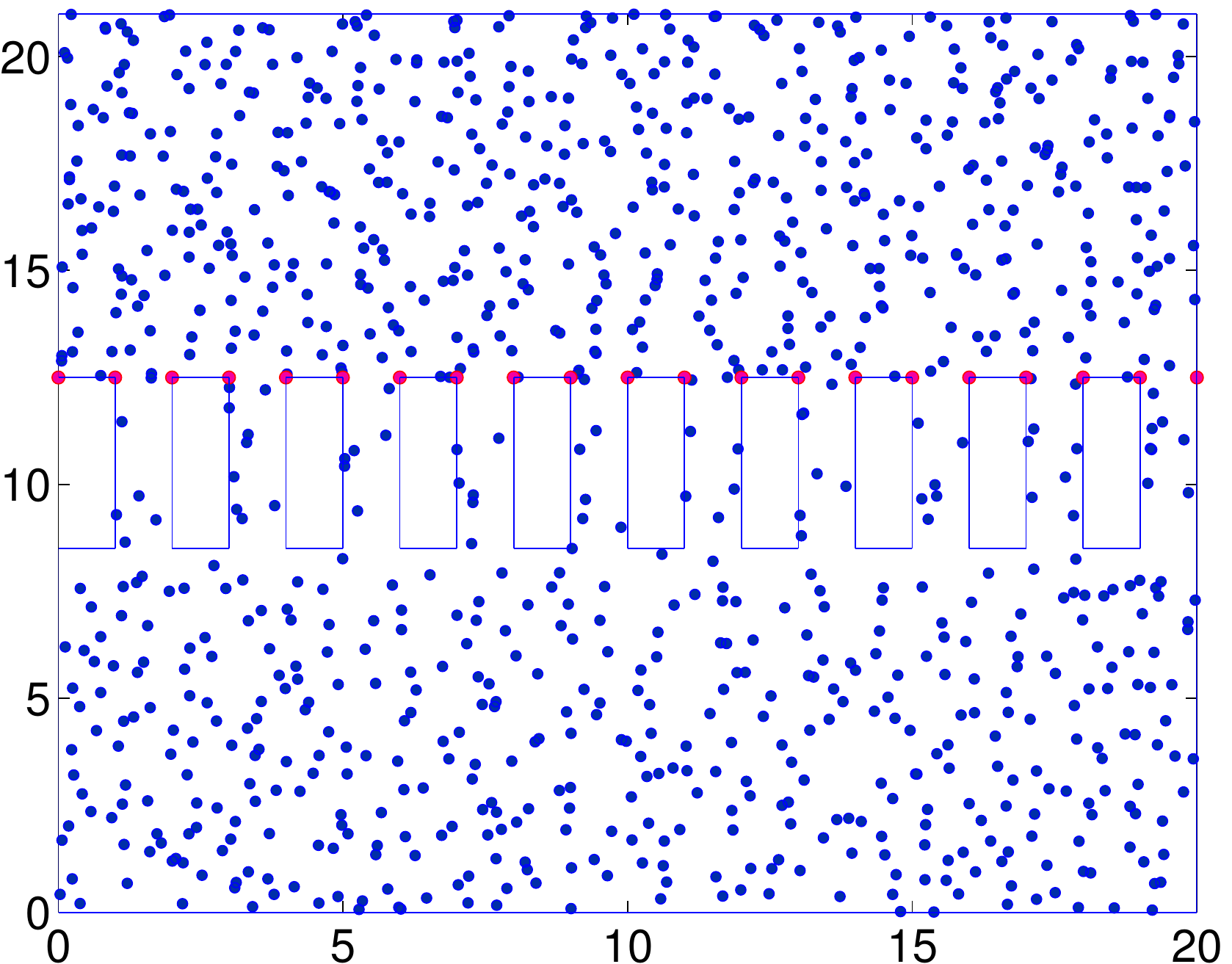}}
\end{center}
\caption{ Initial and final ion locations - Case 1.a. }
\label{fig:1000_ions_initial_end_scaling_9}
\end{figure}

\paragraph*{Case 1.a. Ion radius $\tilde\epsilon\sim  \epsilon_1/4$ }
This case is the typical situation of the nanopore.   Figure
\ref{fig:1000_ions_initial_end_scaling_9} shows the initial
distribution of the ions and the state of the system at time
$t=3$. It is  evident few ions may enter into a channel
at the same time.

\begin{figure}[h!]
\begin{center}
\subfigure[t=0]{\includegraphics[scale=0.25]{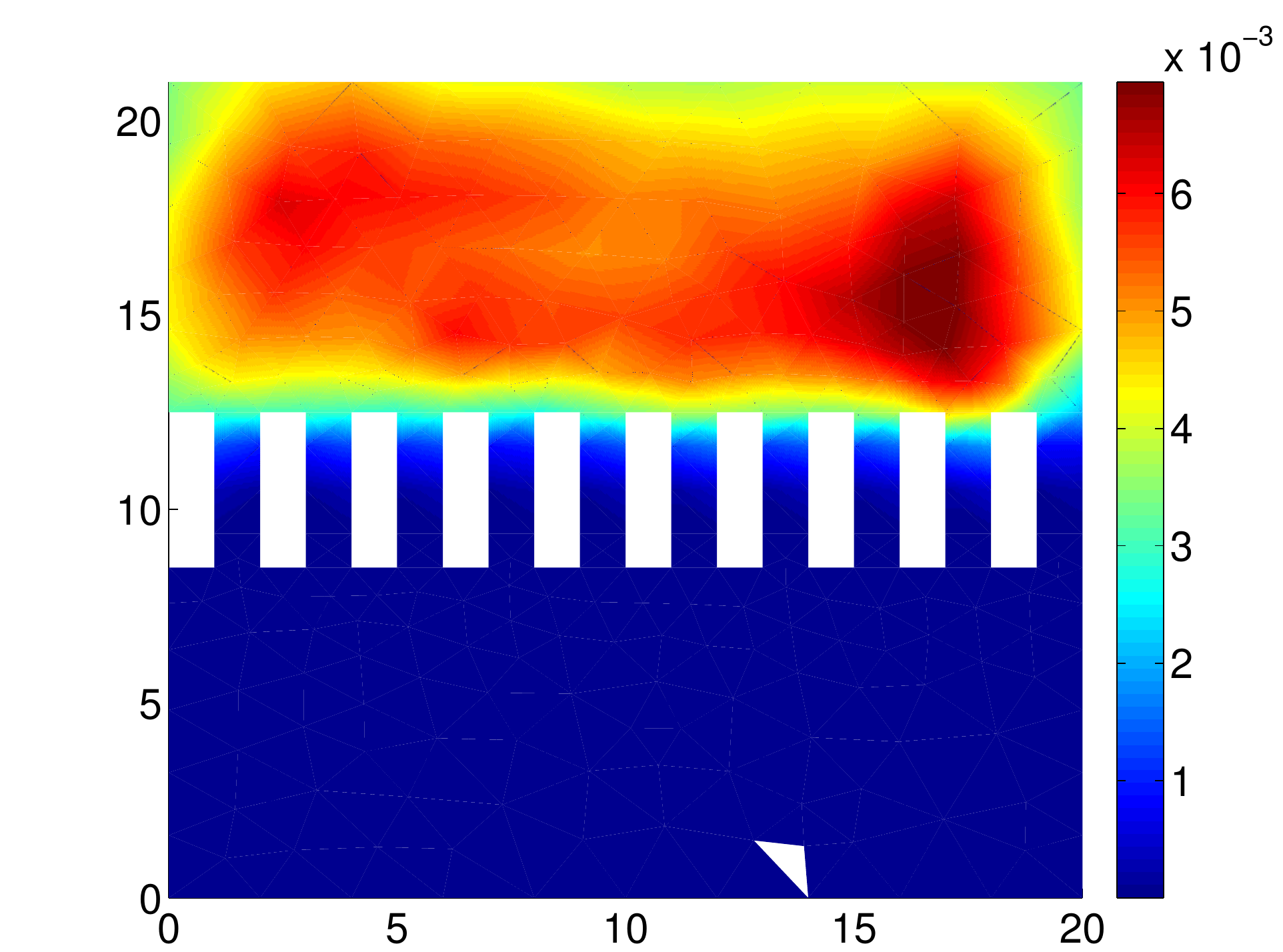}}
\subfigure[t=1]{\includegraphics[scale=0.25]{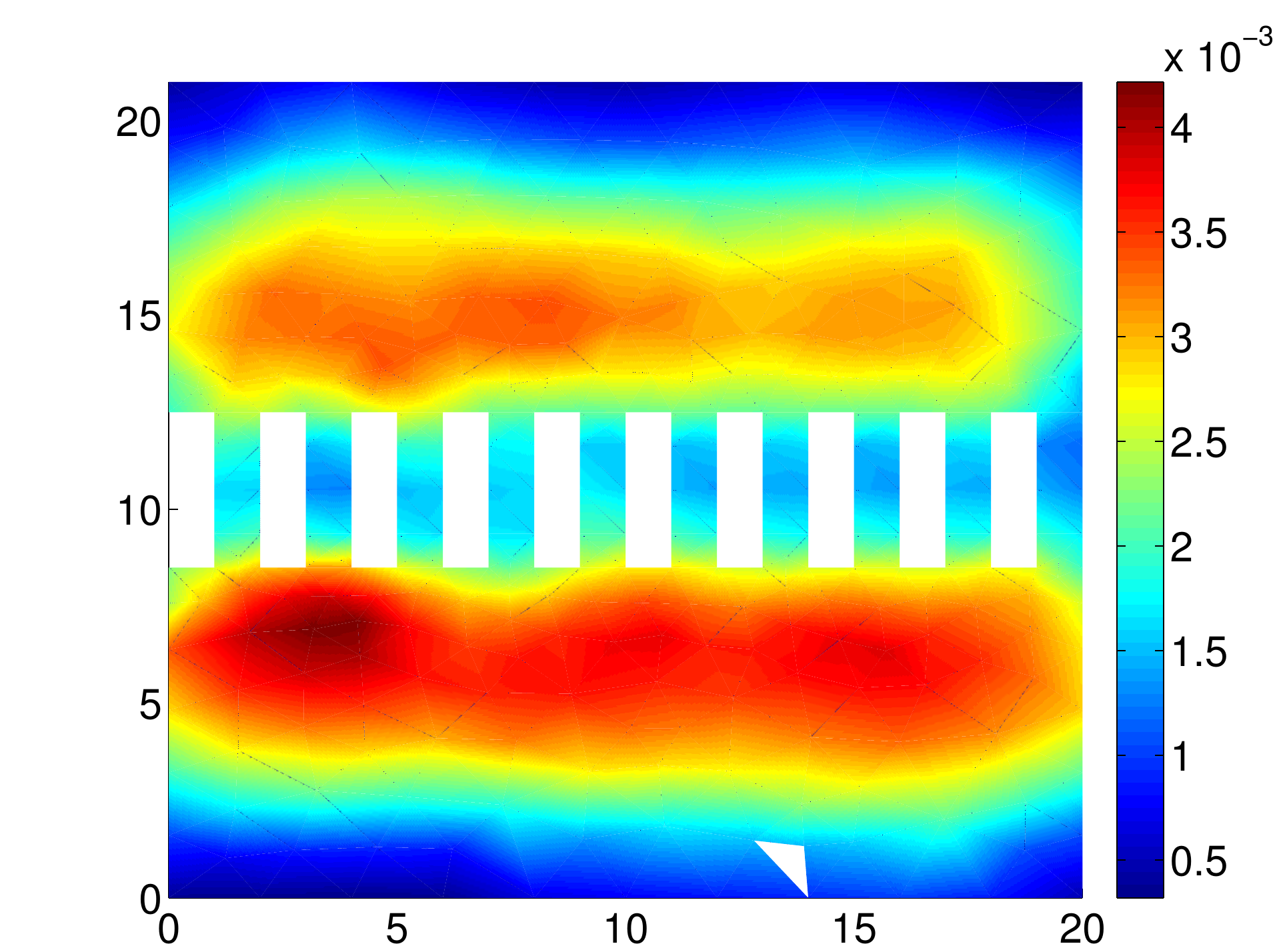}}

\subfigure[t=2]{\includegraphics[scale=0.25]{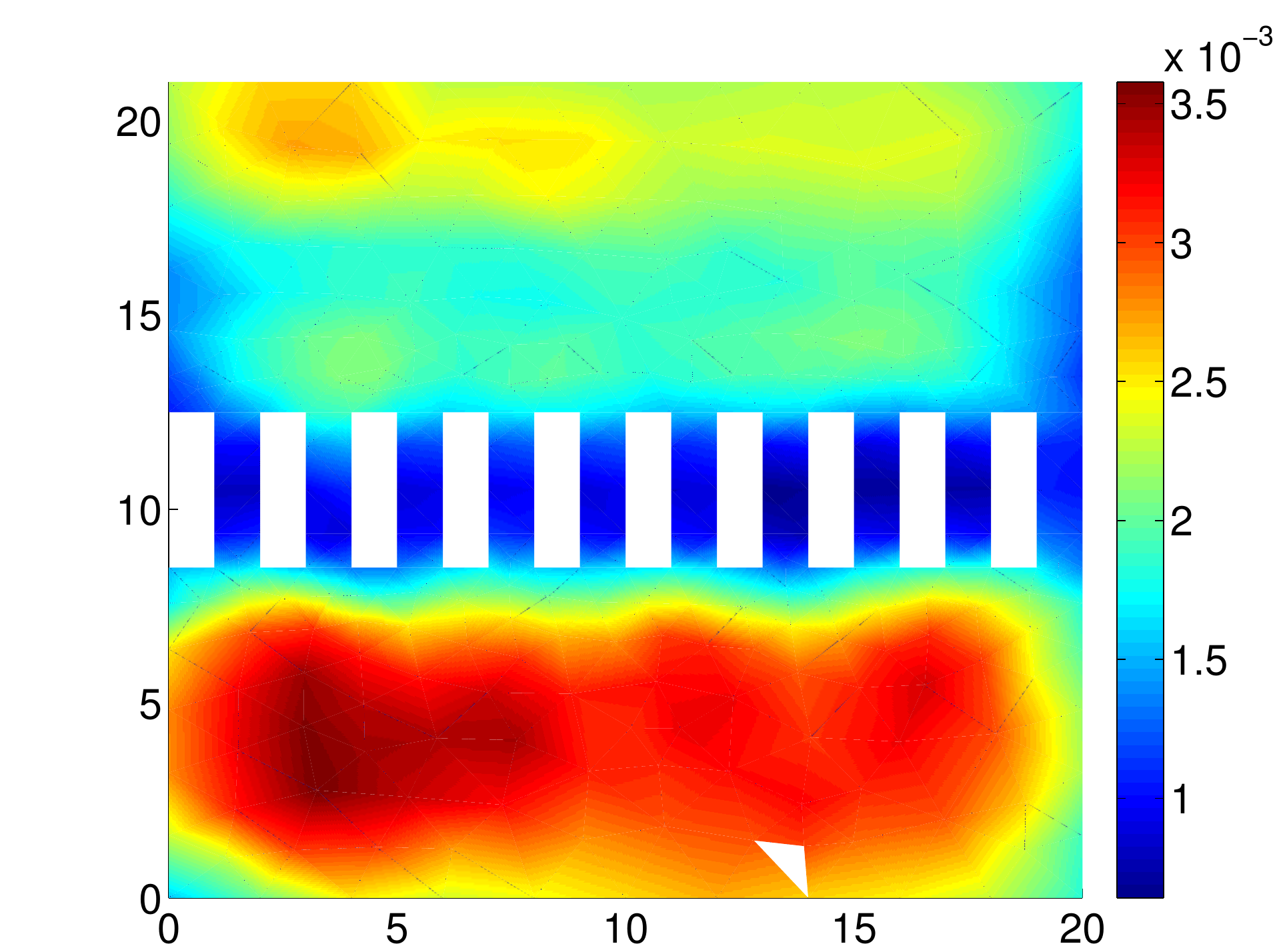}}
\subfigure[t=3]{\includegraphics[scale=0.25]{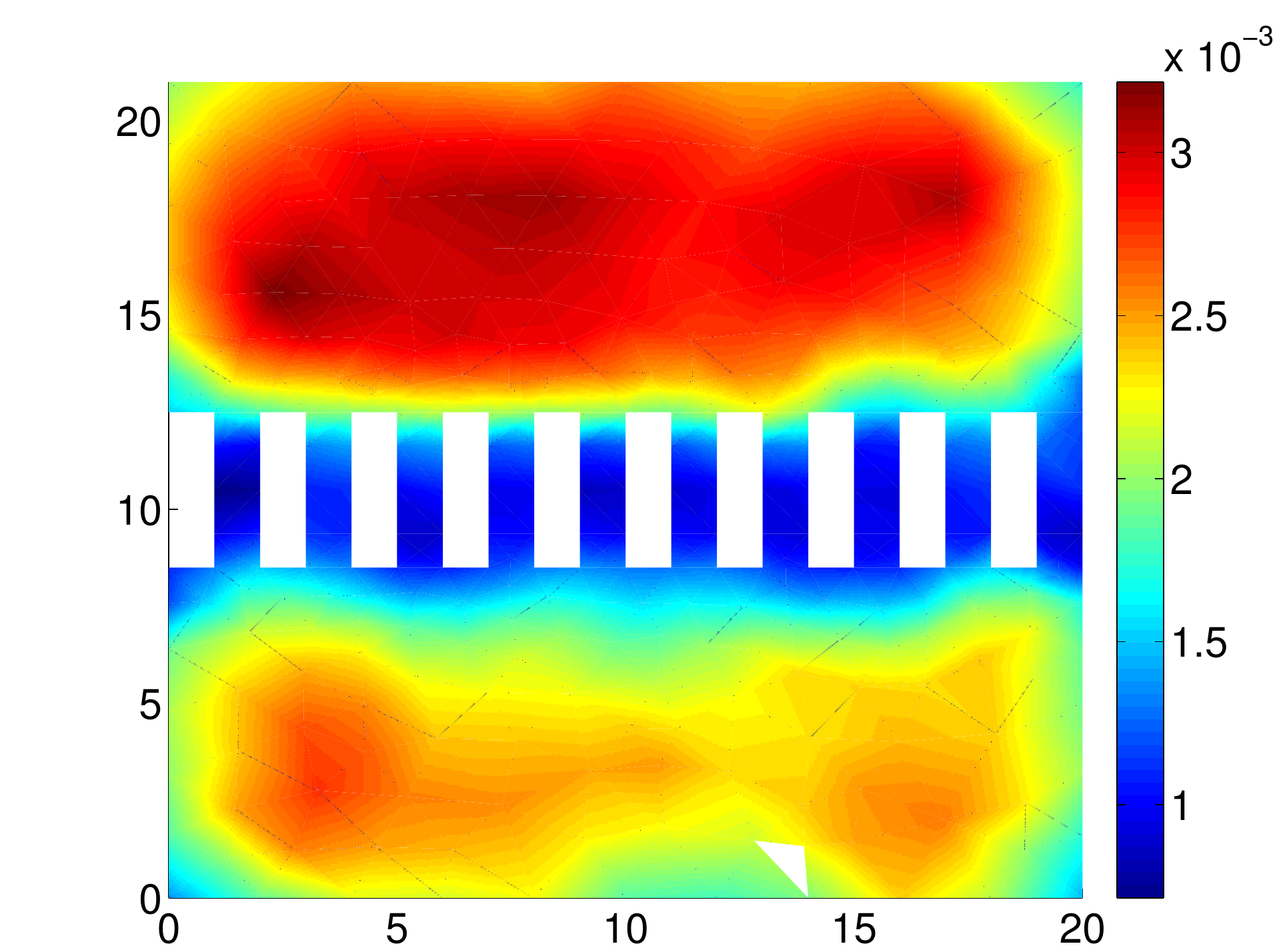}}
\end{center}\caption{Time evolution of the ion density - Case 1.a. }
\label{fig:density_1_scaling_9}
\end{figure}

Figure \ref{fig:density_1_scaling_9} shows the time evolution of
the ion spatial distribution. Since the dynamics are rather fast,
after a small time interval at the very beginning of the
simulation, the number of ions present in each region is varying
but not too much.

\begin{figure}[h!]
\begin{center}
\subfigure{\includegraphics[scale=0.25]{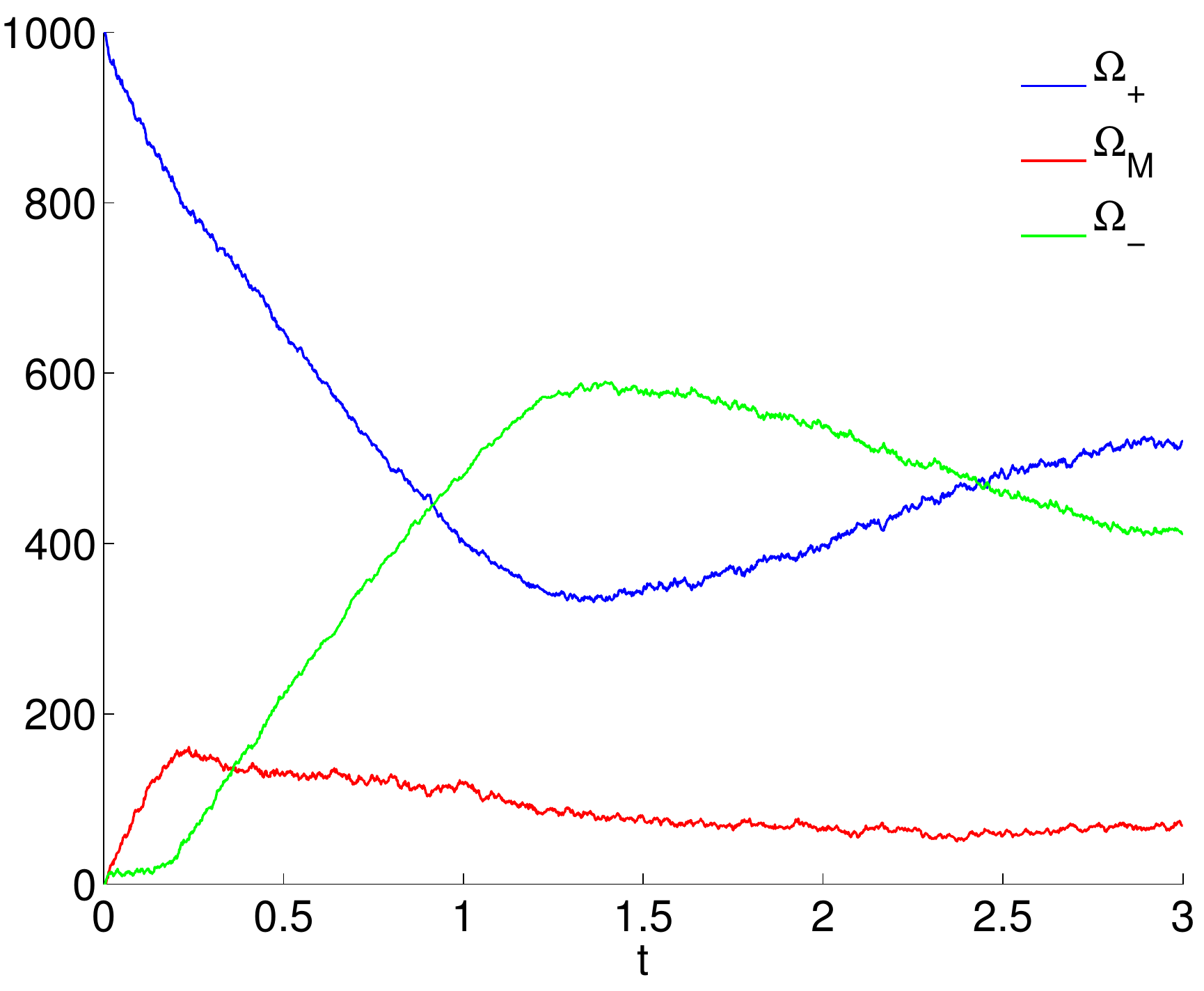}}
\subfigure{\includegraphics[scale=0.25]{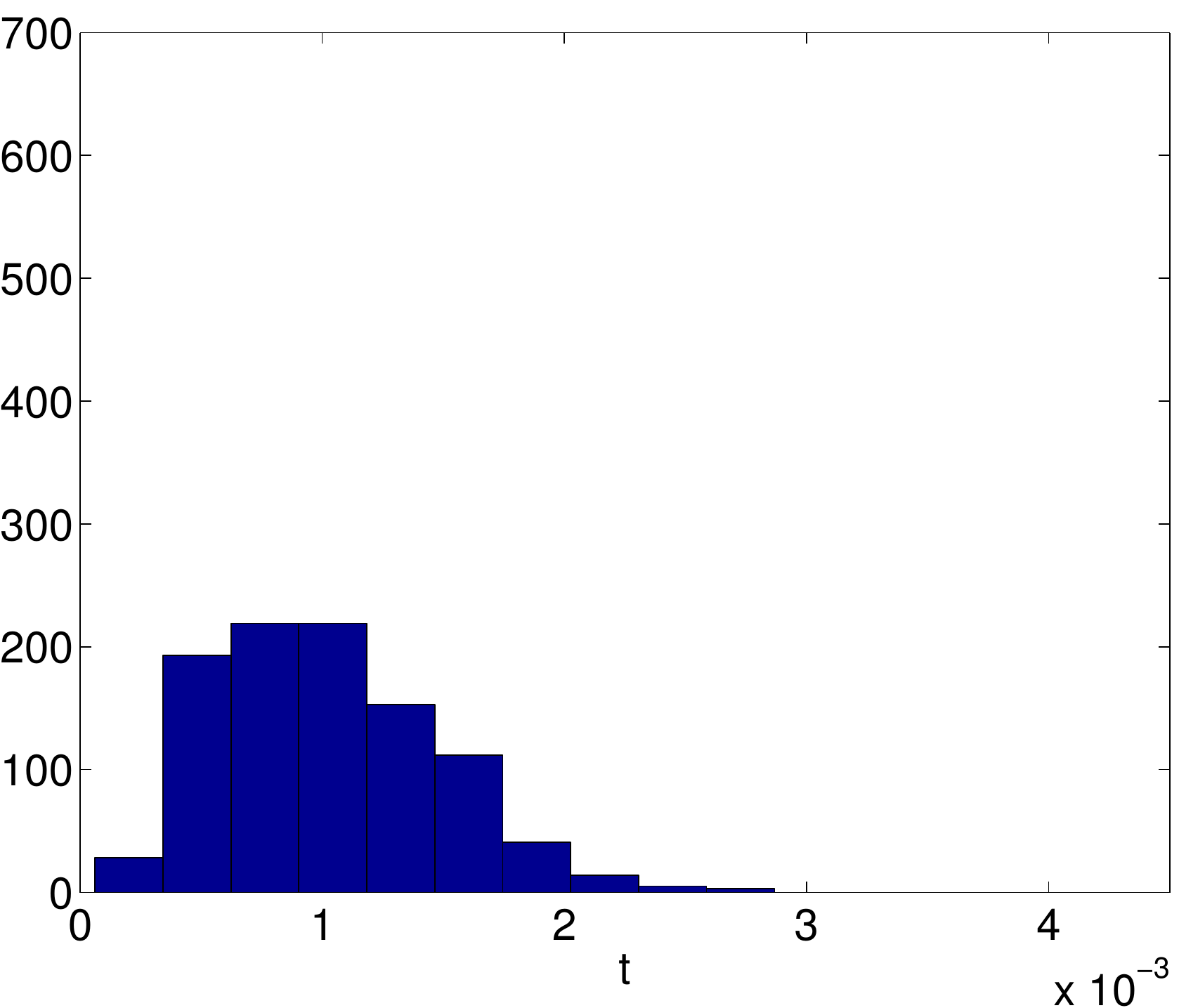}}
\end{center}
\caption{Left: evolution of ion number in
$\Omega_+,\Omega_-,\Omega_M$. Right: histogram of the average time
spent by ions in  $\Omega_M$. Case 1.a. }
\label{fig:counting_ions_scaling_9}
\end{figure}

This is also visualized in Figure
\ref{fig:counting_ions_scaling_9}, left, showing the time
evolution of the number of ions in each region ($\Omega_+,
\Omega_-$, and $\Omega_M$). Ions are rather dispersed outside the
membrane, while inside the ion number is small and not varying too
much. On the right-hand side of  Figure
\ref{fig:counting_ions_scaling_9} the distribution of the time
spent by a particle into a channel is shown.

 \medskip

\paragraph*{Case 1.b.  Ion radius $\tilde\epsilon\sim \epsilon_1 \cdot 10^{-2}$}

Here we consider the simulation results in the same conditions
with respect to Case 1.a., but the radius of ions, which is here
much smaller. This case corresponds more to the case of micropore.

\begin{figure}[h!]
\begin{center}
\subfigure[ t=0]{\includegraphics[scale=0.25]{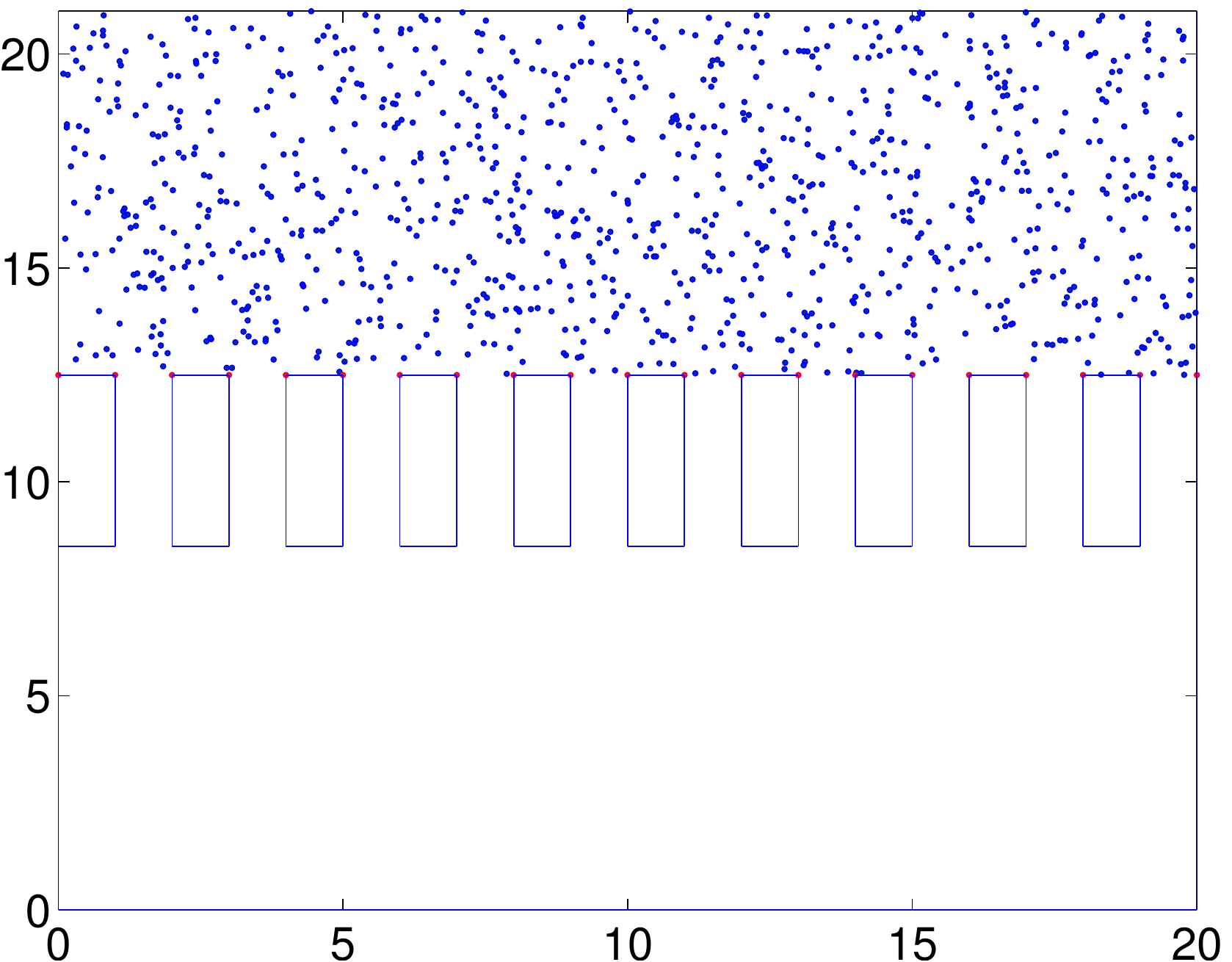}}
%\subfigure[t=1.5]{\includegraphics[scale=0.2]{7/Ions/Ions=1501.eps}}
\subfigure[ t=3]{\includegraphics[scale=0.25]{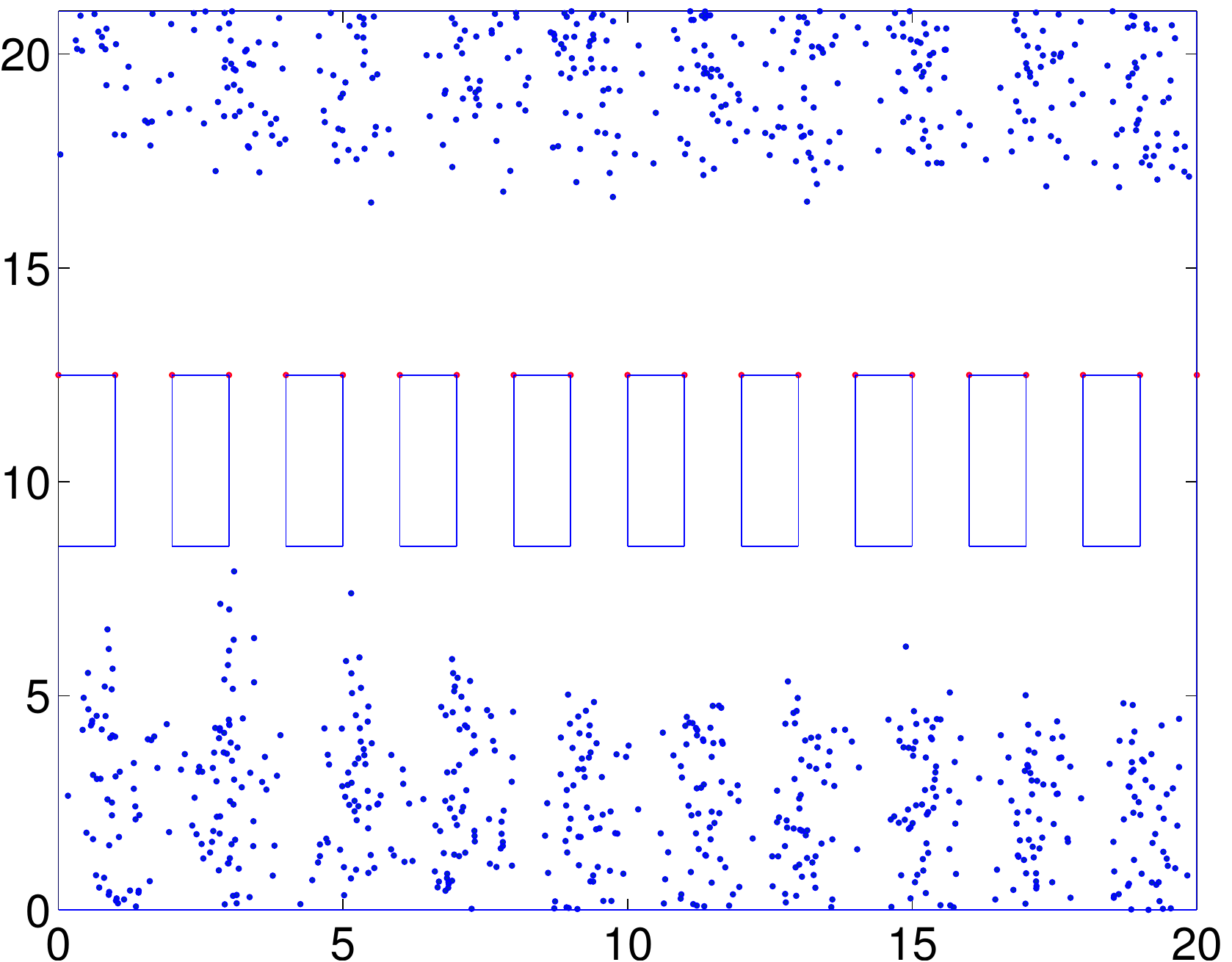}}
\end{center}
\caption{ Initial and final ion locations - Case 1.b. }
\label{fig:1000_ions_initial_end_scaling_9_casob}
\end{figure}

\begin{figure}[h!]
\begin{center}
\subfigure[t=0]{\includegraphics[scale=0.25]{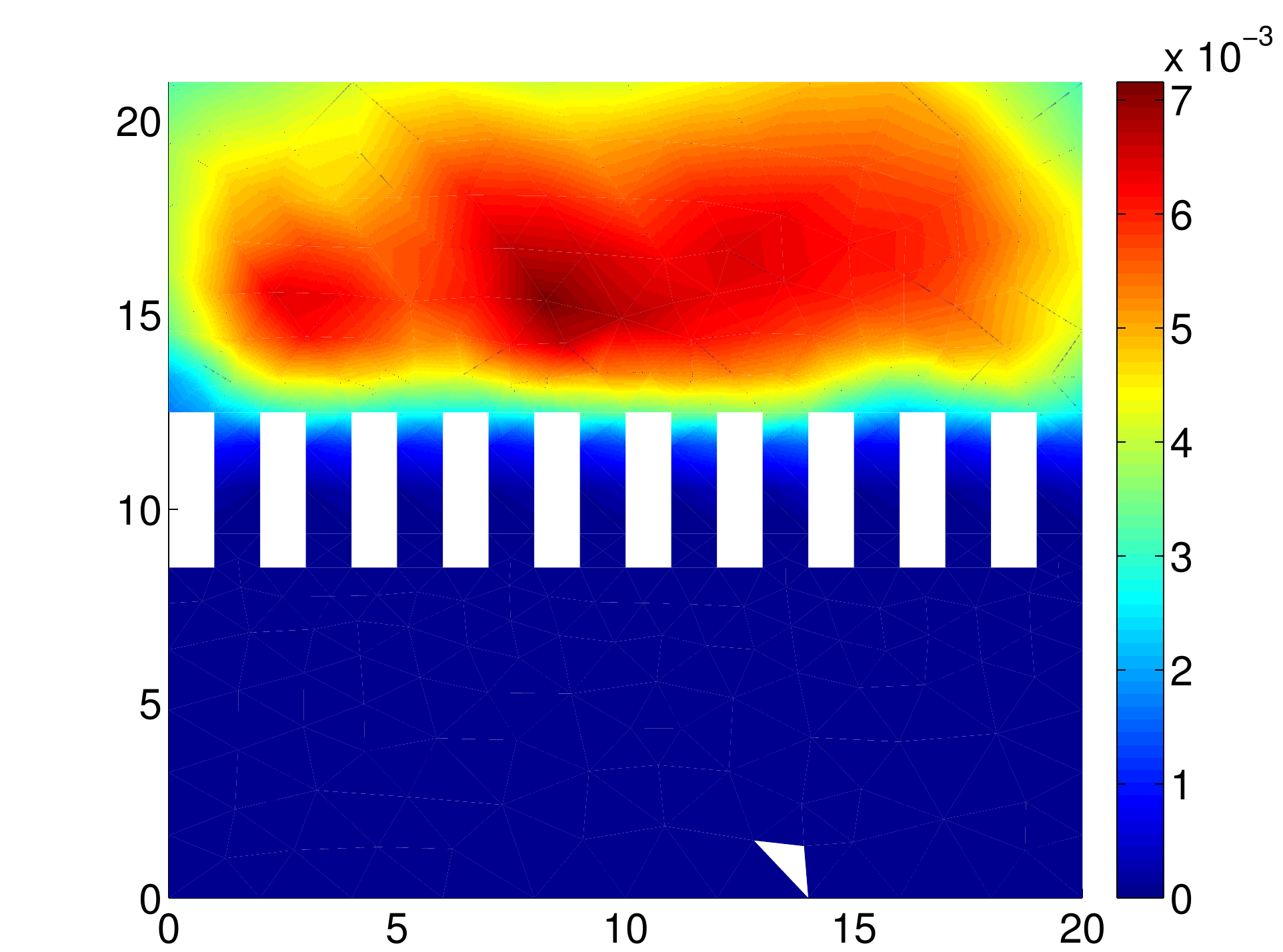}}
\subfigure[t=1]{\includegraphics[scale=0.25]{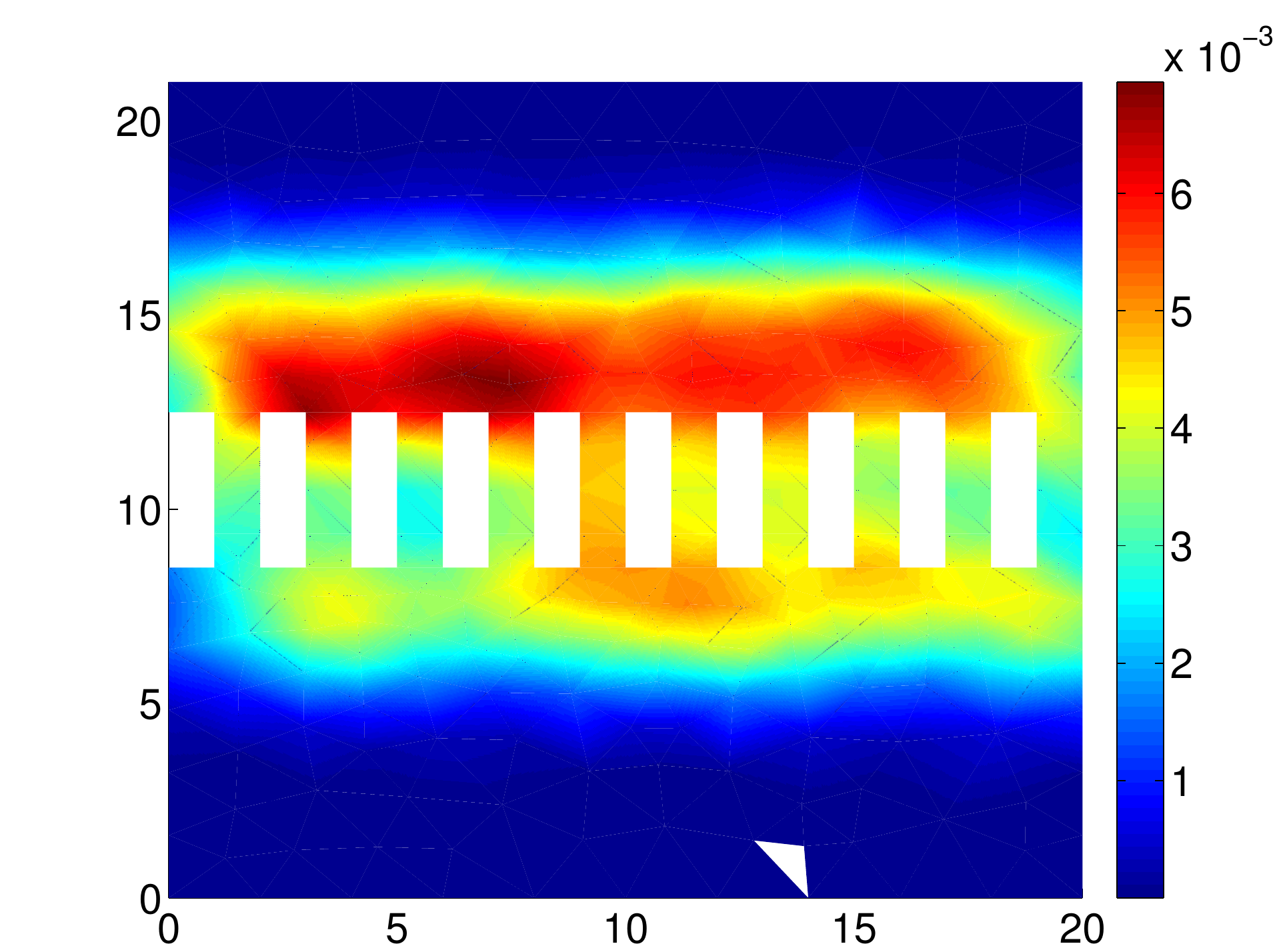}}

\subfigure[t=2]{\includegraphics[scale=0.25]{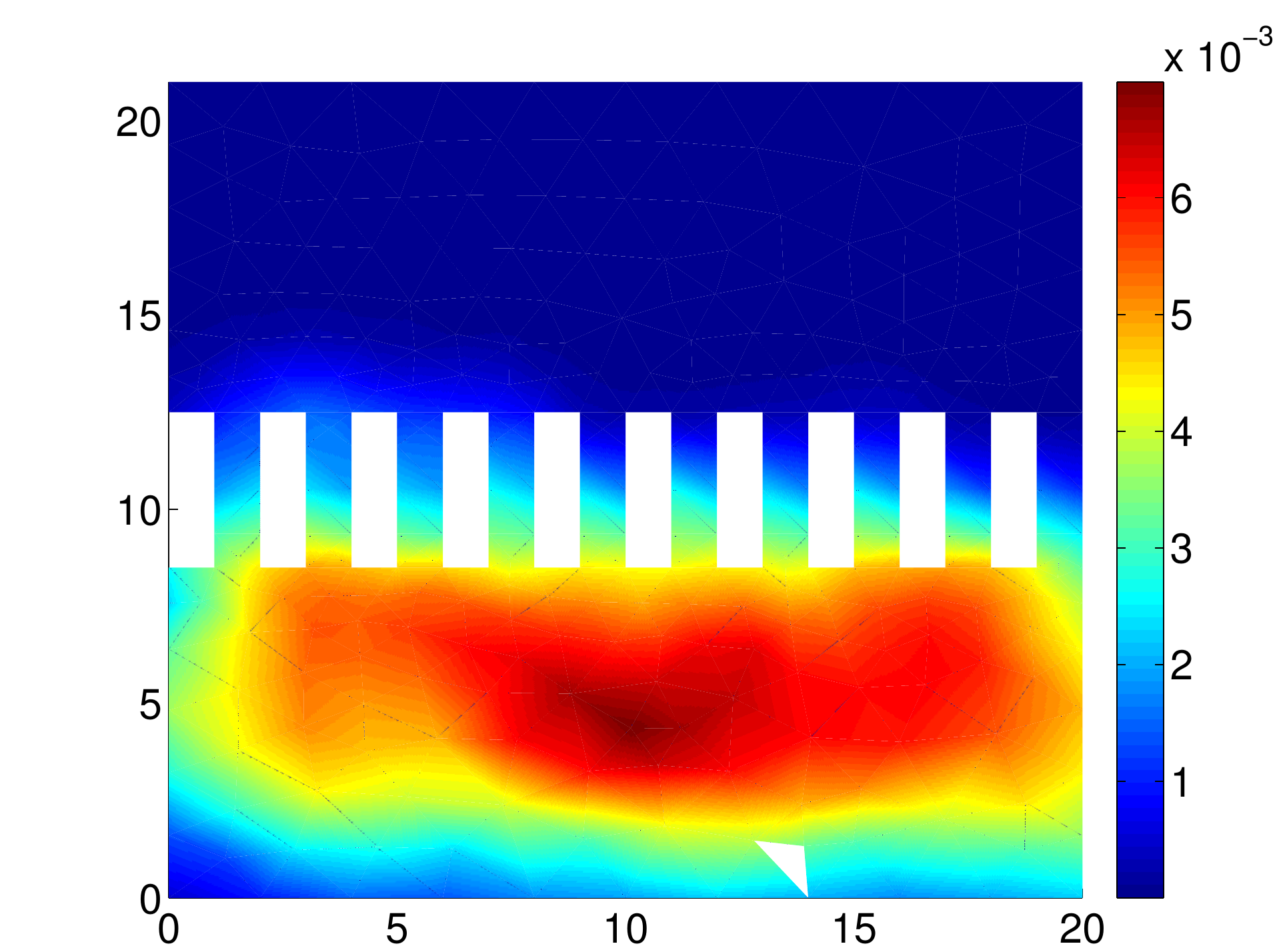}}
\subfigure[t=3]{\includegraphics[scale=0.25]{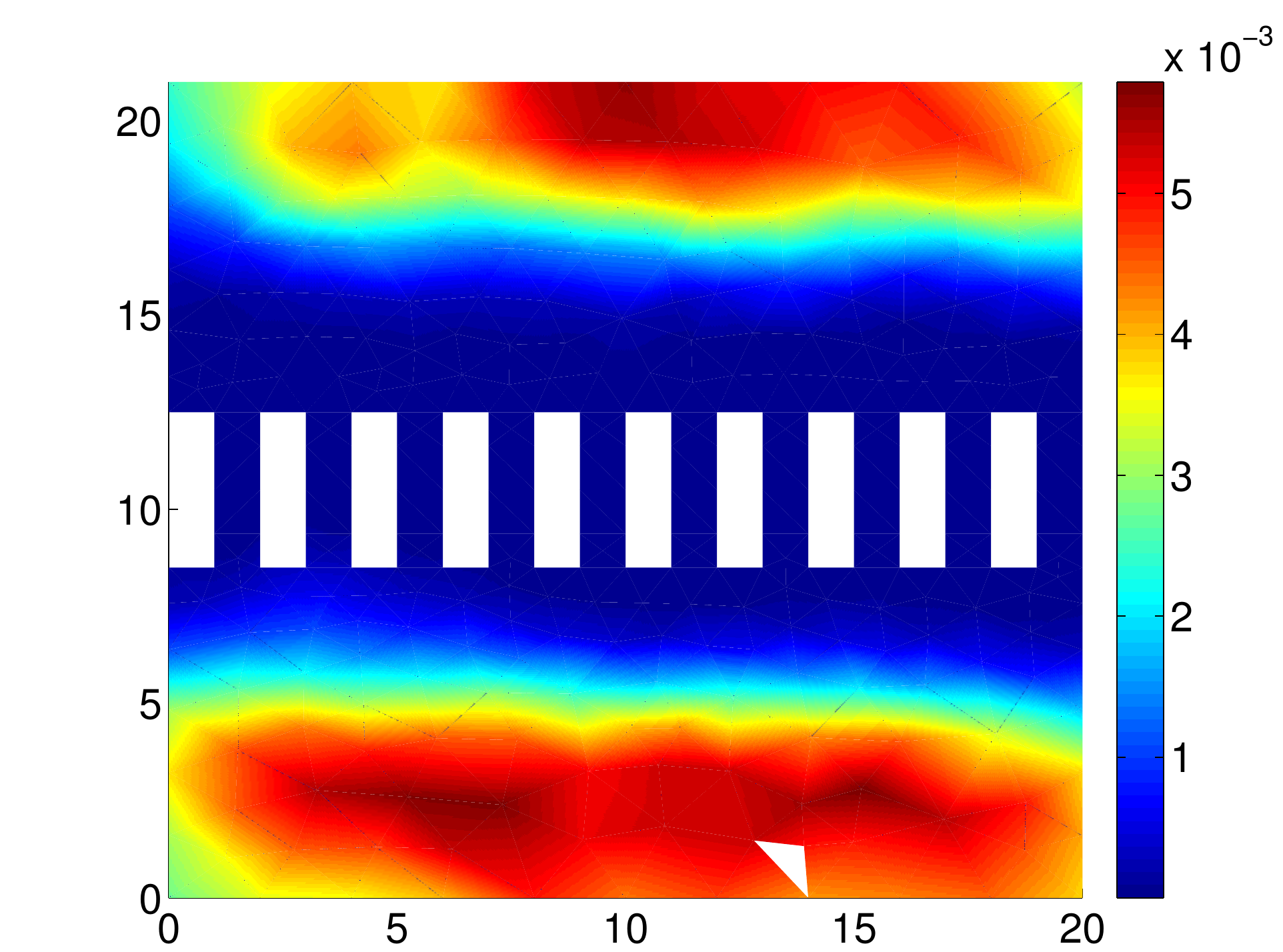}}
\end{center}
\caption{Time evolution of the ion density - Case 1.b.}
\label{fig:density_1_scaling_9_radius2}
\end{figure}

\medskip
 Figure \ref{fig:1000_ions_initial_end_scaling_9_casob} shows how
 ions are slower than the previous case. Furthermore,
 in Figure \ref{fig:density_1_scaling_9_radius2} one may observe
that in this case the spatial density of ions in a channel may be
significantly greater than zero, a case not observed previously.
This is due to fact that here we are reproducing the case in which
 in each channel one may enter a large number of ions at the same
time.

\medskip

\begin{figure}[h!]
\begin{center}
\subfigure{\includegraphics[scale=0.25]{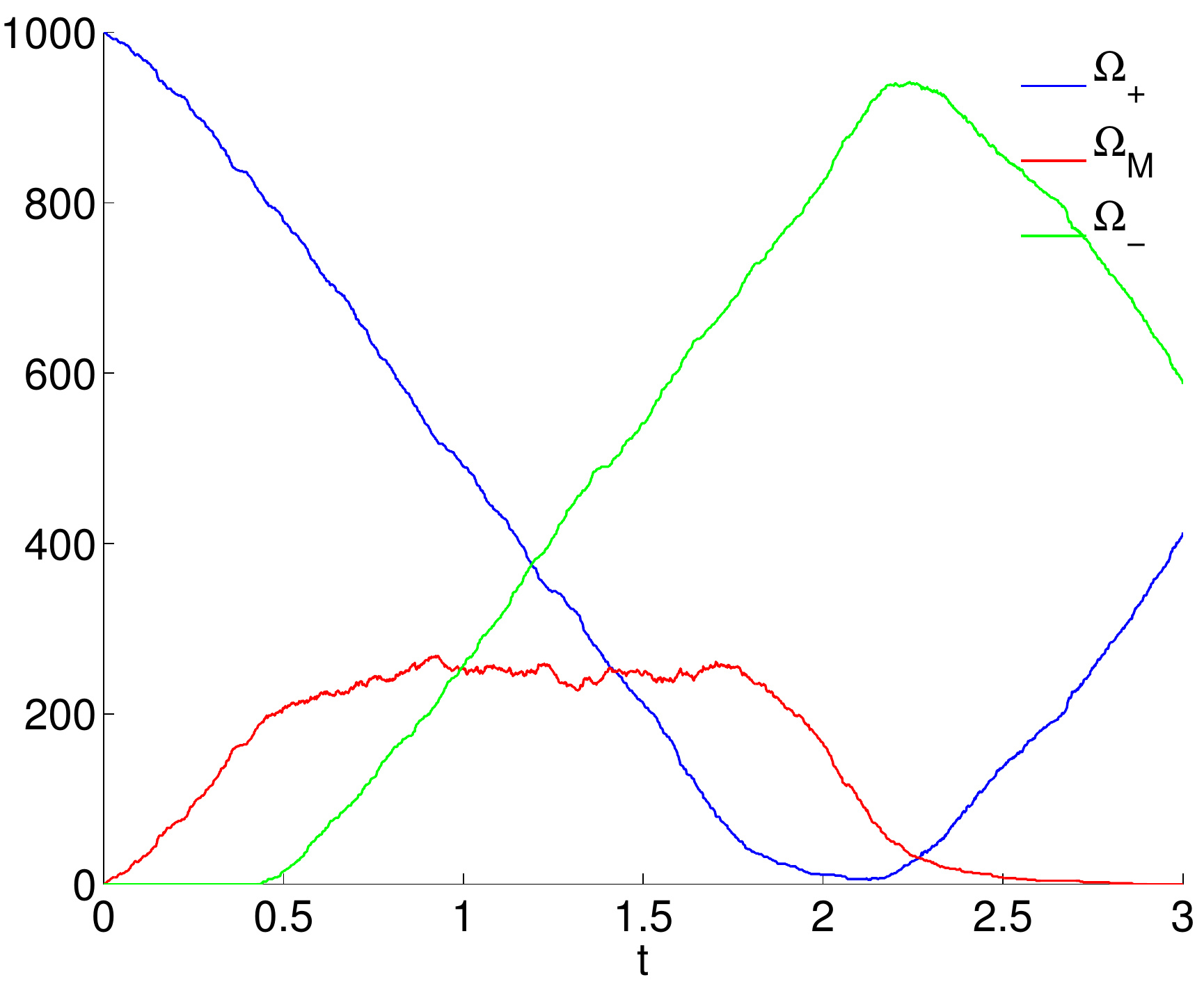}}
\subfigure{\includegraphics[scale=0.25]{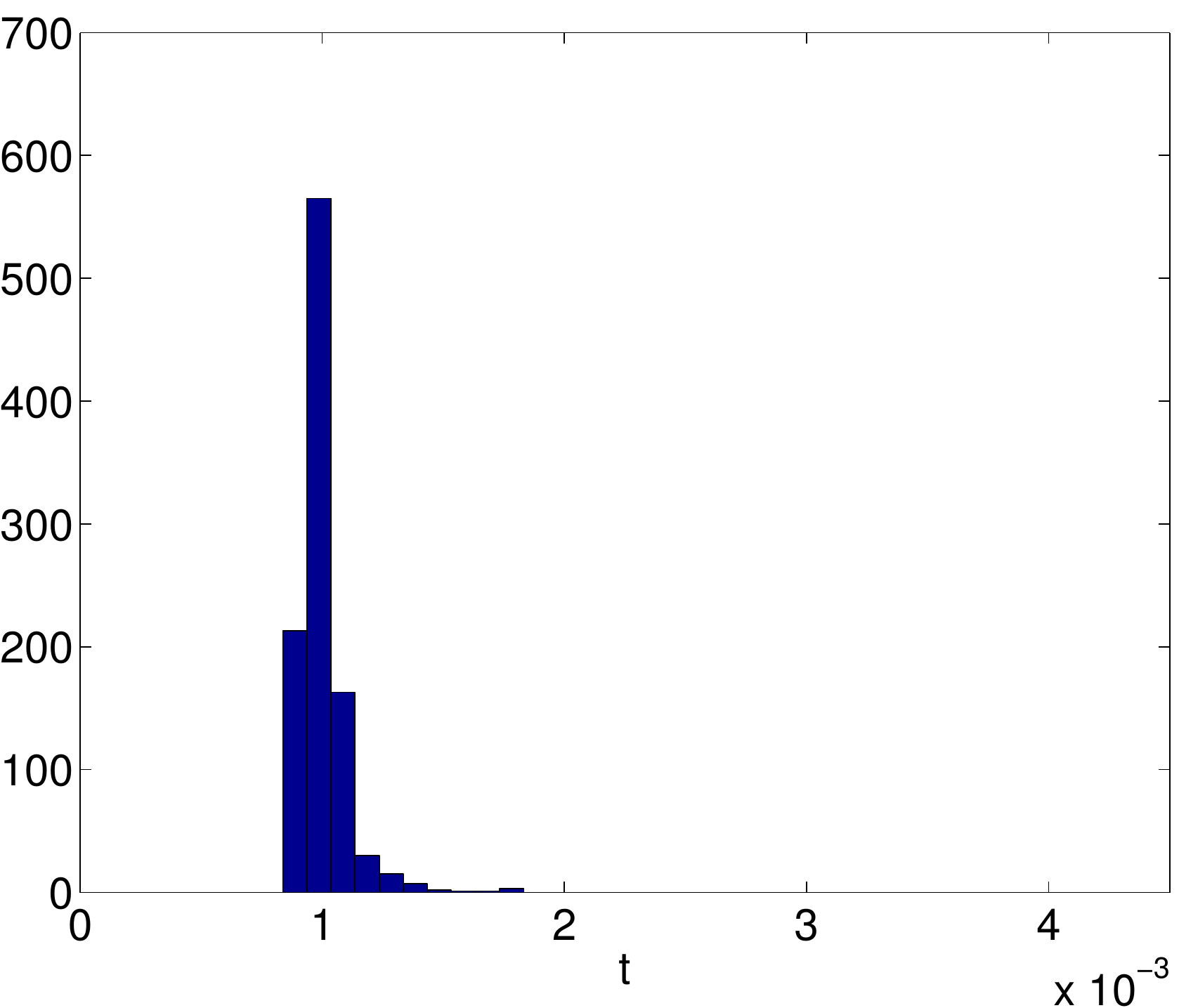}}
\end{center}
\caption{Left: evolution of ion number in $\Omega_+,\Omega_-,$ and
$\Omega_M$. Right: histogram of the average time spent by ions in
$\Omega_M$. Case 1.b.} \label{fig:counting_ions_scaling_9_radius2}
\end{figure}

This may be seen   in Figure \ref{fig:counting_ions_scaling_9_radius2}, left, where the number of ions in the membrane is clearly higher. As regards to the time needed for
crossing a channel, we see how the distribution is less dispersed
and is significantly higher in the mean. See also  Table
\ref{table:estimates_number_ions}.

\subsection{Case 2:  dimension of the channel  $\mathbf{\epsilon_1\sim O(10^{-7} m) }$}

\medskip

Now we simulate the coupled system for  the scaling parameter
$\epsilon=\epsilon_1=10^{-7}$, i.e. with parameters
\eqref{eq:parameters_L_vanderwalls_-7}.

\medskip

\paragraph*{Case 2.a.  Ion radius  $\tilde\epsilon\sim  \epsilon_1/4$}
Again first we consider the case in which few ions may
enter in a channel at the same time, due to the fact  that the
diameter  $\tilde\epsilon$ is taken  as $  \epsilon_1/4$

\begin{figure}[h]
\begin{center}
\subfigure[t=0]{\includegraphics[scale=0.25]{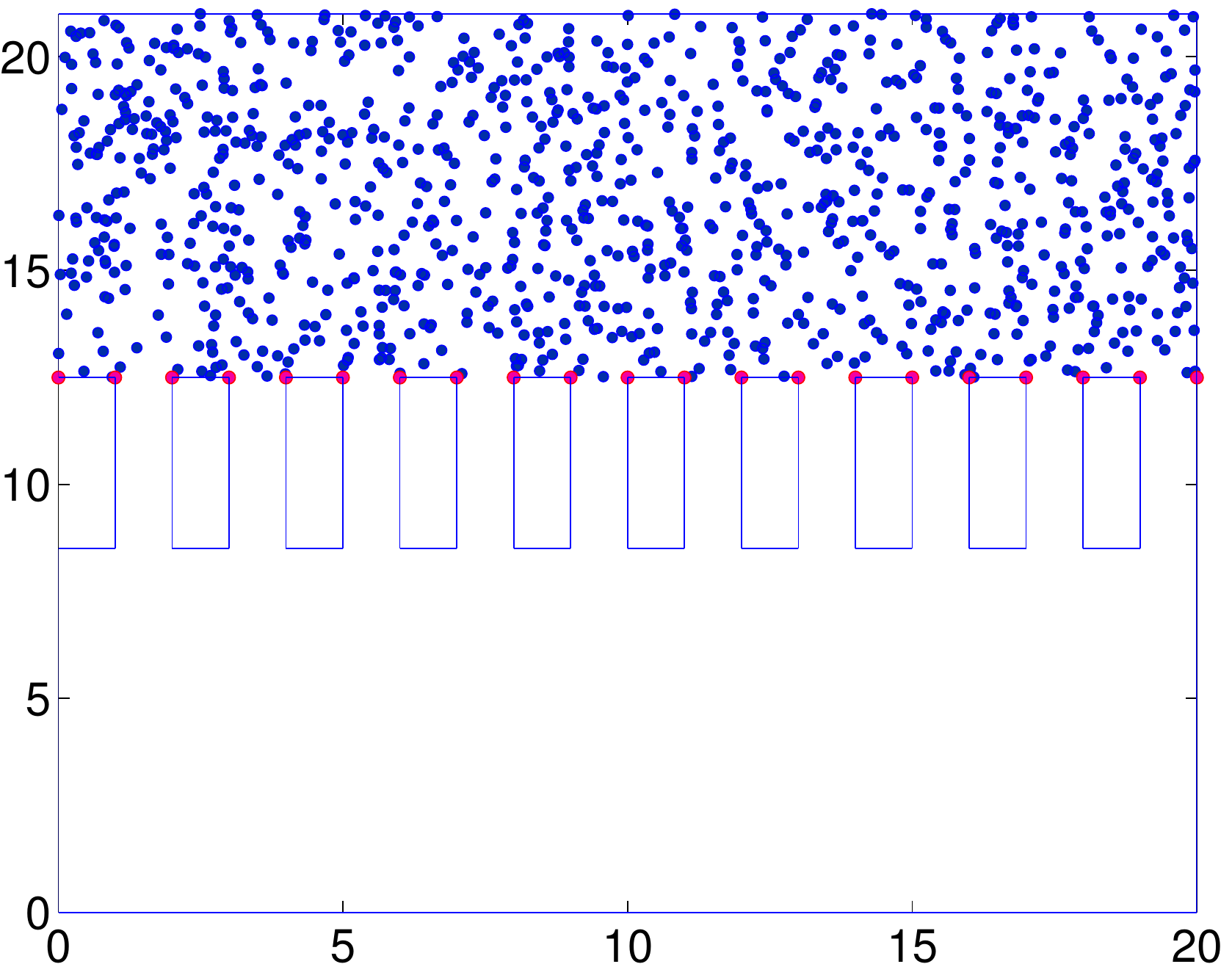}}
%\subfigure[t=1.5]{\includegraphics[scale=0.2]{7-ion_100/Ions/Ions=1501.eps}}
\subfigure[t=3]{\includegraphics[scale=0.25]{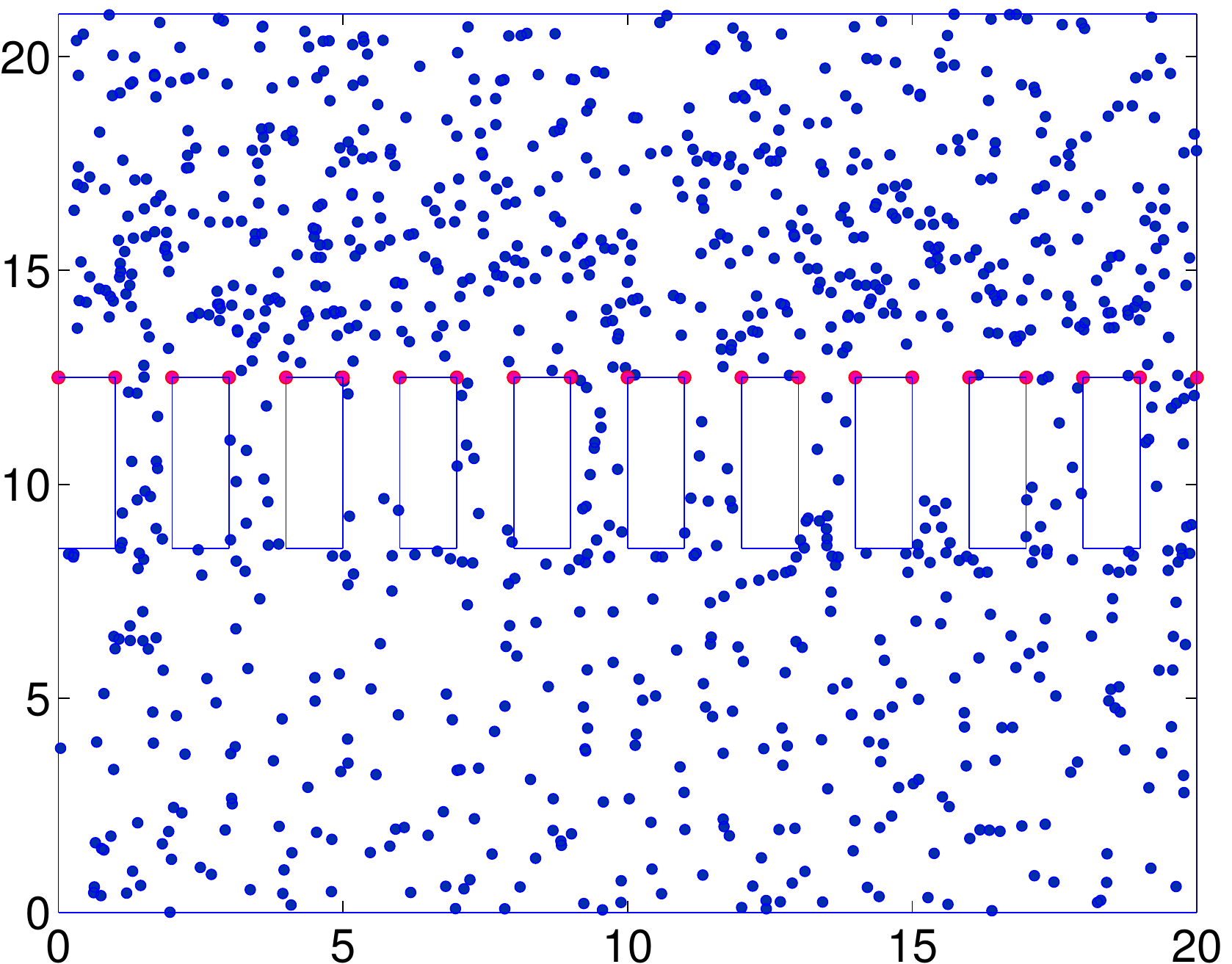}}
\end{center}
\caption{ Initial and final ion locations - Case 2.a. }
\label{fig:1000_ions_initial_end_scaling_7}
\end{figure}

\begin{figure}[h!]
\begin{center}
\subfigure[t=0]{\includegraphics[scale=0.25]{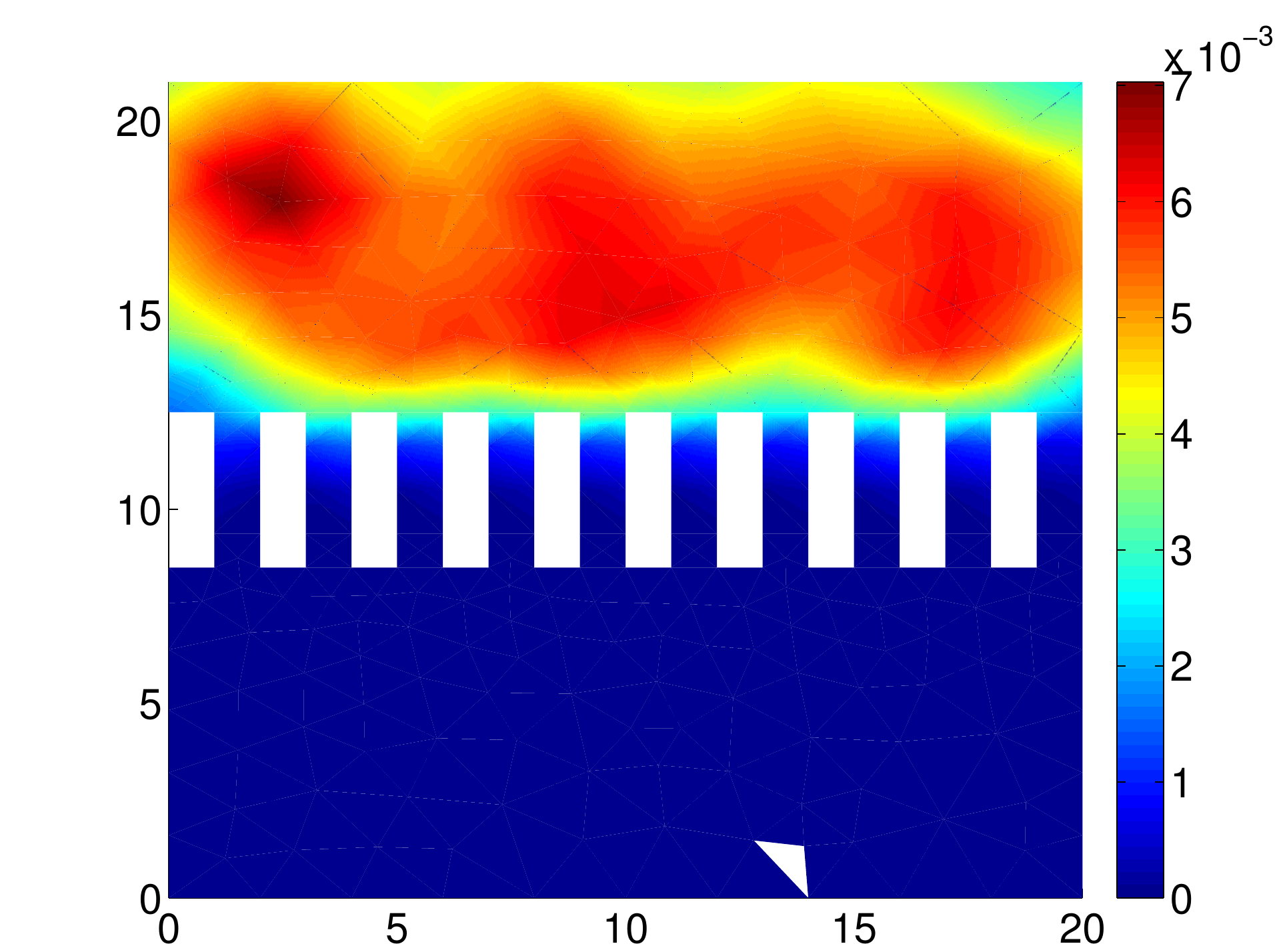}}
\subfigure[t=1]{\includegraphics[scale=0.25]{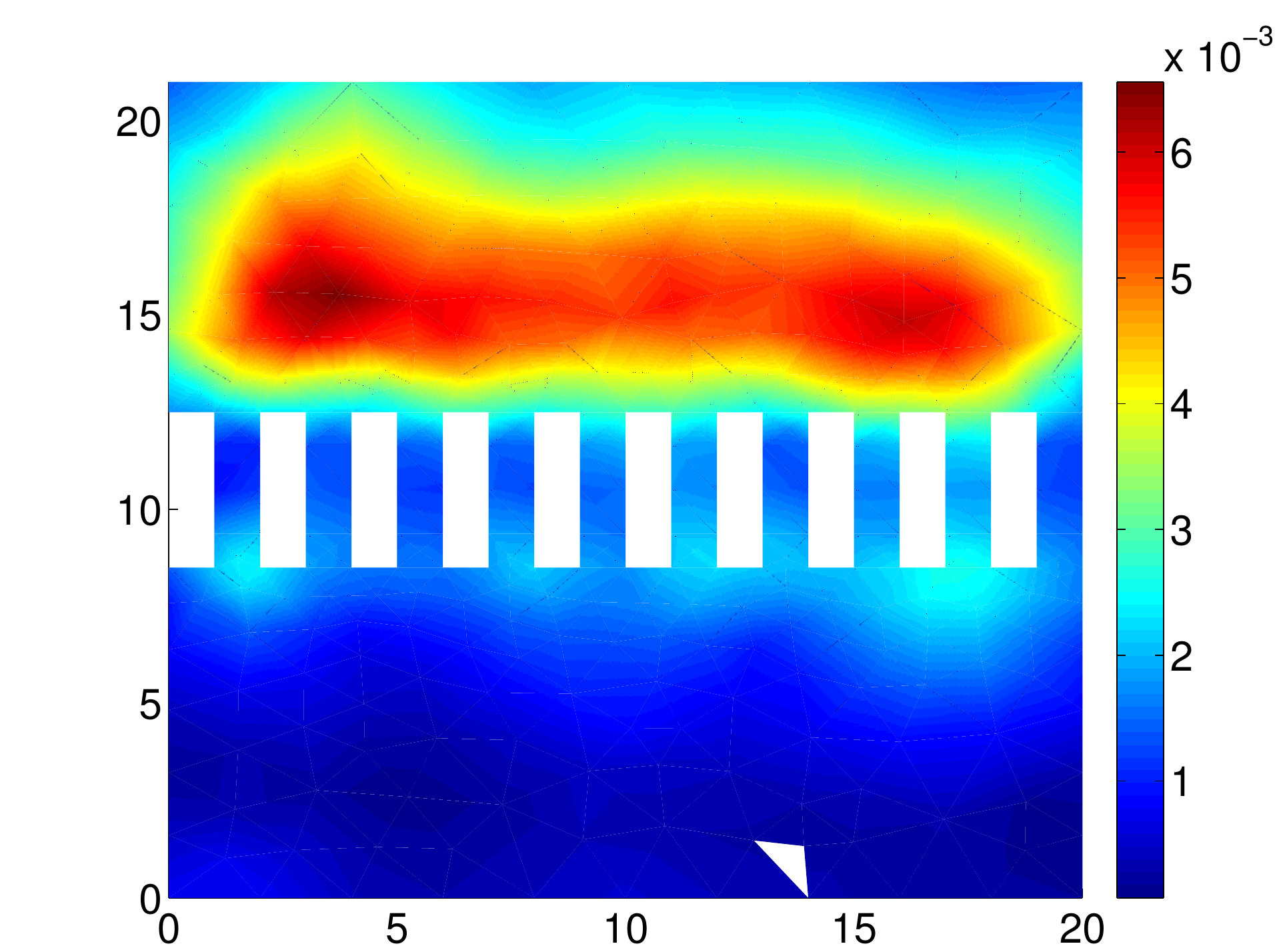}}

\subfigure[t=2]{\includegraphics[scale=0.25]{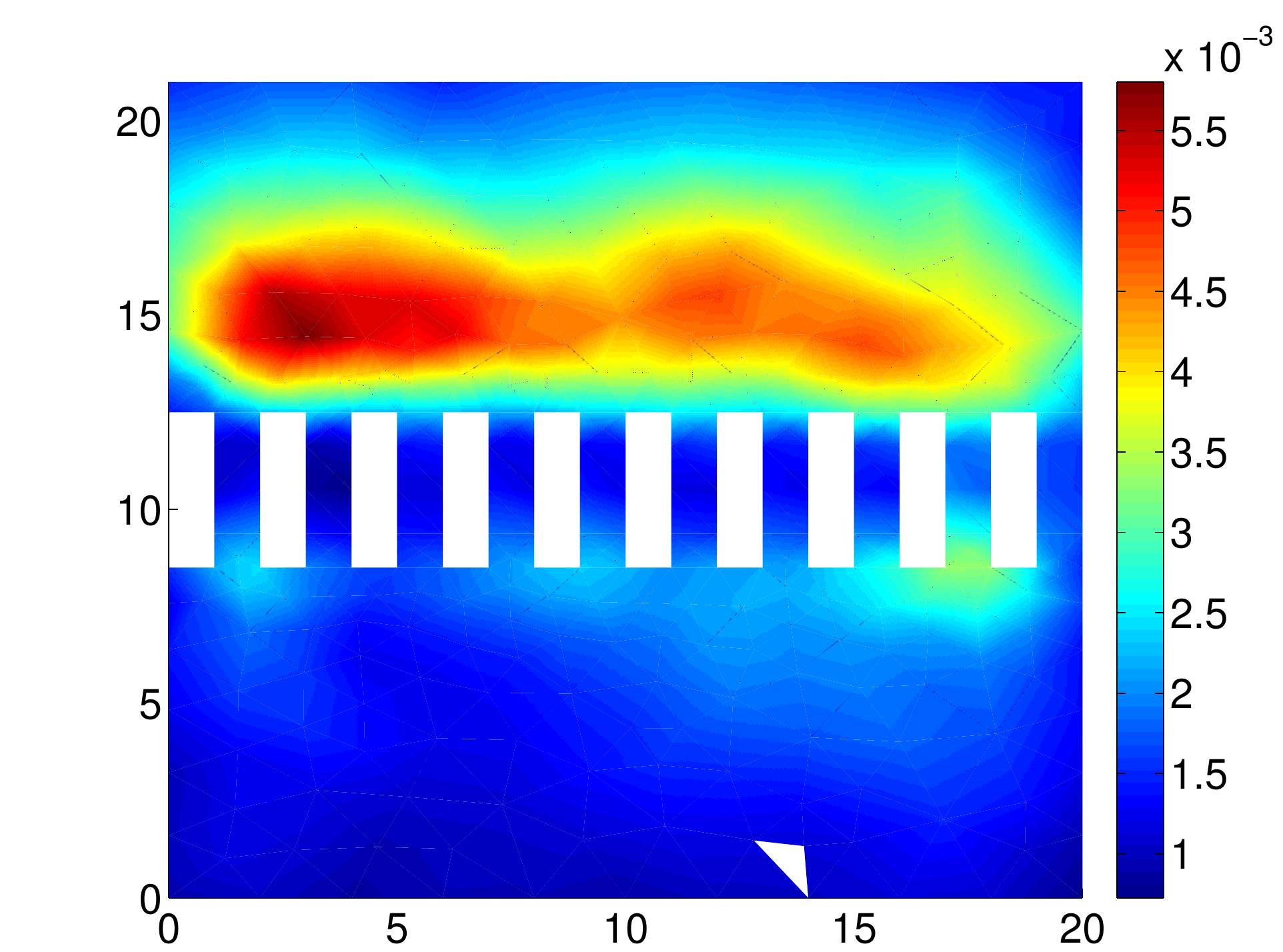}}
\subfigure[t=3]{\includegraphics[scale=0.25]{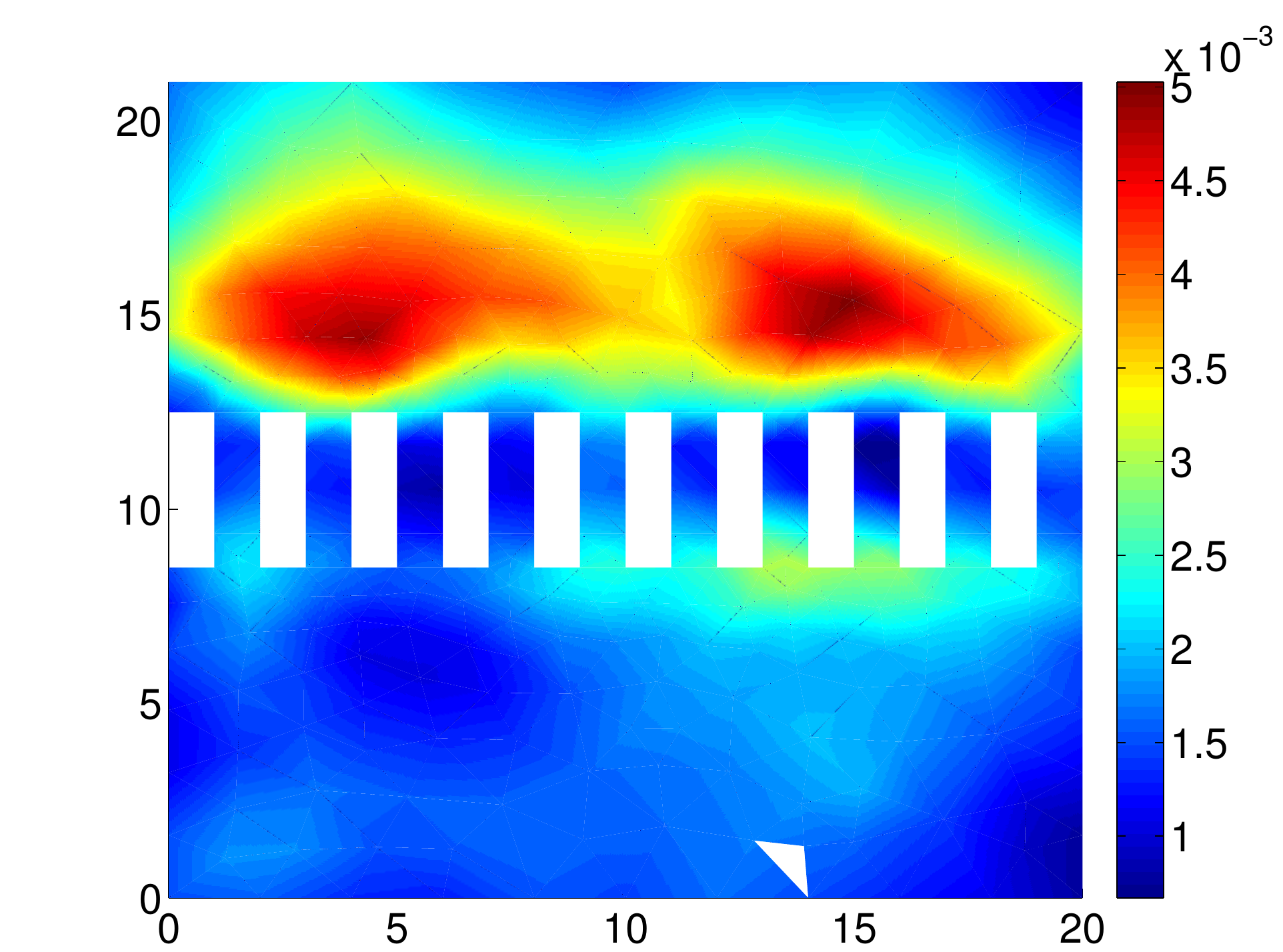}}
\end{center}
\caption{Time evolution of the ion density - Case 2.a.}
\label{fig:density_1_scaling_7_case2a}
\end{figure}

\begin{figure}[h!]
\begin{center}
\subfigure{\includegraphics[scale=0.25]{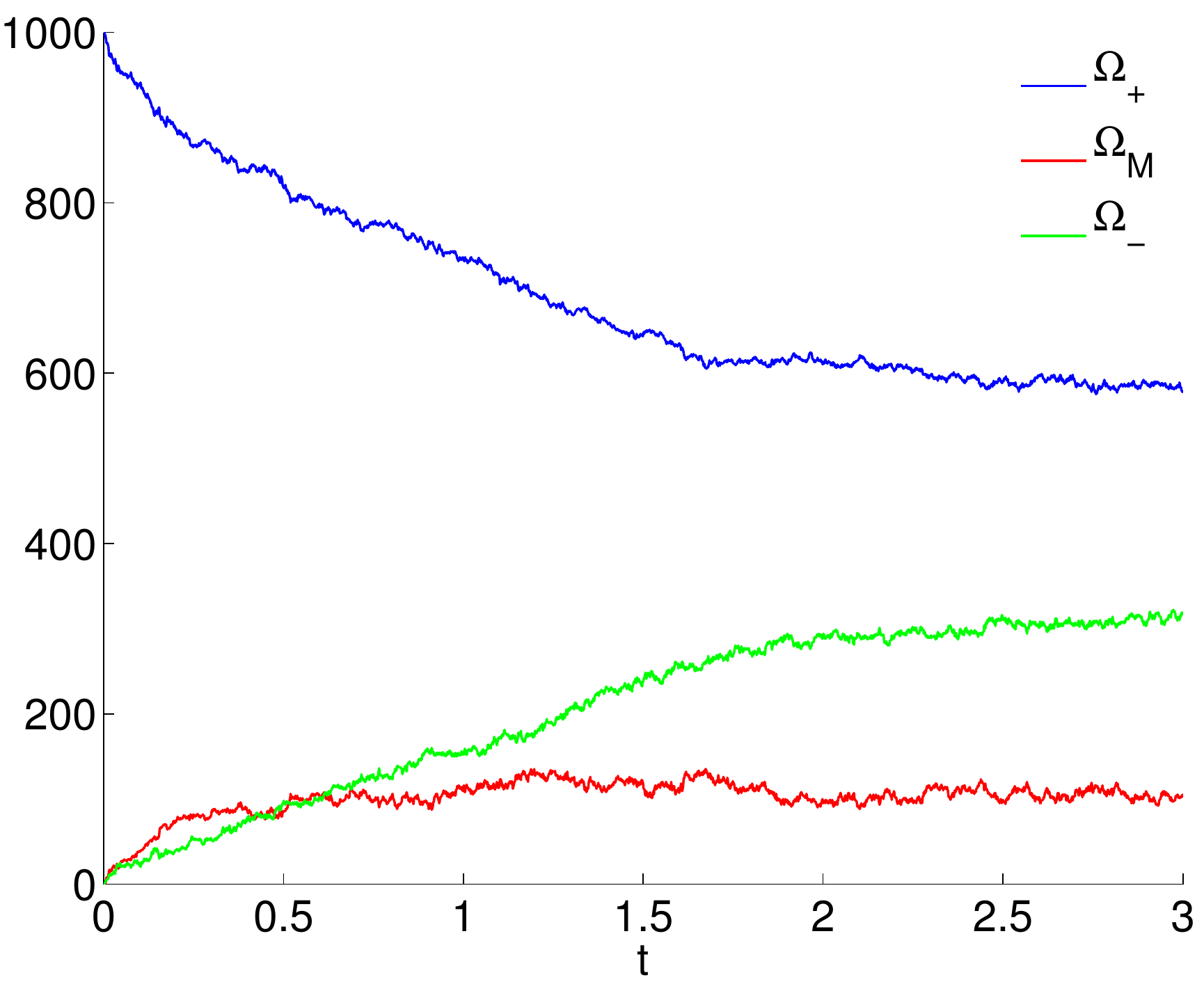}}
\subfigure{\includegraphics[scale=0.25]{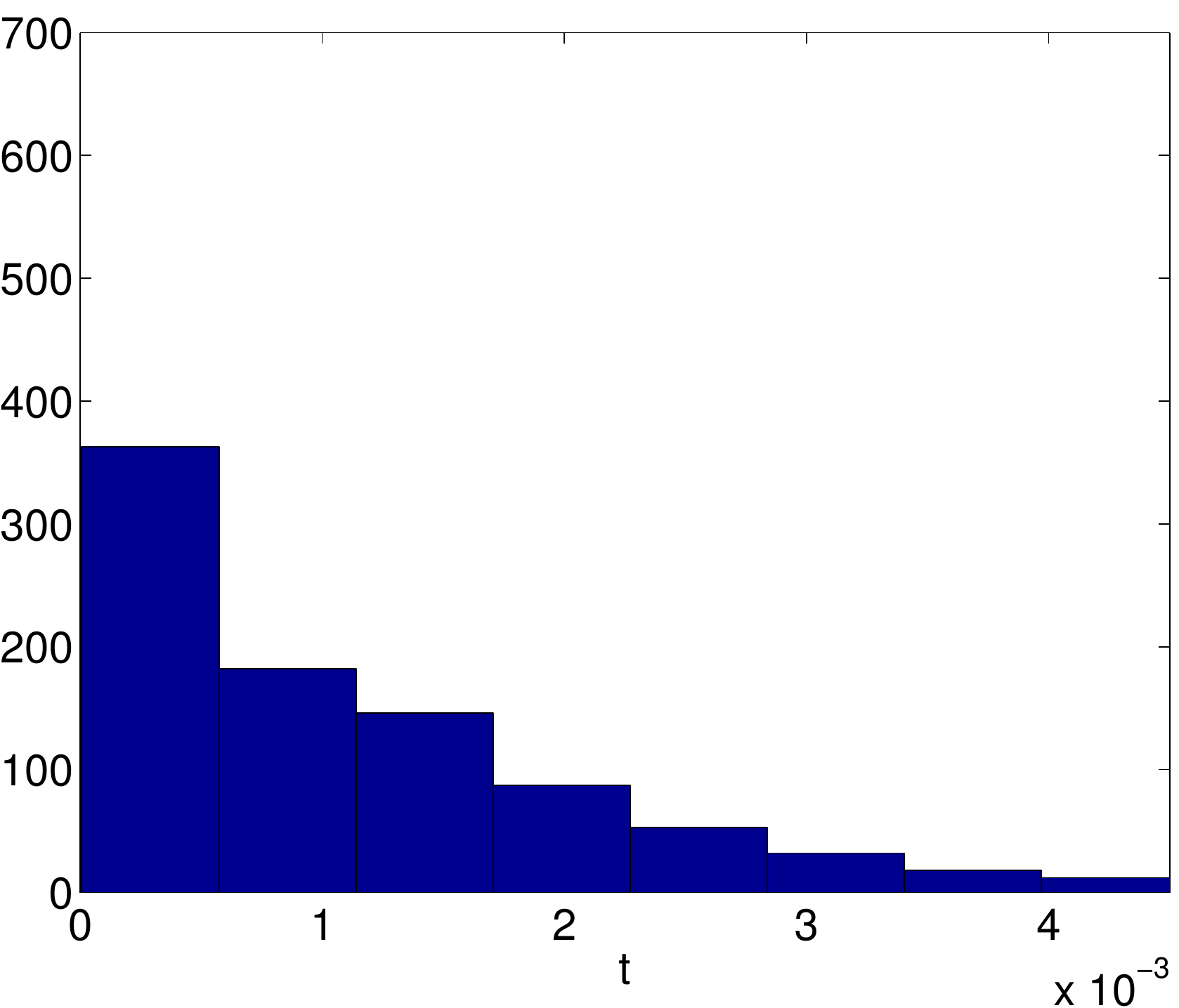}}
\end{center}
\caption{Left: evolution of ion number in
$\Omega_+,\Omega_-,\Omega_M$. Right: histogram of the average time
spent by ions in  $\Omega_M$. Case
2.a.}\label{fig:counting_ions_7_case2a}
\end{figure}

\medskip

Due to the fact that the time rescaling
\eqref{eq:scaling_parametr_t_0} is proportional to $\epsilon_1$,
we expect slower dynamics. This is seen in Figure
\ref{fig:counting_ions_7_case2a}:   again the number of ions in
the channel region are few, and   the velocity of crossing the
membrane is smaller. Such dynamics are confirmed by Figure
\ref{fig:1000_ions_initial_end_scaling_7}--\ref{fig:density_1_scaling_7_case2a}.

\medskip

\paragraph*{Case 2.b. Ion radius $\tilde\epsilon\sim \epsilon_1 \cdot 10^{-2}$}

\begin{figure}[h]
\begin{center}
\subfigure[t=0]{\includegraphics[scale=0.25]{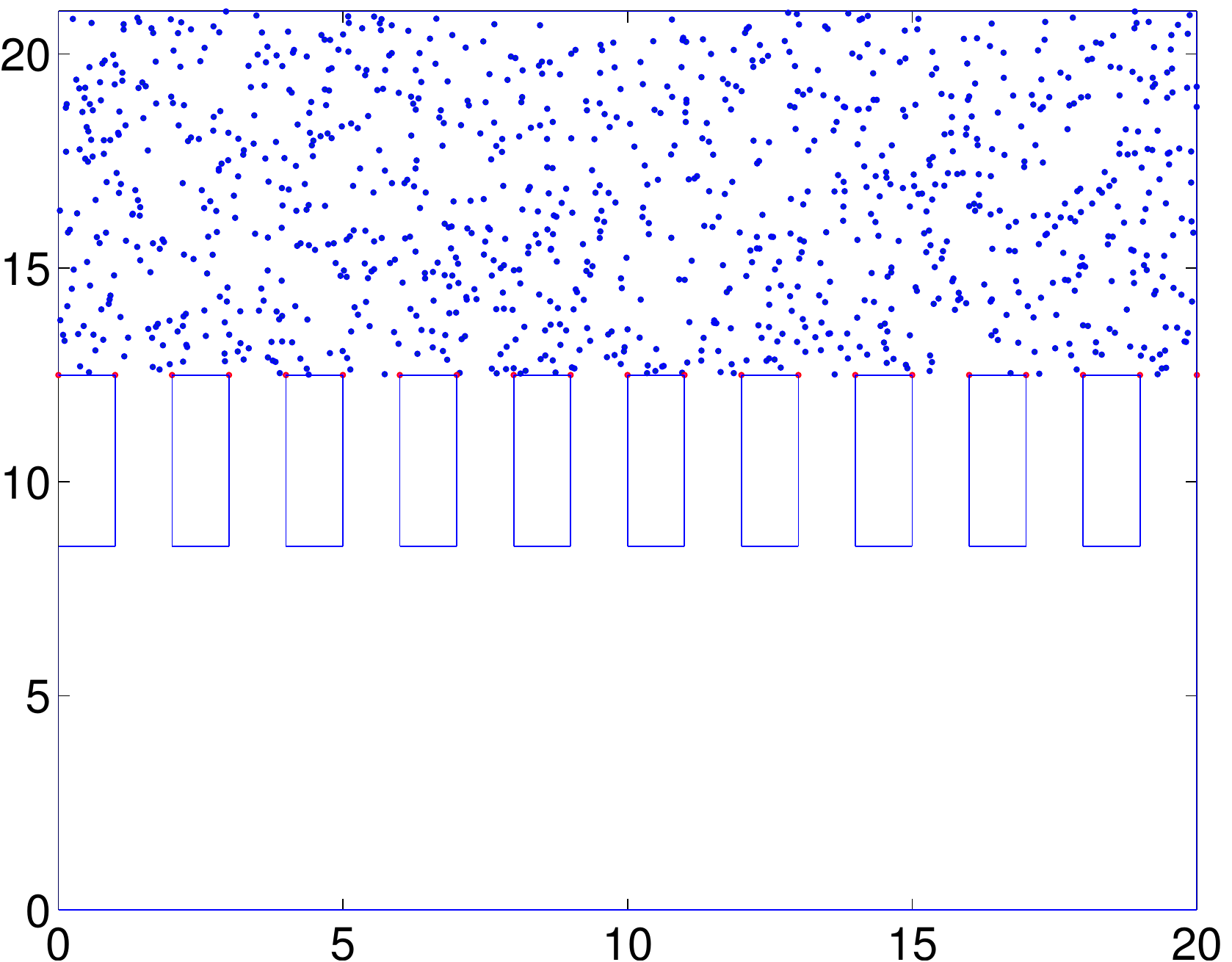}}
%\subfigure[t=1.5]{\includegraphics[scale=0.2]{7-ion_100/Ions/Ions=1501.eps}}
\subfigure[t=3]{\includegraphics[scale=0.25]{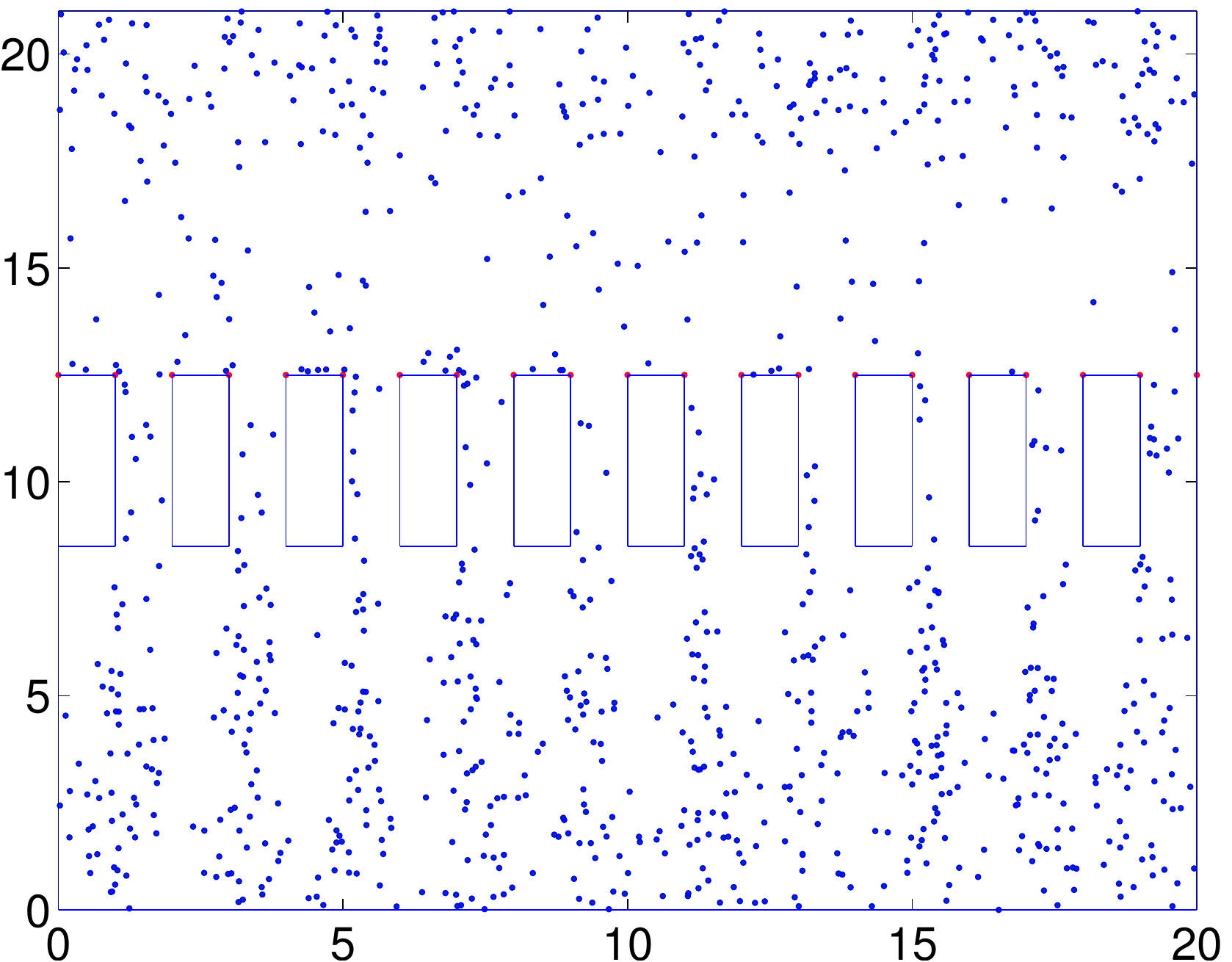}}
\end{center}
\caption{ Initial and final ion locations - Case 2.b. }
\label{fig:1000_ions_initial_end_scaling_7caseb}
\end{figure}

Now again we consider ions much smaller than the channel. Figure
\ref{fig:1000_ions_initial_end_scaling_7caseb} shows the initial
distribution of the ions and the state of the system at time
$t=3$.

\begin{figure}[h!]
\begin{center}
\subfigure[t=0]{\includegraphics[scale=0.25]{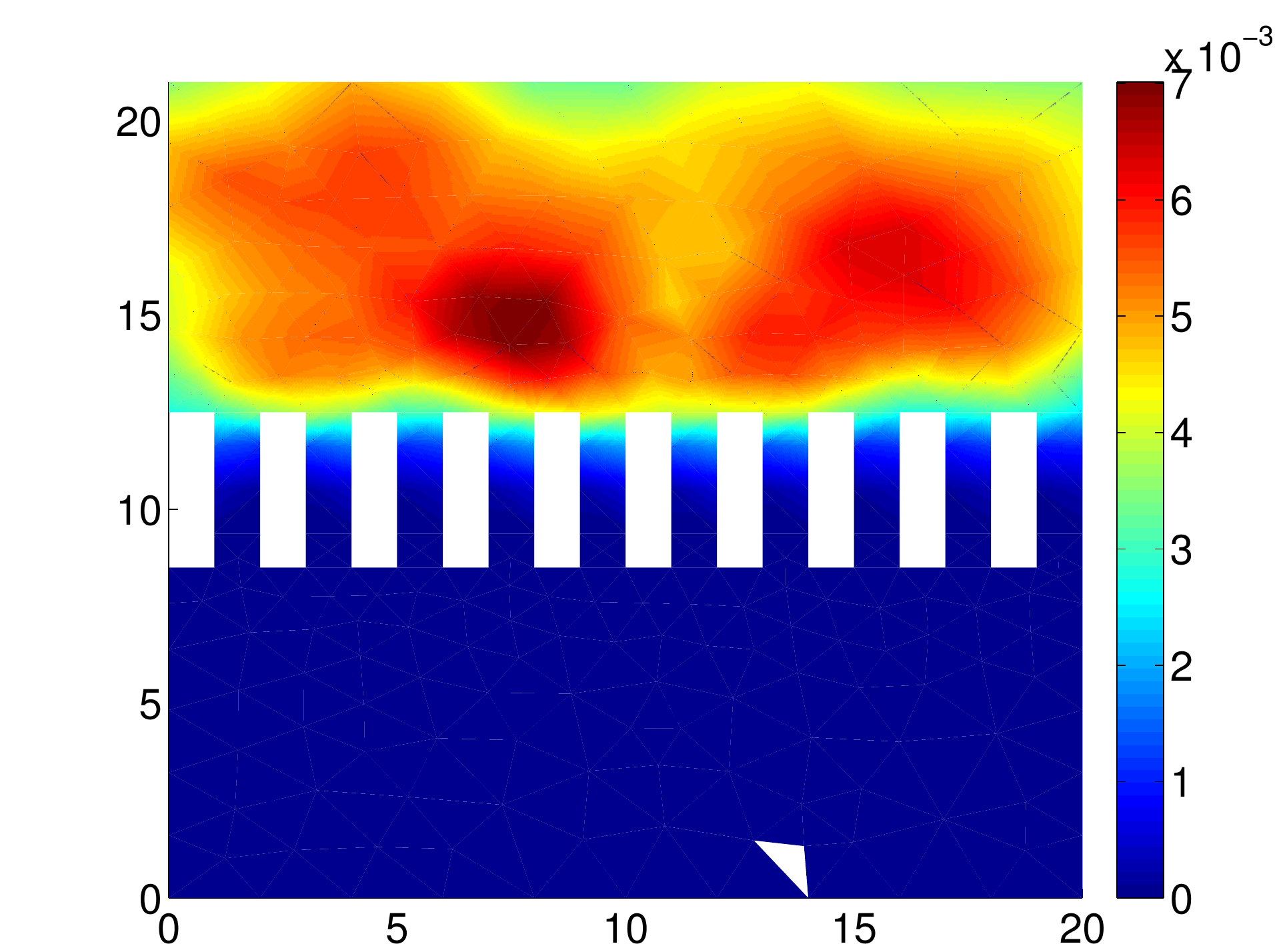}}
\subfigure[t=1]{\includegraphics[scale=0.25]{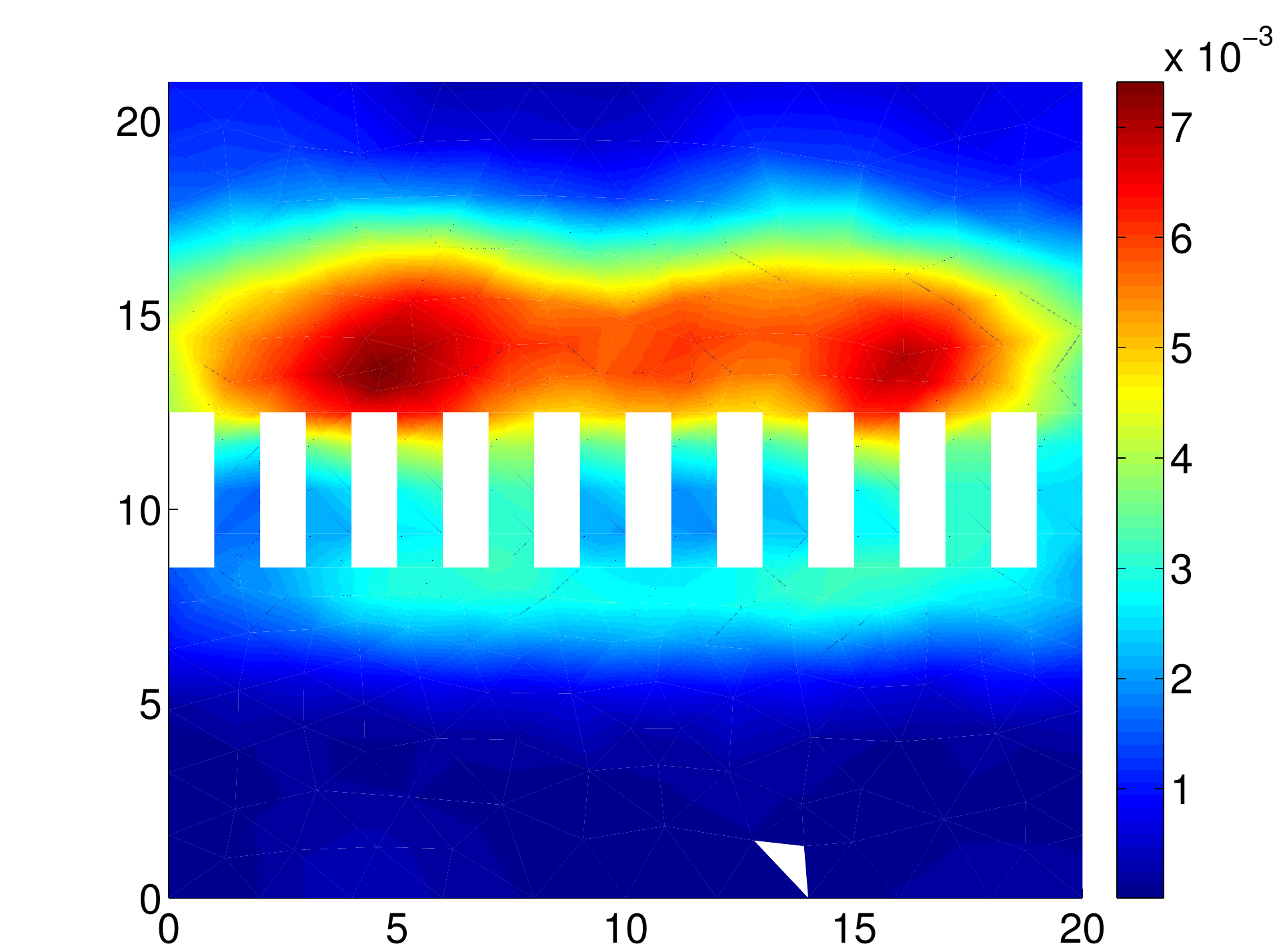}}

\subfigure[t=2]{\includegraphics[scale=0.25]{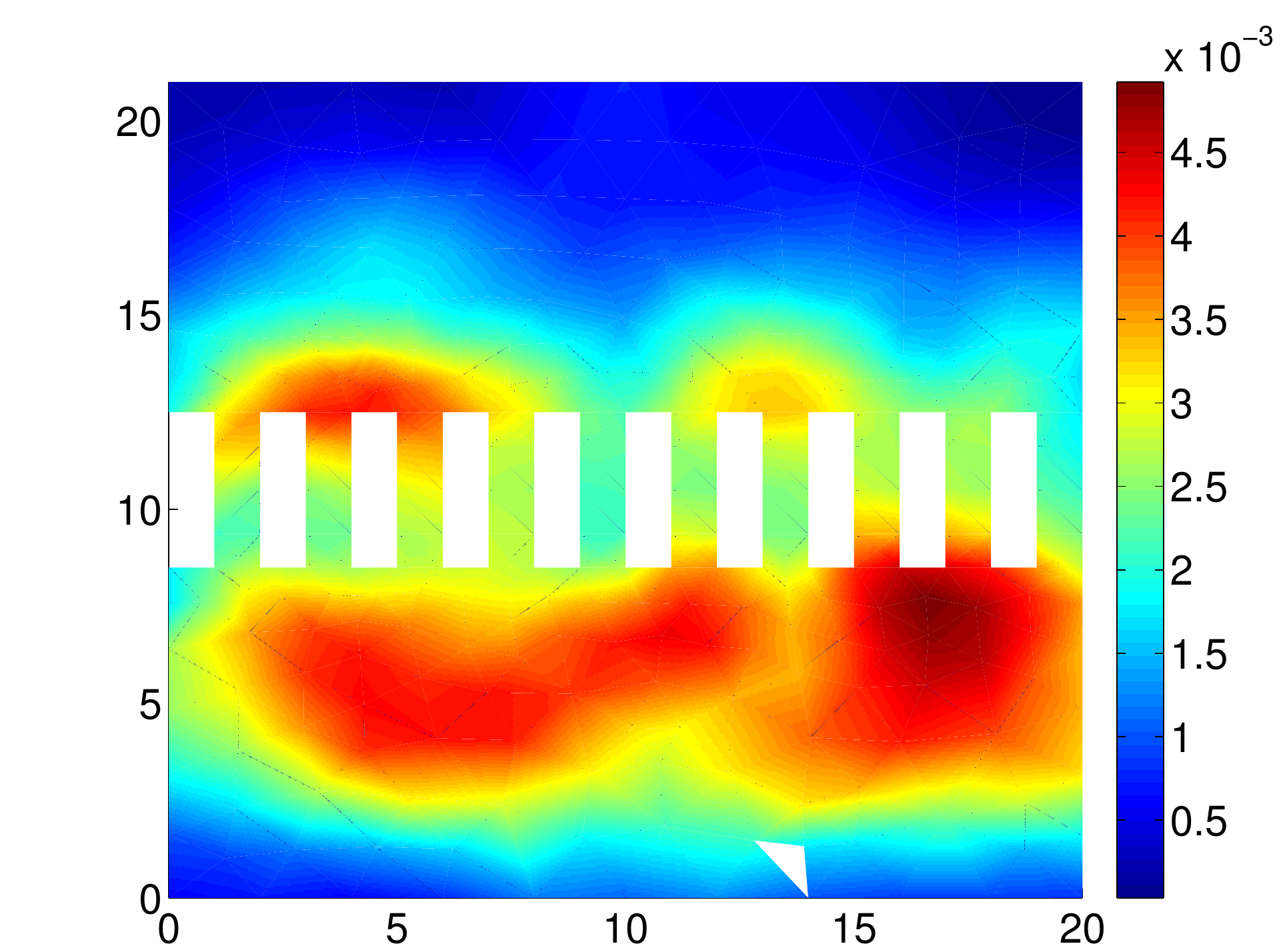}}
\subfigure[t=3]{\includegraphics[scale=0.25]{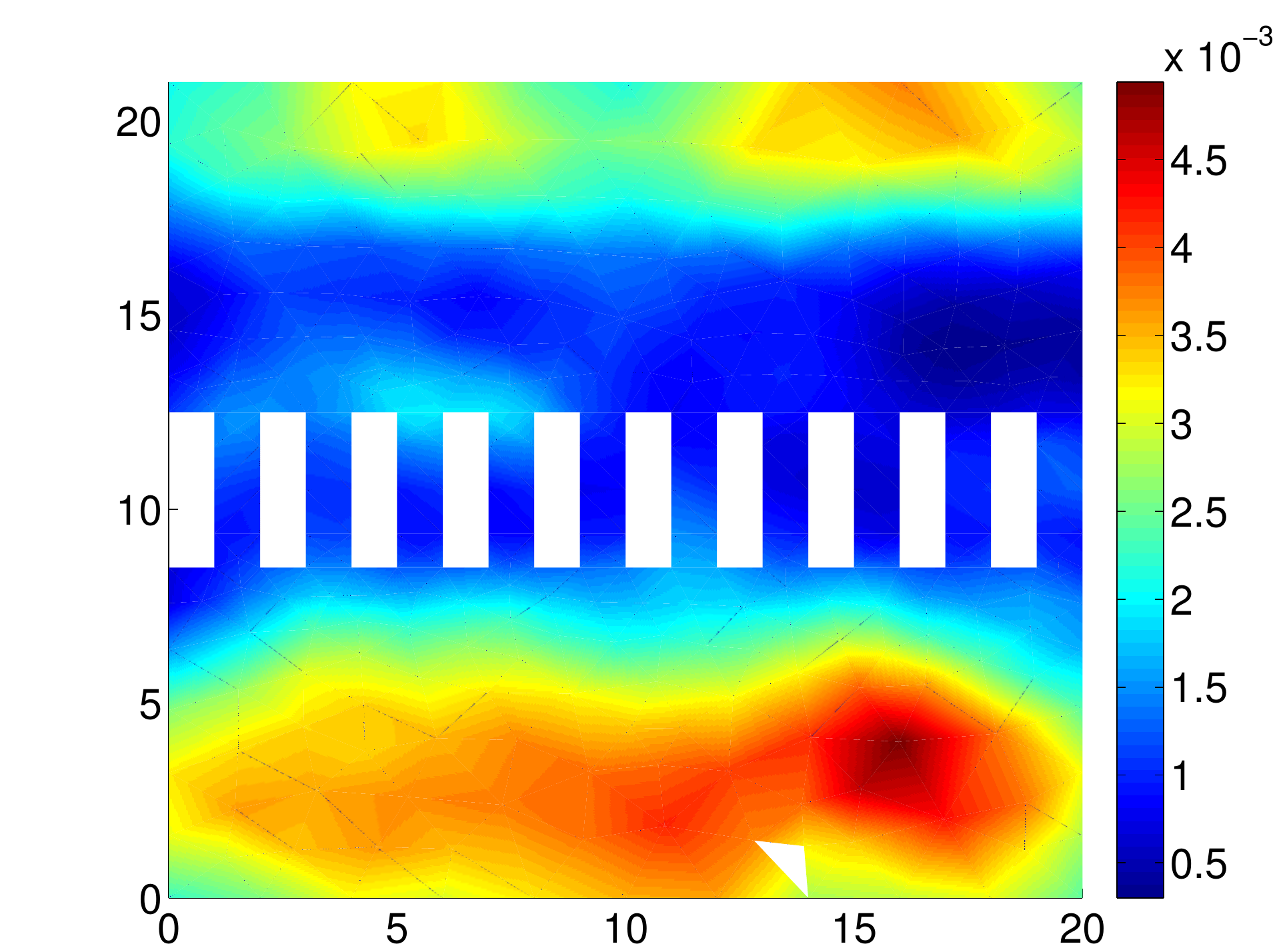}}
\end{center}
\caption{Time evolution of the ion density - Case 2.b.}
\label{fig:density_1_scaling_7}
\end{figure}
\begin{figure}[h!]
\begin{center}
\subfigure{\includegraphics[scale=0.25]{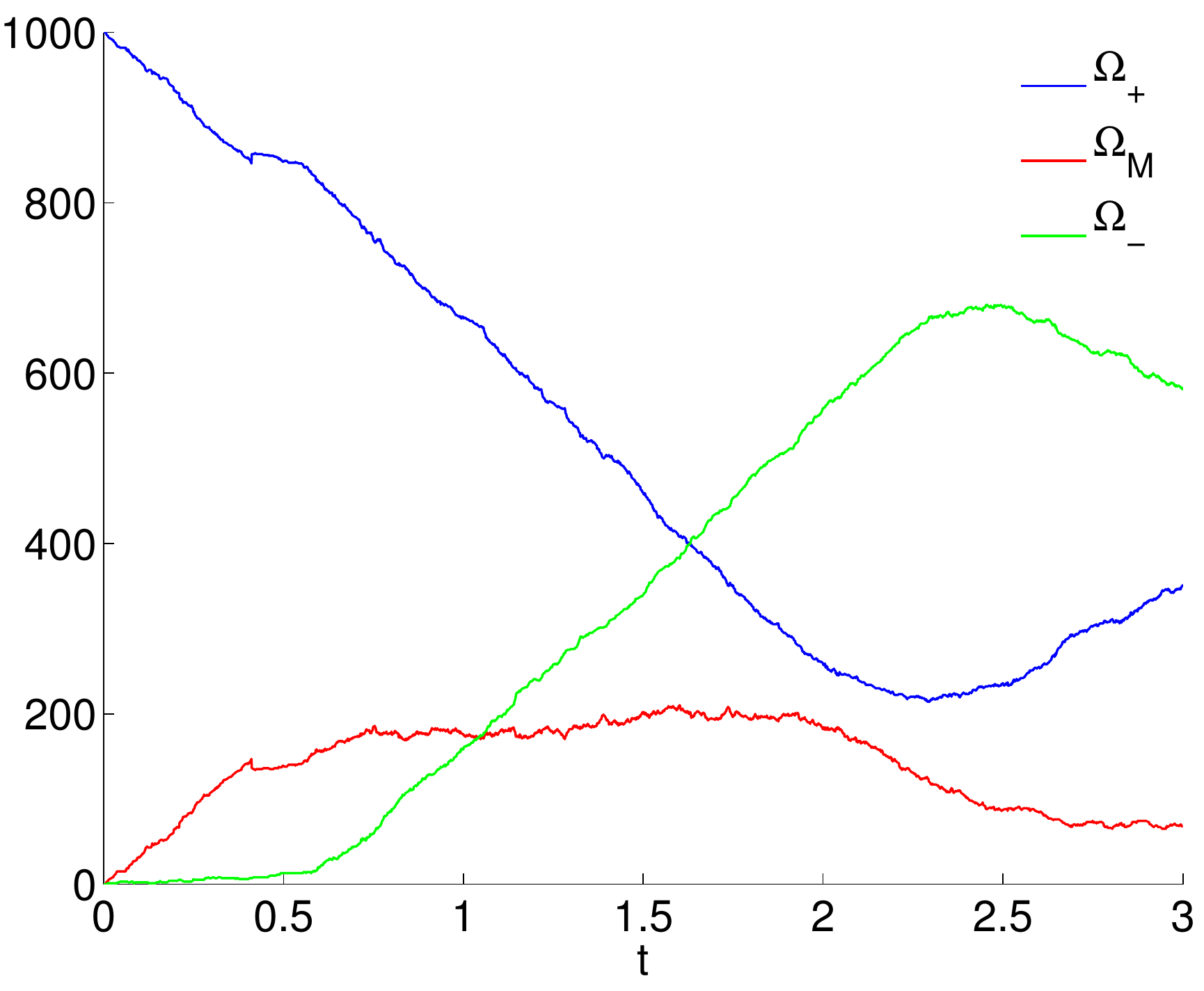}}
\subfigure{\includegraphics[scale=0.25]{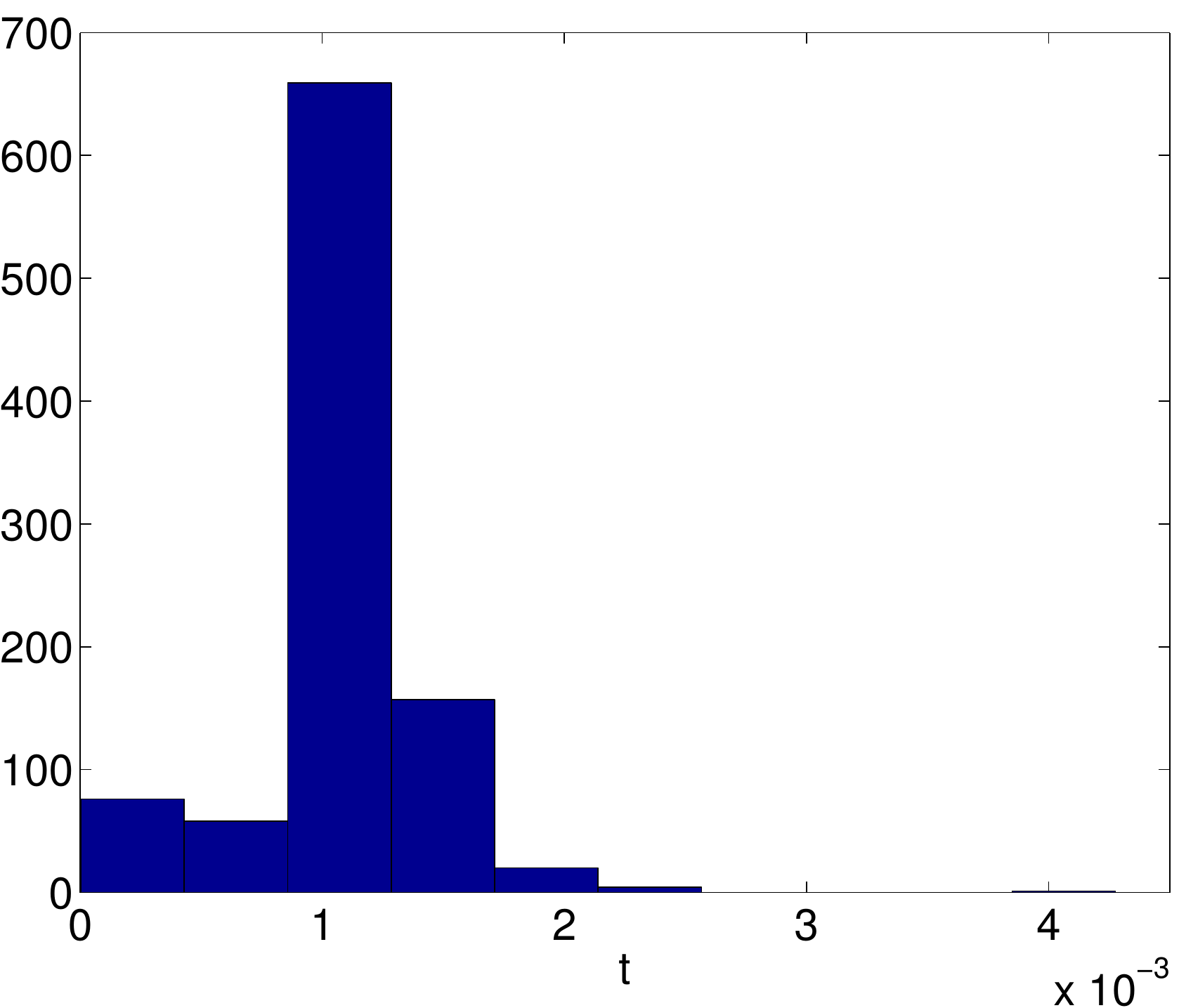}}
\end{center}
\caption{Left: evolution of ion number in
$\Omega_+,\Omega_-,\Omega_M$. Right: histogram of average time
spent by ions in  $\Omega_M$. Case 2.a.}
\label{fig:counting_ions_7}
\end{figure}

By comparing Figure \ref{fig:1000_ions_initial_end_scaling_9_casob} with Figure \ref{fig:1000_ions_initial_end_scaling_7caseb} and Figure \ref{fig:density_1_scaling_7} with Figure \ref{fig:density_1_scaling_9_radius2} it is
evident that again the density in the channel is higher with
respect to the case with bigger ions, but, as already mentioned,
speed is higher.

The qualitative behavior of the time evolution of the number of
ions in each region in the bath and in the membrane as shown in
Figure \ref{fig:counting_ions_7}, is similar to the one of case
1.b. in Figure \ref{fig:counting_ions_7_case2a}.

\medskip

 Quantitative estimates for the four case studies
  are shown in Tables \ref{table:crossing_channel_time} and \ref{table:estimates_number_ions}. The mean time needed by
ions for crossing the channel is higher    in the case 1.a. and
always the mean number of ions in the membrane region are smaller.

\begin{table}[h!]\footnotesize
\begin{center}
\begin{tabular}{ c|c c }
  \hline
  Case & Mean & Standard deviation \\
  \hline
  1.a. & 0.0885 & 0.0400 \\
  1.b. & 0.1450 & 0.0142 \\
  2.a. & 0.1122 & 0.1065 \\
  2.b. & 0.1434 & 0.0503 \\
  \hline
\end{tabular}\end{center}\vspace{3mm}
 \caption{Crossing channel time: mean
and standard deviation in the four cases. }
\label{table:crossing_channel_time}
\end{table}

\medskip

\begin{table}[h!]\footnotesize
\begin{center}
\begin{tabular}{ c|l| r r r }
  \hline
  Case &  & $\Omega_+$ & $\Omega_-$ & $ \Omega_M $\\
  \hline
  1.a. & Mean &  498.03      &   414.59     & 87.38 \\
       & Std    &   156.97   &   170.99     & 30.13  \\
  1.b. & Mean &  377.71      &   477.29     & 144.99 \\
       & Std    &  312.09    &   334.92      & 103.75  \\
  2.a. & Mean &  692.46      &  206.42      & 101.11 \\
       & Std    & 109.69     &   20.93     &  53.23 \\
  2.b. & Mean &    518.83    &   341.35     &  139.81\\
       & Std    &   255.30   &    254.16    &  96.14 \\
  \hline
\end{tabular}
\end{center}\vspace{3mm}\caption{Estimates of the number of ions in each region: mean and
standard deviation (Std) in the four cases. }
\label{table:estimates_number_ions}
\end{table}

\subsection{The effective  diffusion coefficient }

As discussed in \cite{smithsansom}, the effective diffusion
coefficient is different inside and outside the membrane. There
are several ways proposed in literature for computing such an estimation
\cite{AKC,pokern,smithsansom}. A simple way to estimate the
effective diffusion is via the mean square displacement of an ion.
In our system, even though the diffusion coefficient governing the
randomness of the velocity is constant, the effective diffusion
estimated as the mean square displacement of an ion  also depends
upon   on all the exerted nonlinear forces. In particular in the
channel, the boundary forces exerted on the ions are much stronger
then in the bath, and they also influence the particle mean square
displacement.

\medskip

Split the time interval $[0,T]$ in $L$ time intervals of length
$\Delta t$ and let $t^n=n\Delta t, n=1, \dots,L$, where we defined
for each $\Delta t=10^{-3}$ the Euclidean distance $\delta_n^{i}$
as follows:
\begin{equation*}
\delta_{n+1}^{i}=||X^i(n+1)-X^i(n)||_2
\end{equation*}
where both $X^i(n+1)$ and $X^i(n)$, the positions of the $i$th
ion at time $t^n$ and $t^{n+1}$, lie in a region of the same area
$1\times k$ in $\Omega_M$, or $\Omega_+$; see Figure 2. Let
$\delta$ be the sample vector of such distances
$$\delta=(\delta_1^{1},\dots,\delta_1^{i_1}, \dots,\delta_{L}^{1},\dots,\delta_L^{i_L} ),$$
where $i_n$ is the number of ions feasible for the sample at time
$t^n$. Let $d=i_1+\ldots+ i_L$ be the size of the sample. We
estimated the mean square displacement by unit time by means of
the sample variance $V$ of the vector $\delta$ divided by $\Delta
t$.

\medskip

 Table \ref{table:effective_diffusion} shows the
estimation of the mean square displacement per unit time  in
$\Omega_+$ and in the membrane $\Omega_M$.   We may notice
how  the order of the mean square displacement is smaller in the
channel, as expected.

\medskip

\begin{table}[h!]\footnotesize
\begin{center}
\begin{tabular}{l*{6}{c}r}
\hline
& Region & $V$ & $V/\Delta t$ & $V_y$ & $V_y/\Delta t$ \\
\hline
\multirow{3}{*}{Case 1a}
&$\Omega_M$      & $9.8284~10^{-4}$ & $0.982$ & $0.0383$ & $38.3020$ \\
%&$\Omega_M^{nb}$ & $4.1849~10^{-1}$ & $418.4937$ & $0.2786$ & $278.6120$ \\
&$\Omega_+$ & $2.9685~10^{-1}$ & $296.8525$ & $0.1532$ & $153.2315$ \\
\hline
\multirow{3}{*}{Case 1b}
& $\Omega_M$ & $5.5878~10^{-6}$ & $0.0056$ & $5.4420~10^{-6}$&$0.0054$ \\
%&$\Omega_M^{nb}$&$0.0819$&$81.9460$&$0.0718$&$71.8056$ \\
&$\Omega_+$&$0.2576$&$257.6380$&$0.2428$&$242.8094$\\
\hline
\multirow{3}{*}{Case 2a}
&$\Omega_M$&$0.0025$&$2.4511$&$0.0016$&$1.5633$ \\
%&$\Omega_M^{nb}$&$0.7075$&$707.4922$&$0.4583$&$458.3457$\\
&$\Omega_+$&$0.7206$&$720.6112$&$0.4286$&$428.8674$ \\
\hline
\multirow{3}{*}{Case 2b}
&$\Omega_M$&$0.0054$&$5.3956$&$3.7941~10^{-4}$&$0.3794$\\
%&$\Omega_M^{nb}$&$0.1278$&$127.7718$&$0.0846$&$84.6272$\\
&$\Omega_+$&$0.1381$&$138.0646$&$0.0644$&$64.3650$
\end{tabular}
\end{center}\vspace{3mm}
\caption{Estimation of the mean square displacement of ions and
per unit time in the $(x,y)$-plane ($V$ and $V/\Delta t$) and
along the $y$-direction ($V_y$ and $V_y/\Delta t$).   }
\label{table:effective_diffusion}
\end{table}

\section{Discussion}

We  have considered  a fully stochastic mathematical  model
describing  the main characteristics of a multiple channel system,
in which ion movement in the bath and throughout membrane channels
is modeled in terms of a system of stochastic differential
equations, coupled with a Poisson  equation, which becomes itself
stochastic due to the consequent randomness   of the ion
positions.

\medskip

By considering real parameters as  in Tables \ref{table:vandewalls}
and \ref{table:parameters}, we  have faced the problem of
identifying  a  right scale for  the nondimensionalization of
the system.

\medskip

Through a direct comparison of the simulation results of the
rescaled system via typical scales  of the channels, we see that a
first rescaling by a channel size of order $\epsilon_1\sim
10^{-10}$ is  fast enough to make   the dynamics evident. Instead,
from Table \ref{table:crossing_channel_time}, we may see
 that the mean time  required to cross the membrane is significantly higher in the case
of the second rescaling  by  $\epsilon_1\sim 10^{-7}$.

Furthermore, we  may notice that with the same rescaling, smaller
particles are slower, and a change of velocity is less probable.
This is due to the fact that with equal parameters, the smaller
the diameter of  the  ions, the weaker  are  the forces  of
Pauling and Lennard--Jones acting on the velocity, since they
depend on the ion diameter,  as shown by
\eqref{eq:rescaled_pauling} and \eqref{eq:rescale_LJ}.

\medskip

The second issue we are  interested in   is the coupling of the
dynamics  inside and outside the membrane. This is particularly
important in the case of nanopores (cases 1.a. and 2.a). Indeed,
as is well shown in  Figure \ref{fig:density_1_scaling_9},
while there is a very high number of ions in the bath region, this
is not so  in the membrane, since the number of ions within  each
channel becomes very small.

For the  time being we propose  case 1.a. as a satisfactory
rescaling of the system. In such a case, the dynamics have to be
maintained to always be stochastic and discrete.

\medskip

A    future step, interesting from the mathematical point of view,
is to study the same system when both the number of channels and
ions are increasing.
  The interesting mathematical question, which is  relevant
also for a reduction of the computational costs in the
si\-mu\-lations, concerns   the  coupling of  the  two scales
regarding the dynamics in the bath and  within the channels, with
respect  to the  significant difference  in the number of ions per
unit volume for the two regions.

\medskip

As mentioned  in the introduction,  it is well known that it is
reasonable to consider a continuous model for the ion
concentrations  whenever a  law of large numbers may be applied,
i.e. when the number of ions per unit volume  is sufficiently
large. On the other hand  such laws cannot  be applied in each
channel, so that   the system has to be kept  discrete and
stochastic. As a consequence,   for a  large number  of ions in
the bath the problem of coupling their dynamics in the bath
(averaged continuum ), and inside the channels (discrete and
stochastic) arises. An interesting issue refers, in particular, to
the transition conditions at  the  membrane boundaries. As already
mentioned, some work in this direction has already   been done,
for example by Schuss and his collaborators, but, to the best of our
knowledge, existing literature has not provided  a satisfactory
answer  yet.

\medskip

 {
  The present study may be regarded as a first step for
further investigations that the authors intend to carry on for
what it may concern the asymptotic analysis of the Langevin type
model for ionic permeation. Moreover, the presented model might be
given in a more realistic setting through the introduction of more
realistic physiological details, as discussed in the
introduction.}

%%%%%%%%%%%%%%%%%%%%%%%%%%%%%%%%%%%%%%%%%%%%%%%%%%%%%%%%%%%
%%% APPENDIX
%%%%%%%%%%%%%%%%%%%%%%%%%%%%%%%%%%%%%%%%%%%%%%%%%%%%%%%%

%\Appendix

\section*{Appendix: Computing the rescaled equations}
 \label{app:rescaling}

\medskip

\noindent {\em The Poisson Equation.}\ By considering the
rescaling \eqref{eq:time_space_rescaling} and
\eqref{eq:scaling_potential}, from \eqref{eq:poisson} we
rewrite the rescaled term
$-\textrm{div}_{x_{s}}{\left(\alpha_{r}\nabla_{x_{s}}\Phi_{s}\left(t_{s},x_{s}\right)\right)}
$ involving the rescaled potential field $\Phi_s(t_s,x_s)$, where
$\alpha_{r}$ satisfies \eqref{eq:rescaled_alpha}. Then we
have the following:
\begin{eqnarray*}
-\textrm{div}_{x_{s}}{\left(\alpha_{r}\nabla_{x_{s}}
\Phi_{s}\left(t_{s},x_{s}\right)\right)} &=&
 -\epsilon^{2}\textrm{div}_{x}\left(\dfrac{\alpha_{w}}{\alpha_{0}}
 \nabla_{x}\dfrac{\Phi\left(t,x\right)}{\tilde{\Phi}}\right)\\&=&-\dfrac{\epsilon^{2}}{\alpha_{0}\tilde{\Phi}}\textrm{div}_{x}
\left(\alpha_{w}\nabla_{x}\Phi\left(t,x\right)\right) \\
&=&
\dfrac{\epsilon^{2}q}{\alpha_{0}\tilde{\Phi}}\left(z{\displaystyle
\frac{1}{N\kappa_1 }
\sum_{k=1}^{N}K_{\kappa_1}\left(x-X_s^{k}(t)\right)}
\right. \\
&& \hspace{1cm}\left.+z_{F}{\displaystyle \frac{1}{J\kappa_2 } \sum_{h=1}^{J}K_{\kappa_2}\left(x-X_s^{h}(t)\right)}\right) \\
\end{eqnarray*}
By considering the definition of the interacting kernels
\eqref{eq:V_and_W} and the rescaled ones
\eqref{eq:rescaled_potential_K_W}, one obtains
\begin{eqnarray*}
-\textrm{div}_{x_{s}}{\left(\alpha_{r}\nabla_{x_{s}}
\Phi_{s}\left(t_{s},x_{s}\right)\right)} &=&
-\epsilon^{2}\textrm{div}_{x}\left(\dfrac{\alpha_{w}}{\alpha_{0}}
\nabla_{x}\dfrac{\Phi\left(t,x\right)}{\tilde{\Phi}}\right)  \\
&=& \dfrac{\epsilon^{2}q}{\alpha_{0}\tilde{\Phi}\epsilon^{3}}
\left(z{\displaystyle \frac{1}{N\kappa_1^s}\sum_{k=1}^{N}   U_{1}
\left(\epsilon\left(x_{s}-X_s^{k}(t_s)\right)\right)} \right. \\
&& \hspace{1.3cm}\left.+z_{F}{\displaystyle
\frac{1}{J\kappa_2^s}\sum_{h=1}^{J}
 U_{2}\left(\epsilon\left(x_{s}-Y_s^{h}\right)\right)}\right)\\
 &=&    \dfrac{\epsilon^{2}q}{\alpha_{0}\tilde{\Phi}\epsilon^{3}}
\left( z \lambda {\displaystyle\frac{1}{N\kappa_1^s}
\sum_{k=1}^{N}U_{1}^{s}\left(x_{s}-X_s^{k}(t_s)\right)} \right. \\
 &&\hspace{1.3cm}\left.+
 z_{F} \frac{1}{J\kappa_2^s}\lambda {\displaystyle \sum_{h=1}^{J}U_{2}^{s}\left( x_{s}-Y_s^{h}\right)}\right)
\end{eqnarray*}
So, finally, the right-hand side term becomes
$$ \dfrac{\epsilon^{2}q}{\alpha_{0}\tilde{\Phi}\epsilon^{3}}
 \left(z \left(K_{\kappa_1^s}^{s}*X^{s}(t_{s})\right)(x_{s})+z_{F}
 \left(K_{\kappa_2^s}^{s}*Y^{s} )\right)(x_{s})\right);
$$
hence, by means of the definition \eqref{eq:lambda}, we obtain the
rescaled Poisson equation \eqref{eq:rescaled_poisson}.

\medskip

\noindent {\em The Langevin system.}\ \ The nondimensionalization
of the Langevin equations
\eqref{Langevin_equation_nonscaled_1}-\eqref{Langevin_equation_nonscaled_2}
has been done as follows. We scale first the location equation,
obtaining, for any $j=1,\dots,N$
\begin{eqnarray*}
dX^j_{s}\left(t_{s}\right)&=&\epsilon^{-1}dX^j\left(t\right)=\epsilon^{-1}V^j\left(t\right)dt\\
&=&\epsilon^{-1}\dfrac{\epsilon}{t_{0}}V^j_{s}\left(t_{s}\right)t_{0}dt_{s}=V^j_{s}\left(t_{s}\right)dt_{s},\end{eqnarray*}
that is
$$
dX^j_{s}\left(t_{s}\right)=V^j_{s}\left(t_{s}\right)dt_{s}.
$$
Then, we scale the equation describing the evolution of the
velocity $V^j_s$, for any $j=1,\dots,N$. From
\eqref{scaling_space_time} we have that
\begin{equation}\label{eq:rescaling_velocity_2}
m_{s}d_{t_{s}}V^j_{s}\left(t_{s}\right)=
\dfrac{t_{0}}{M\epsilon}md_{t}V^j\left(t\right).
\end{equation}
Hence,  we need to scale the right term in
\eqref{Langevin_equation_nonscaled_2}. Let us denote by
$\gamma_s$, $\bar{\sigma}$, and $M_S$, the nondimensionalized
friction, diffusion coefficient, and mass, respectively, such that
$$
\gamma=\bar{\gamma}\gamma_{s}; \quad \sigma=\sigma_{s}\bar{\sigma}
\quad m= M m_s.
$$
From the Stokes-Einstein relation and \eqref{eq:mgamma_sigma} we
have $$
 m_{s}\gamma_{s}=\dfrac{K_BT}{DM\bar{\gamma}},
\quad  \sigma=\sqrt{2K_BTMm_{s}\bar{\gamma}\gamma_{s}}; $$ hence,
 $\bar{\sigma}=\sqrt{2K_BTM\bar{\gamma}}$.
From the definition of the rescaled potential
\eqref{eq:scaling_potential}, the electrical field becomes $$
\nabla_{x}\phi\left(t,x\right)=\dfrac{K_BT}{q}\frac{1}{\epsilon}
\nabla_{x_{s}}\phi_s\left(t_{s},x_{s}\right).
$$
The forms of the interacting term are the same as in the
dimensionalized term. Indeed,   the Pauling potential
\eqref{eq:H_paulii} becomes
$$
H_{P}\left({r}\right)=\dfrac{\left(\epsilon(r_{1,s}+r_{2,s})\right)^{10}}{\left(\epsilon
r_{s}\right)^{9}}=\dfrac{\epsilon\left(r_{1,s}+r_{2,s}\right)^{10}}{\left(r_{s}\right)^{9}}
$$
and
\begin{eqnarray*}
\nabla_{x}{ H^*_{P}\left[\nu_X(t)+\nu_Y\right](x)} &=&
\nabla_{x_{s}}\left(\sum_{k=1}^{N}\dfrac{\left(r_{1,s}+r_{2,s}\right)^{10}}{|x_{s}-X_{s}^{k}(t_{s})|^{9}}
+\sum_{h=1}^{J}\dfrac{\left(r_{1,s}+r_{2,s}\right)^{10}}{|x_{s}
- Y_{s}^{h}  |^9}\right)\\
&=& \nabla_{x_{s}}H^{s,*}_{P}
 \left[\nu^s_X(t)+\nu^s_Y\right](x),
\end{eqnarray*}
where, given  a distance $r_s$  in the new coordinate scale, we
have defined the scaled Pauling potential as follows:
$$
H^{s}_{P}\left(r_s\right)=\dfrac{\left(r_{1,s}+r_{2,s}\right)^{10}}{r_s^{9}}.
$$

\medskip

In a same way it is possible to scale the Lennard--Jones potential
\eqref{eq:tsJL_potential}. For the distance from the border
$d_{\partial{\Omega}}$ defined in \eqref{def:rho} we have that
\begin{equation*}\begin{split}
d_{\partial{\Omega}}\left(\epsilon
x_s\right)&=\min_{\epsilon~y_{s}\in{\partial{\Omega}}}{|\epsilon~x_{s}-\epsilon~y_{s}|} \\
&= \epsilon~\min_{y_{s}\in{\partial{\Omega_{s}}}}{|x_{s}-y_{s}|}\\
&=\epsilon~d_{\Omega_{s}}^{s}\left(x_{s}\right),
\end{split}\end{equation*}
where $d_{\Omega_{s}}^{s}\left(x_{s}\right)$ is the distance of
the point $x_s$ from the membrane boundary in the new scale. The
potential $\widetilde{H}_{LJ}$ is scaled as follows:
\begin{equation*}\begin{split}
\widetilde{H}_{LJ}\left(\epsilon r_{s}\right) &= \epsilon\varepsilon_{LJ}^{s}\left[\left(\dfrac{2R_{vdW}}{\epsilon r_{s}}\right)^{12}-2\left(\dfrac{2R_{vdW}}{\epsilon r_{s}}\right)^{6}\right] \\
&= \epsilon\varepsilon_{LJ}^{s}
\left[\left(\dfrac{2R^{s}_{vdW}}{r_{s}}\right)^{12}-2\left(\dfrac{2R^{s}_{vdW}}{r_{s}}\right)^{6}\right]
\end{split}\end{equation*}
where   $\varepsilon_{LJ}^{s}$ is the nondimensionalization of the
size of   $\varepsilon_{LJ}$ which depends on length, and
$R_{vdW}^{s}$ is the scaled van der Waals radii. If we define the
rescaled truncated shifted Lennard--Jones potential as
$$
\widetilde{H}_{LJ}^{s}\left(x_{s}\right)=\varepsilon_{LJ}^{s}
\left[\left(\dfrac{2R^{s}_{vdW}}{x_{s}}\right)^{12}-2\left(\dfrac{2R^{s}_{vdW}}{x_{s}}\right)^{6}\right],
$$
it follows that
\begin{equation*}
\widetilde{H}_{LJ}\left(x\right)=\epsilon
\widetilde{H}_{LJ}^{s}\left(x_{s}\right).
\end{equation*}

For the random term, we rescaled the  Wiener process $W^j_t$ and
obtained $W^j_t=W^j_{t_0 t_s}\sim \sqrt{t_0} W^j_{t_s}$, where $
W^j_{t_s}$ is a Wiener process with respect to time $t_s$.

\medskip

Finally, by gathering together all the previous terms, from
\eqref{eq:rescaling_velocity_2} and
\eqref{Langevin_equation_nonscaled_2} we have the following:
\begin{eqnarray*}
m_{s}d_{t_{s}}V^j_{s}\left(t_{t_{s}}\right) &=&
-\dfrac{t_{0}^{2}}{M\epsilon}
\left[Mm_{s}\bar{\gamma}\gamma_{s}\dfrac{\epsilon}{t_{0}}V^j_{s}(t_{s}) +z~q\dfrac{K_BT}{\epsilon q}\nabla_{x_{s}}\Phi_{s}\left(t_{s},X^j_{s}\right)\right. \\
&&\hspace{1.2cm}  +F_{I}\nabla_{x_{s}}H_{P}^{s,*} \left[\nu_X^{s}(t_{s})+\nu_Y^{s}\right](X^j(t_s))  \\
&&\hspace{1.2cm}  +F_{I}\nabla_{x_{s}}H_{LJ}^{s}\left(d_{\partial{\Omega_{s}}}^{s}(X^j(t_s))\right) \bigg] dt_{s} \\
&&+ \dfrac{t_{0}\sqrt{t_{0}}}{M\epsilon}
\sqrt{2K_BTM\bar{\gamma}m_{s}\gamma_{s}}dW^j_{t_{s}}.
 \end{eqnarray*}
By considering the relation \eqref{eq:sigma_rescaled} between
$\sigma_s$ and $\gamma_s$,
\begin{eqnarray*}
m_{s}d_{t_{s}}V^j_{s}\left(t_{t_{s}}\right) &=&-\left[
{\bar{\gamma}}{t_{0}}m_{s}\gamma_{s}V^j_{s}
\left(t_{s}\right)  -\dfrac{t_{0}K_BT}{M\epsilon^{2}}\nabla_{x_{s}}\Phi_{s}\left(t_{s},X^j_{s}\right)\right. \\
&&\hspace{0.5cm}+\dfrac{F_{I}t_{0}^{2}}{M\epsilon}\left(\nabla_{x_{s}}{H_{P}^{s,*}}
\left[\nu_X^{s}(t_s)+ \nu_Y^{s}(X^j_{s}(t_s))\right] + \nabla_{x_{s}}{H_{LJ}^{s}}\left(d_{\partial{\Omega_{s}}}\right)\right)\bigg]dt_s\\
&& + \dfrac{t_{0}\sqrt{t_{0}}}{M\epsilon}\sqrt{K_BTM\bar{\gamma}
}\sigma_s dW^j_{t_s}.
\end{eqnarray*}
Thus, we have obtained the Langevin system
\eqref{eq:labgevin_rescale_1_first},
\eqref{eq:labgevin_rescale_2_first}, and
\eqref{eq:rescaled_parameter}.

%\newpage

\end{document}